\documentclass[a4paper,11pt]{article}
\pdfoutput=1 

\usepackage{jcappub} 
\usepackage[toc,page]{appendix}
\usepackage{subcaption}
\usepackage{amsmath}

\usepackage{graphicx}                
\usepackage{multirow}
\usepackage[T1]{fontenc} 
\usepackage[table]{xcolor}
\usepackage{float}
\def\noi{{\noindent}}

\definecolor{verde}{rgb}{0,0.5,0}

\def\be{\begin{equation}}
\def\ee{\end{equation}}
\def\bea{\begin{eqnarray}}
\def\eea{\end{eqnarray}}
\def\be{\begin{equation}}
\def\ee{\end{equation}}
\def\ba{\begin{eqnarray}}
\def\ea{\end{eqnarray}}

\hypersetup{pdftitle={Non-gaussianity in curved field-space inflationary models with large scalar fluctuations}
}
\usepackage{bm}

\title{\boldmath Multi-field inflation with large scalar fluctuations: non-Gaussianity and perturbativity}

\author[a,b]{Laura Iacconi}
\author[a]{and David J. Mulryne}
\affiliation{$^{a}$Astronomy Unit, Queen Mary University of London, \\Mile End Road, London, E1 4NS, UK}
\affiliation{$^{b}$Institute of Cosmology \& Gravitation, University of Portsmouth, \\Burnaby Road, Portsmouth, PO1 3FX, UK}

\emailAdd{l.iacconi@qmul.ac.uk}
\emailAdd{d.mulryne@qmul.ac.uk}

\abstract{Recently multi-field inflation models that can produce large scalar fluctuations on small scales have drawn a lot of attention, primarily because they could lead to primordial black hole production and generation of large second-order gravitational waves. In this work, we focus on models where the scalar fields responsible for inflation live on a hyperbolic field space. In this case, geometrical destabilisation and non-geodesic motion are responsible for the peak in the scalar power spectrum. We present new results for scalar non-Gaussianity and discuss its dependence on the model's parameters. On scales around the peak, we typically find that the non-Gaussianity is large and close to local in form. We validate our results by employing two different numerical techniques, utilising the transport approach, based on full cosmological perturbation theory, and the $\delta N$ formalism, based on the separate universe approximation. We discuss implications of our results for the perturbativity of the underlying theory, focusing in particular on versions of these models with potentially relevant phenomenology at interferometer scales.}

\begin{document}
	\maketitle
	\flushbottom
	
\section{Introduction}
\label{sec: intro}
\noi Cosmological inflation has become the leading paradigm for describing the very early universe. Not only does it solve the main issues connected to the standard Hot Big Bang cosmology, but also explains the origin of the large-scale structure in the universe. 
The strongest bounds on inflation come from large-scale observations of the cosmic microwave background (CMB) \cite{Planck:2018jri}, and up to now they are consistent with the simplest inflationary scenario, single-field slow-roll (SFSR) inflation. In this case, a single, canonical scalar field slowly rolling down its own potential produces the background accelerated expansion and seeds primordial scalar perturbations that are almost scale invariant and approximately Gaussian. 

Large-scale probes test the inflaton evolution when the CMB scales crossed the horizon, e.g. approximately 50-60 e-folds before the end of inflation. Constraints on the rest of observable inflation, i.e. on the small-scale statistics of the scalar perturbations, are much looser and deviations from the simple SFSR large-scale behavior are possible.
The scalar power spectrum could, for example, strongly deviate from approximate scale invariance and exhibit a large peak on scales smaller than those probed in the CMB. The interest in this class of inflationary models has gained a lot of momentum in recent years, as enhanced scalar perturbations could lead to the production of black holes of primordial origin (PBHs) \cite{10.1093/mnras/168.2.399} (see e.g. the reviews \cite{Sasaki:2018dmp, Carr:2020gox}), which could possibly explain a fraction/the totality of dark matter \cite{Bird:2016dcv,Bertone:2018krk, Bartolo:2018evs}. A large peak in the scalar power spectrum would also lead to enhanced gravitational waves (GWs) sourced at second order in perturbation theory \cite{Ananda:2006af, Baumann:2007zm}, even in the absence of PBH formation. The detection and characterisation of this cosmological signal with current and future GWs observatories would provide us with a unique window on the very early universe (see e.g. \cite{Saito:2008em,Saito:2009jt, Fumagalli:2021cel, Witkowski:2021raz, Braglia:2020taf}), especially concerning the portion of observable inflation that is beyond large-scale probes.

Usually, the production of a peak in the scalar power spectrum is associated with departures from slow-roll \cite{Motohashi:2017kbs}. In the context of single-field inflation this can be achieved when the field slows down rapidly on its potential, for example because an inflection point leads to a phase of ultra-slow roll \cite{Garcia-Bellido:2017mdw, Germani:2017bcs, Ballesteros:2017fsr, Cicoli:2018asa, Dalianis:2018frf, Passaglia:2018ixg, Bhaumik:2019tvl, Balaji:2022rsy, Ragavendra:2023ret}\footnote{See \cite{Geller:2022nkr, Qin:2023lgo} for a multi-field model that effectively yields ultra-slow-roll, single-field behavior.}. 

In this work we focus on an alternative mechanism, based on the (likely) possibility that many fields were playing a role during inflation \cite{Baumann:2014nda}. In this case, geometrical effects and non-geodesic motion could be responsible for enhanced scalar fluctuations \cite{Fumagalli:2020adf, Palma:2020ejf, Braglia:2020eai}. Recently \cite{Iacconi:2021ltm, Kallosh:2022vha, Braglia:2022phb} this possibility has been investigated in the context of $\alpha$-attractor models of inflation \cite{Kallosh:2013hoa, Kallosh:2013daa, Ferrara:2013rsa, Kallosh:2013pby, Kallosh:2013lkr,Kallosh:2013maa, Kallosh:2013tua, Kallosh:2013yoa}. Excellent agreement with current CMB constraints, large-scale predictions that are independent of the specific shape of the inflaton potential, and possible embeddings of these models in supergravity theories are all features that make $\alpha$-attractors very compelling models for inflation. 

Cosmological $\alpha$-attractors are usually formulated in terms of a complex field, $Z$, living on the Poincaré hyperbolic disc \cite{Kallosh:2015zsa, Carrasco:2015uma}, with potential $V(Z,\, \bar Z)$ regular everywhere on the disk. The complex field $Z$ can be parametrised in terms of two scalar fields, the radial and angular fields $r$ and $\theta$ as
\begin{equation}
    \label{Z}
    Z\equiv r \,\text{e}^{i\theta} \equiv \tanh{\left(\frac{\phi}{\sqrt{6\alpha}}\right)} \text{e}^{i\theta} \;,
\end{equation}
where in the last step we have transformed $r$ into the canonical radial field $\phi$. The fields $\phi$ and $\theta$ live on a hyperbolic field space, with field-space curvature 
\begin{equation}
    \label{curvature alpha}
    \mathcal{R}_\text{fs}=-\frac{4}{3\alpha} \;.
\end{equation}
Usually $\theta$ is strongly stabilised during inflation, in which case $\phi$ plays the role of the inflaton in a single-field version of these models \cite{Carrasco:2015uma}. Importantly, the transformation to the canonical radial field $\phi$ renders the original potential a generic function of $\tanh{\left({\phi}/{\sqrt{6\alpha}}\right)}$. This provides a natural mechanism for producing a single-field potential with a plateau region (at $\phi\gg \sqrt{\alpha}$ in this case), which could sustain slow-roll evolution, from a generic potential $V(r,\,\theta)$. In addition, this leads to \textit{universal} predictions for the large-scale observables \cite{Kallosh:2013hoa, Galante:2014ifa, Fumagalli:2016sof}, see below eqs.\eqref{exponential univ}-\eqref{polynomial univ}, meaning that they are stable against different choices of the functional form of the potential. 

The explicit form of the large-scale universal predictions depend on the behavior of the derivative of $V(r,\,\theta)$ at the boundary of the hyperbolic disk: when the potential and it's derivative are not singular at the boundary of the hyperbolic disk we classify the resulting $\alpha$-attractors as \textit{exponential}\footnote{The adjectives \textit{exponential} and \textit{polynomial} refer to the way in which the plateau in the potential is approached at large values of the canonical field.} models, which include the well known E- and T-models \cite{Kallosh:2015zsa}, while \textit{polynomial} $\alpha$-attractors admit potentials with a singular derivative at the boundary of the hyperbolic disk \cite{Kallosh:2022feu}. There are many possible realisations of polynomial models \cite{Kallosh:2022feu}, but in this work we focus on the particular one which yields the following potential written in terms of the canonical variable\footnote{Note that the polynomial $\alpha$-attractor potentials (and their derivative) are not sigular when expressed in terms of the canonical variable.}
\begin{equation}
\label{polynomial we use}
    V(\phi) = V_0 \frac{\phi^2}{\phi^2+{\phi_0}^2}\;,
\end{equation}
where the parameter $\phi_0$ describes the polynomial approach to the plateau at large field values, $V(\phi)\simeq \left(1-{\phi_0}^2/\phi^2 +\cdots \right)$.  

The large-scale predicitions for the scalar spectral tilt, $n_s-1$, and
the tensor-to-scalar ratio, $r_\text{CMB}$, given at leading order in $\left(\Delta N_\text{CMB}\right)^{-1}$, are 
\begin{align}
\label{exponential univ}
&\text{exponential}\; \alpha\text{-attractors:}\quad n_s = 1-\frac{2}{\Delta N_\text{CMB}}\;, \quad r_\text{CMB}=\frac{12 \alpha}{{\Delta N_\text{CMB}}^2} \;, \\
\label{polynomial univ}
&\text{polynomial}\; \alpha\text{-attractor \eqref{polynomial we use}:}\quad n_s = 1-\frac{3}{2 \Delta N_\text{CMB}}   \;,\quad  r_\text{CMB}= \frac{\sqrt{2}\phi_0}{\Delta N_\text{CMB}^{3/2}}\;,
\end{align}
where $\Delta N_\text{CMB}$ is the number of e-folds elapsed between the horizon crossing of the CMB scale to the end of inflation. Polynomial models lead to slightly higher $n_s$ values with respect to exponential $\alpha$-attractors, and the predictions for the two classes of models (when including all types of polynomial attractors) cover almost completely the $68\%$ C.L. area in the $(n_s,\, r)$ plane singled out by the results of the \textit{Planck} mission \cite{Kallosh:2022feu}. 

For some models the angular field $\theta$ is also dynamical during inflation, and the full multi-field evolution has to be taken into account \cite{Brown:2017osf,Mizuno:2017idt,Achucarro:2017ing, Linde:2018hmx,Christodoulidis:2018qdw, Iacconi:2021ltm, Kallosh:2022vha}. For models with strongly-curved field spaces, the background trajectory deviates from geodesic motion, the isocuravature perturbation experiences a transient tachyonic instability and, due to the turning trajectory, sources a peak in the scalar power spectrum. This has been studied in the context of exponential $\alpha$-attractors in \cite{Iacconi:2021ltm} and of polynomial $\alpha$-attractors in \cite{Braglia:2020eai, Kallosh:2022vha}. In \cite{Iacconi:2021ltm} it has been shown that exponential $\alpha$-attractor models with a strongly-curved field space, $\alpha\ll1$, can lead to enhanced scalar fluctuations on small scales. A first connection between the model of \cite{Braglia:2020eai} and $\alpha$-attractors was made in \cite{Iacconi:2021ltm}, where the authors explicitly show the correspondence between polar and planar coordinates on a 2D-hyperbolic field space, where the former are usually employed in the context of $\alpha$-attractors and the latter in the model of \cite{Braglia:2020eai}. In particular, the kinetic Lagrangian (given explicitly in eq.\eqref{action}) is the same as for an $\alpha$-attractor model provided $b_1=\sqrt{2/(3\alpha)}$ and using appropriate coordinate transformations. This connection was further formalised in \cite{Kallosh:2022vha}, where the radial field potential is identified as that of a polynomial $\alpha$-attractor and a supergravity realisation of the model is given. For appropriate choices of the geometrical parameter $b_1$, these models can deliver a large peak in the scalar power spectrum \cite{Braglia:2020eai,Kallosh:2022vha}. 

Due to the presence of a peak in the scalar power spectrum, the large-scale universal predictions for the tilt, eqs.\eqref{exponential univ}-\eqref{polynomial univ}, are modified \cite{Iacconi:2021ltm, Kallosh:2022vha} such that the scalar spectral tilt becomes
\begin{align}
\label{exponential modified}
&\text{exponential}\; \alpha\text{-attractors:}\quad n_s \approx  1-\frac{2}{ \left(\Delta N_\text{CMB}-\Delta N_\text{peak}\right)}\;, \\
\label{polynomial modified}
&\text{polynomial}\; \alpha\text{-attractor \eqref{polynomial we use}:}\quad n_s \approx 1-\frac{3}{2 \left(\Delta N_\text{CMB}-\Delta N_\text{peak}\right)} \;,
\end{align}
where $\Delta N_\text{peak}$ is the e-folding separation between the horizon crossing of the peak scale in the scalar power spectrum and the end of inflation. Requiring that the scalar spectral tilt is consistent with the current large-scale measurement \cite{Planck:2018jri}
\begin{equation}
    \label{Planck ns 95}
    n_s = 0.9649 \pm 0.0042 \quad (68\%\;\text{C.L.})\,,
\end{equation}
in turn constrains the position of the peak in the scalar power spectrum \cite{Iacconi:2021ltm}. Assuming $H\simeq \text{const}$ during inflation, the peak scale can be expressed as 
\begin{equation}
    \label{k peak}
    k_\text{peak}\simeq \text{e}^{\Delta N_\text{CMB}-\Delta N_\text{peak}}\times 0.05\,\text{Mpc}^{-1} \;.
\end{equation} 
By requiring eqs.\eqref{exponential modified}-\eqref{polynomial modified} to yield values compatible at least at $95\%$ C.L. with \eqref{Planck ns 95}, and employing eq.\eqref{k peak} yields 
\begin{align}
\label{k peak exponential}
&\text{exponential}\; \alpha\text{-attractors:}\quad k_\text{peak}\gtrsim 4.6\times 10^{18}\,\text{Mpc}^{-1}\;, \\
\label{k peak polynomial}
&\text{polynomial}\; \alpha\text{-attractor \eqref{polynomial we use}:}\quad k_\text{peak}\gtrsim 4.7\times 10^{13}\,\text{Mpc}^{-1} \;.
\end{align}
Due to the constraint in eq.\eqref{k peak exponential}, exponential models can only lead to the formation of PBHs with very light masses and production of second-order GWs that peak at ultra-high frequencies, beyond the reach of current and planned GWs observatories \cite{Iacconi:2021ltm}. On the other hand, peaks on slightly larger scales are allowed for the polynomial model, see \eqref{k peak polynomial}, whose phenomenology could potentially be tested by upcoming GWs obvservations at interferometer scales (see e.g. \cite{Braglia:2020eai,Braglia:2020taf}). For this reason in the following we will focus on the polynomial $\alpha$-attractor model of \cite{Braglia:2020eai, Kallosh:2022vha}. 

Strong enhancements of the scalar perturbations on small scales might be expected to be associated with significant non-Gaussianity, which can impact both the production of PBHs \cite{Bullock:1996at,Yokoyama:1998pt,  Saito:2008em, Byrnes:2012yx, Young:2013oia,Bugaev:2013vba, Young:2014oea, Young:2015cyn, Franciolini:2018vbk, Atal:2018neu, DeLuca:2019qsy, Passaglia:2018ixg,Ozsoy:2021qrg, Taoso:2021uvl, Davies:2021loj, Ferrante:2022mui, Gow:2022jfb} and induced second-order GWs \cite{Domenech:2021ztg, Cai:2018dig, Unal:2018yaa, Yuan:2020iwf, Atal:2021jyo, Adshead:2021hnm, Ragavendra:2020sop, Garcia-Saenz:2022tzu}. Large non-Gaussianities also imply that the tree-level scalar power spectrum itself, i.e. the power spectrum calculated using the linear equations of motion for the perturbations, 
could get large corrections when expanding the Lagrangian to higher-order in the fluctuations, with the risk of the theory becoming non-perturbative. For these reasons, in this work we calculate the scalar 3-point correlation function, the bispectrum, for the models of \cite{Braglia:2020eai, Kallosh:2022vha}, and evaluate the amplitude of non-Gaussianity in models with large peaks in the scalar power spectrum. We then employ our results as a diagnostic tool to assess  perturbativity\footnote{Alternatively one could calculate the loop-corrections to the tree-level scalar power spectrum using the In-In formalism. This method can be employed for cases where the non-Gaussianity is not of the local type. Loop corrections in single-field models delivering large scalar fluctuations have been calculated recently \cite{Inomata:2022yte,Kristiano:2022maq,Riotto:2023hoz, Choudhury:2023vuj,
Choudhury:2023jlt, Kristiano:2023scm, Riotto:2023gpm, Firouzjahi:2023aum, Motohashi:2023syh, Choudhury:2023rks, Choudhury:2023hvf, Firouzjahi:2023ahg, Firouzjahi:2023btw} (for an earlier work see \cite{Cheng:2021lif}), while to our knowledge this attempt has not been done for multi-field models.} of the underlying theory for scales around the peak, focusing in particular on versions of these models with potentially relevant phenomenology at interferometer scales. 

\medskip
\textit{Content:} We summarise general features of multi-field inflation in section \ref{subsec: multi field warm up}. In section \ref{subsec: model and background} we focus on the polynomial $\alpha$-attractor of \cite{Braglia:2020eai, Kallosh:2022vha} and introduce three realisations of it, distinguished by different values of the field-space curvature, which will be our working examples. In section \ref{sec: non Gaussianities PT} we calculate the amplitude of the scalar 3-point correlation function by employing the numerical code \texttt{PyTransport} \cite{Mulryne:2016mzv, Ronayne:2017qzn}, based on the transport equations for the correlators which follow from full cosmological perturbation theory. In section \ref{sec: delta N}, we develop a second numerical tool, based on a numerical realisation of the $\delta N$ formalism, to cross check the \texttt{PyTransport} results for the scalar power spectrum and bispectrum amplitudes. In section \ref{sec: perturbativity} we explore further the details of the dependence of non-Gaussianity on the field-space geometry and discuss the perturbativity of these models. We comment on our results in section \ref{sec: discussion} and provide additional materials in a series of appendices. In particular, in appendices \ref{app: vary chi in} and \ref{app: Pz fNL many b1} we systematically analyse the dependence of the scalar power spectrum and bispectrum amplitudes on the model parameter space, namely variations in the value of the initial condition for the second field and in the field-space curvature. Finally, in appendix \ref{app: analytic 2pt corr} we provide analytic expressions for the fields and velocities 2-point correlators at horizon crossing, required to develop the numerical $\delta N$ approach of section \ref{sec: delta N}. 

\medskip
\textit{Conventions:} Throughout this work, we consider a spatially-flat Friedmann--Lema\^{i}tre--Robertson--Walker universe, with line element $\text{d}s^2=-\text{d}t^2+a^2(t)\delta_{ij}\text{d}x^i\text{d}x^j$, where $t$ denotes cosmic time and $a(t)$ is the scale factor. The Hubble rate is defined as $H\equiv {\dot a}/{a}$,  where a derivative with respect to cosmic time is denoted by $\dot f \equiv {\mathrm{d}f}/{\mathrm{d}t}$. The number of e-folds of expansion is defined as $N\equiv \int\,H(t)\mathrm{d}t$ and $f'\equiv {\mathrm{d}f}/{\mathrm{d}N}$. We use natural units and set the reduced Planck mass, $M_\text{Pl}\equiv(8\pi G_N)^{-1/2}$, to unity unless otherwise stated.

\section{Calculating non-Gaussianities: the transport approach}
\label{sec: non-gaussianities}

\subsection{Multi-field inflation warm up}
\label{subsec: multi field warm up}
The action of multi-field inflationary models can be written as
\begin{equation}
\label{multi field generic action}
    \mathcal{S}=\int \mathrm{d}^4x \sqrt{-g} \left[\frac{{M_{Pl}}^2}{2}R -\frac{1}{2} \mathcal{G}_{IJ}\left(\phi^K \right)\partial_\mu \phi^I \partial^\mu \phi^J -V\left(\phi^K \right)\right] \;,
\end{equation}
where $\mathcal{G}_{IJ}\left(\phi^K\right)$ is the metric on the field space and $V\left(\phi^K\right)$ is the multi-field potential. In this work we focus on a two-field model. We therefore  restrict the number of fields described by the action \eqref{multi field generic action} to two, and consider a FRLW universe. The fields and background evolution are described by the equations 
\begin{gather}
\label{H eq multi field t}
3 H^2=\frac{1}{2} \dot\sigma^2 +V \;,\\
\label{Hdot eq multi field t}
\dot{H}=-\frac{1}{2}  \dot{\sigma}^2\;, \\
\label{fields eq multi field t}
\mathcal{D}_t \dot{\phi^I} +3H \dot{\phi^I} +\mathcal{G}^{IJ} V_{,J}=0 \;,
\end{gather}
where $V_{,J}\equiv {\mathrm{d}V}/{\mathrm{d}\phi^J}$, $\dot{\sigma}^2 \equiv \mathcal{G}_{IJ}\dot{\phi}^I\dot{\phi}^J$ is the kinetic energy of the fields, $\mathcal{D}_t A^I=\dot{A^I} +\Gamma^I_{JK} \dot\phi^{J} A^K$, and $\Gamma^I_{JK}$ are the Christoffel symbols on the field space. The first Hubble slow-roll parameter is defined as 
\begin{equation}
    \label{epsilon H}
    \epsilon_H \equiv -\frac{\dot H}{H^2}\;, 
\end{equation}
and inflation ends when $\epsilon_H=1$.

When studying the perturbations around the inflating background, it is useful to project the covariant perturbation in the spatially-flat gauge \cite{Gong:2011uw}, $\mathcal{Q}^I$, along the instantaneous adiabatic and entropic directions \cite{Gordon:2000hv,GrootNibbelink:2001qt}. The adiabatic direction is (instantly) coincident with the field-space background trajectory direction, while the entropic direction is orthogonal to it. More precisely, the new basis is defined by the unit-norm vectors $\hat{\sigma}^I \equiv \dot{\phi}^I/\dot{\sigma}$ and $\hat{s}^I\equiv {\omega^I}/{\omega}$, where $\omega$ is the turn rate in field space and is defined by $\omega^I\equiv \mathcal{D}_t\hat{\sigma}^I$. In the following, we will also refer to the dimensionless bending parameter  
\begin{equation}
    \label{eta perp}
    \eta_\perp\equiv \frac{\omega}{H} \;,
\end{equation}
which measures the deviation of the background trajectory from a geodesic in field space. 

The adiabatic and entropic perturbations are defined respectively as $\mathcal{Q}_\sigma\equiv \hat{\sigma}_I \mathcal{Q}^I$ and $\mathcal{Q}_s\equiv \hat{s}_I \mathcal{Q}^I$, and from these the dimensionless comoving curvature and entropic perturbations are given by
\begin{equation}
\label{adiabatic and entropic perturbations}
    \zeta\equiv \frac{H}{\dot{\sigma}} \mathcal{Q}_\sigma, \;\;\;\;\; \mathcal{S}\equiv \frac{H}{\dot{\sigma}} \mathcal{Q}_s \;.
\end{equation}
The 2- and 3-point correlation functions for the curvature perturbation $\zeta$ are given by 
\begin{align}
    \label{2pt zeta}
    \langle \zeta(\mathbf{k_1}) \zeta(\mathbf{k_2}) \rangle &\equiv (2\pi)^3 \delta \left( \mathbf{k_1} + \mathbf{k_2}\right)P_\zeta(k) \;,\\
    \label{3pt zeta}
    \langle \zeta(\mathbf{k_1}) \zeta(\mathbf{k_2})  \zeta(\mathbf{k_3})\rangle &\equiv (2\pi)^3 \delta \left( \mathbf{k_1} + \mathbf{k_2}+\mathbf{k_3}\right){B}_\zeta(k_1,\, k_2,\,k_3) \;.
\end{align}
From these we can define the dimensionless scalar power spectrum 
\begin{equation}
\label{Pzeta definition}
    \mathcal{P}_\zeta(k) \equiv \frac{k^3}{2\pi^2} {P}_\zeta(k) \;,
\end{equation}
and the reduced bispectrum
\begin{equation}
    \label{fNL defintion}
    f_\text{NL} \equiv \frac{5}{6} \frac{{B}_\zeta(k_1, \, k_2,\, k_3)}{{P}(k_1){P}(k_2) + {P}(k_2){P}(k_3) + {P}(k_1){P}(k_3)} \;.
\end{equation}
In multi-field inflation, the adiabatic and entropic perturbations \eqref{adiabatic and entropic perturbations} are coupled in the presence of a non-zero bending of the field-space trajectory, i.e. non-geodesic motion in field-space \cite{Gordon:2000hv}. For example, in the super-horizon regime $(k\ll aH)$ the curvature perturbation is not necessarily constant as in single-field inflation, and obeys (see e.g. \cite{Renaux-Petel:2015mga})
\begin{equation}
    \dot{\zeta} \simeq 2\eta_\perp \frac{H^2}{\dot{\sigma}}\mathcal{Q}_s
    \;.
\end{equation}
In this regime, the entropic perturbation $\mathcal{Q}_s$ evolves according to 
\begin{equation}
    \Ddot{\mathcal{Q}}_s+3H\dot{\mathcal{Q}}_s+{m_{s,\, \text{eff}}}^2 \mathcal{Q}_s \simeq 0 \;,
\end{equation}
where the entropic effective squared-mass in the super-horizon regime is
\begin{equation}
\label{eff mass isocurvature}
    \frac{ {m_{s,\,\text{eff}}}^2}{H^2}\equiv \frac{V_{;ss}}{H^2}+\epsilon_H \mathcal{R}_{\text{fs}}+3\eta_\perp^2 \;.
\end{equation}
In the equation above, the first term represents the Hessian of the multi-field potential, $V_{;ss}\equiv \hat{s}^I \hat{s}^J \left( V_{,IJ}-\Gamma^{K}_{IJ}V_{,K}\right)$, defined by means of a covariant derivative in field space in order to take into account the non-trivial geometry. 
The second term is proportional to $\mathcal{R}_{\text{fs}}$, the intrinsic scalar curvature of the field space, and the last term to the bending of the field-space trajectory, see eq.\eqref{eta perp}. 

In the models under analysis, the field space is hyperbolic, $\mathcal{R}_{\text{fs}}<0$ (for more on geometrical destabilisation see also \cite{Brown:2017osf, Mizuno:2017idt, Turzynski:2014tza,Renaux-Petel:2015mga, Renaux-Petel:2017dia,Garcia-Saenz:2018ifx,Garcia-Saenz:2018vqf, Grocholski:2019mot}). As demonstrated in \cite{Braglia:2020eai}, for suitable choices of the model's parameters the combination $\epsilon_H \mathcal{R}_{\text{fs}}$ is large enough to overcome the other contributions to eq.\eqref{eff mass isocurvature} and ${m_{s,\,\text{eff}}}^2/H^2<0$ transiently. The entropic perturbation becomes tachyonic, grows and sources the curvature perturbation due to the turning trajectory. This produces a peak in the scalar power spectrum \eqref{Pzeta definition} on scales smaller than the CMB scales, which could lead to primordial black hole production and to large GWs produced at second-order in perturbation theory \cite{Braglia:2020eai}. 

\subsection{The model and its background evolution}
\label{subsec: model and background}
Specialising to the model in \cite{Braglia:2020eai}, the multi-field action \eqref{multi field generic action} becomes
\begin{equation}
    \label{action}
    \mathcal{S} = \int \mathrm{d}^4x\, \sqrt{-g} \left[ \frac{{M_{Pl}}^2}{2}R -\frac{1}{2}\left( \partial \phi\right)^2 -\frac{1}{2}e^{2b_1\phi}\left(\partial \chi \right)^2 -V\left(\phi, \,\chi \right) \right] \;.
\end{equation}
Here $\{\phi,\,\chi\}$ are planar\footnote{See Appendix E in \cite{Iacconi:2021ltm} for a discussion about planar and polar coordinates in 2D-hyperbolic field spaces.} coordinates in the field space, the field-space curvature is $\mathcal{R}_\text{fs}=-2{b_1}^2$ and 
the only non-zero Christoffel symbols are 
\begin{equation}
    \label{christoffel}
    \Gamma^\phi_{\chi \chi} = -b_1 e^{2b_1\phi} \quad \text{and} \quad \Gamma^\chi_{\chi \phi} = b_1 \;.
\end{equation}
Following \cite{Braglia:2020eai}, we study the multi-field potential
\begin{equation}
    \label{potential}
    V(\phi, \, \chi) = V_0 \frac{\phi^2}{{\phi_0}^2 + \phi^2} + \frac{1}{2} {m_\chi}^2 \chi^2 \;,
\end{equation}
where we recognise the polynomial $\alpha$-attractor potential of eq.\eqref{polynomial we use} for the canonical scalar field $\phi$ \cite{Kallosh:2022vha}. Similarly to \cite{Braglia:2020eai} we dub the second field $\chi$, but we could have equally called it $\theta$, as we did in section \ref{sec: intro}. In eq.\eqref{potential}, we fix $\phi_0 = \sqrt{6}$ and ${m_\chi}^2 = V_0/500$ \cite{Braglia:2020eai}. 

For appropriate choices of the model's parameters, the background evolution of these models is characterised by two phases. The first is driven by the field $\phi$, which slowly rolls down its potential while $\chi$ is effectively frozen due to the effect of the hyperbolic field-space geometry. When ${\phi\sim b_1}^{-1}$, the suppression is lifted, there is a turn in field space, and the second field starts rolling towards the minimum of its potential, up until the end of inflation \cite{Braglia:2020eai}. At the transition between the two phases, $\epsilon_H\sim1$ and the (negative) term $\epsilon_H \mathcal{R}_\text{fs}$ in eq.\eqref{eff mass isocurvature} causes a transient tachyonic instability for the entropic perturbation, which, as we have described, in turn sources a large $\mathcal{P}_\zeta(k)$ on small scales \cite{Braglia:2020eai}. 

In each model realisation, two parameters are particularly relevant, the value of $b_1$, which determines the field-space curvature and therefore (roughly) the growth of $\mathcal{P}_\zeta(k)$, and the initial condition on the second field, $\chi_\text{in}$, which sets the duration of the second phase of evolution and therefore the position of the peak in the scalar power spectrum \cite{Braglia:2020eai}. 

In section \ref{sec: non Gaussianities PT} we investigate the effect that changes in $b_1$ have on the scalar power spectrum and non-Gaussianity, and do a similar analysis in appendix \ref{app: vary chi in} for models where we vary $\chi_\text{in}$. First, we present the background evolution for three working examples that we use throughout, namely three models with $\{\phi_\text{in}=7,\, \chi_\text{in}=7.31\}$ and different values of the geometrical parameter, $b_1=\{6.4,\, 7.091,\,7.8\}$. 
\begin{figure}
    \centering
    \includegraphics[width = 0.7\textwidth]{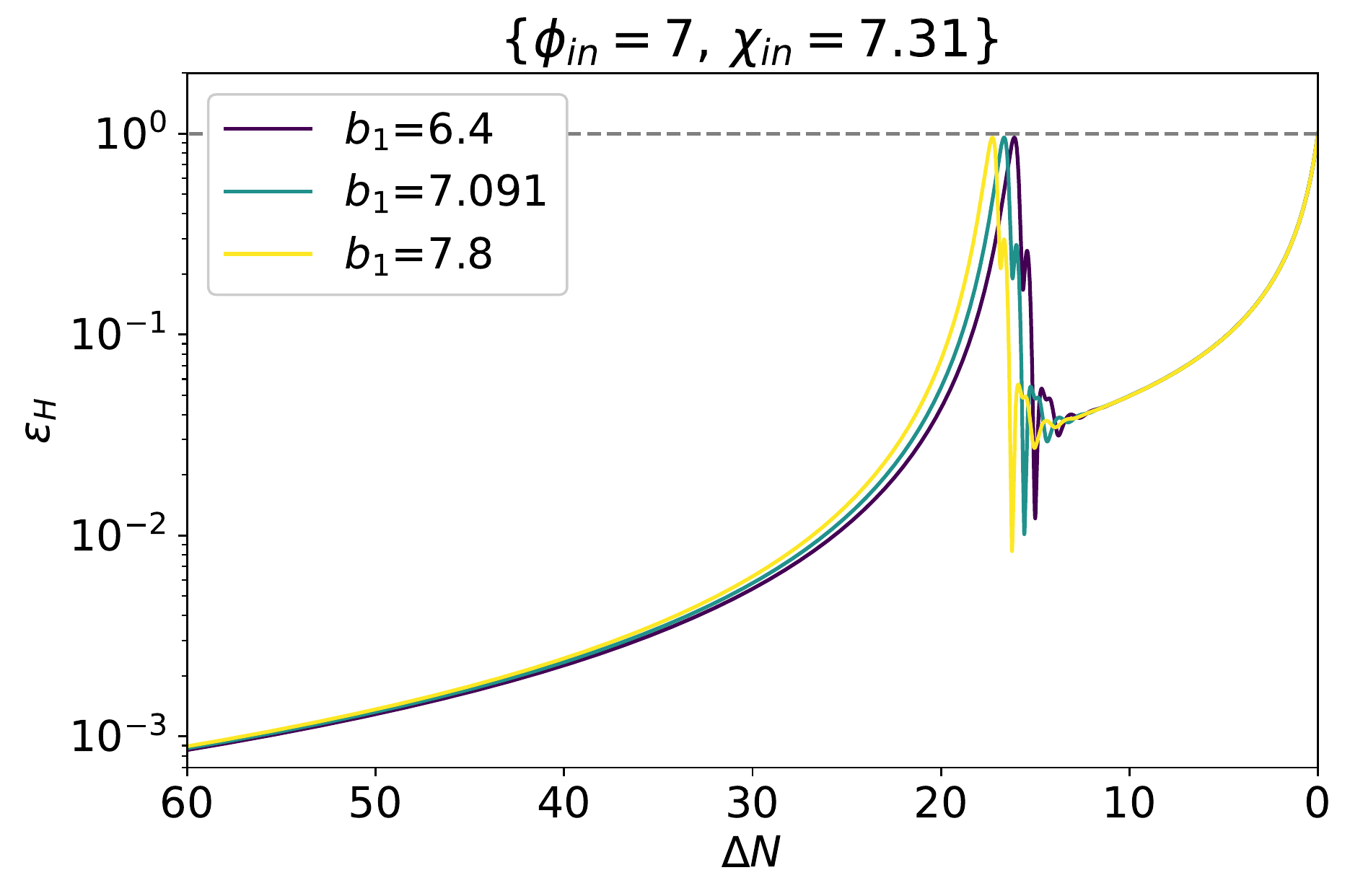}
    \caption{Numerical evolution of $\epsilon_H$ against $\Delta N\equiv N_\text{end}-N$ for three models with initial conditions $\{\phi_\text{in}=7,\, \chi_\text{in}=7.31\}$ and different values of the geometrical parameter $b_1$.}
    \label{fig:background b1}
\end{figure}
We plot in figure \ref{fig:background b1} the resulting evolution of $\epsilon_H$, eq.\eqref{epsilon H}, represented against $\Delta N \equiv N_\text{end}-N$. In all cases, the time-dependence of $\epsilon_H$ shows two phases of evolution, separated by a transition with $\epsilon_H\sim1$. 

\subsection{Scalar non-Gaussianities}
\label{sec: non Gaussianities PT}
In this section, we calculate the scalar power spectrum and the primordial scalar non-Gaussianity, parametrised by the reduced bispectrum \eqref{fNL defintion}, for the working examples introduced in section \ref{subsec: model and background}. The numerical results presented in this section have been obtained using the publicly available code \texttt{PyTransport} \cite{Dias:2016rjq,Mulryne:2016mzv, Ronayne:2017qzn}\footnote{See \cite{Mulryne:2013uka,Seery:2012vj,Mulryne:2010rp,Mulryne:2009kh,Dias:2015rca} for earlier related work, and \cite{Seery:2016lko,Butchers:2018hds} for an alternative open source package, \texttt{CppTransport},  based on the same approach. \texttt{PyTransport} is available at \href{https://github.com/jronayne/PyTransport}{github.com/jronayne/PyTransport}.}. The transport code 
works by evolving the 2- and 3-point function of 
covariant field fluctuations and their conjugate momenta from initial conditions set in the quantum regime on sub-horizon scales. It then converts these correlations into the power spectrum and bispectrum of $\zeta$ \cite{Dias:2014msa}.

An important quantity needed to connect inflationary dynamics with large-scale observables is the number of e-folds elapsed between the horizon crossing of the CMB scale, $k_\text{CMB}=0.05\,\text{Mpc}^{-1}$, and the end of inflation, defined as \cite{Planck:2018jri}
\begin{equation}
\begin{split}
\label{Nstar}
    \Delta N_\text{CMB} 
    &\equiv N_\text{end}-N_\text{CMB} \\
    &\simeq 66.4-\ln{\left(\frac{k_\text{CMB}}{a_0 H_0} \right)}+\frac{1}{4}\ln{\left(\frac{V_\text{CMB}^2}{ \rho_\text{end}} \right) }-\frac{1-3w}{4} \Delta \tilde N_\text{rh} \,.  
\end{split}
\end{equation}
In this expression $a_0H_0$
is the present comoving Hubble rate, $\rho_\text{end}$ is the energy density at the end of inflation, $V_\text{CMB}$ is the value of the potential when the comoving wavenumber $k_\text{CMB}$ crossed the horizon during inflation. The parameters $w$ and $\Delta \tilde N_\text{rh}$ respectively represent the equation of state parameter during reheating and its duration
\begin{equation}
    \label{def rh}
    \Delta \tilde N_\text{rh} \equiv N_\text{rh}-N_\text{end}= \frac{1}{3(1+w)} \log{\left( \frac{\rho_\text{end}}{\rho_\text{th}}\right)} \;,
\end{equation}
where $\rho_\text{th}$ is the energy scale at the end of it. For reheating to be completed before the onset of the Big Bang Nucleosynthesis \cite{Planck:2018jri}, its duration is bounded from above:
\begin{equation}
    \label{max rh}
    \Delta \tilde N_\text{rh}\leq \Delta \tilde N_\text{rh,max} =\frac{1}{3} \log{\frac{\rho_\text{end}}{\left(1\text{TeV} \right)^4}}\;.
\end{equation}
When the equation-of-state parameter for reheating\footnote{For simplicity in the following we assume that $w=0$.} is $-1<w<1/3$, $\Delta N_\text{CMB}$ is maximised for instantaneous reheating,  $\rho_\text{end}=\rho_\text{th}$ or $\Delta \tilde N_\text{rh}=0$. By assuming instant reheating and values of $V_0$ compatible with CMB observations, we iteratively solve eq.\eqref{Nstar} for models with $\phi_\text{in}=7$ and $\chi_\text{in}=7.31$, finding $\Delta N_\text{CMB,inst rh}\simeq 57.3$, regardless of $b_1$. We also find that the maximum duration of reheating is $\Delta \tilde N_\text{rh,max}\simeq 37.8$. These quantities, together with the e-folding separation between the horizon crossing of the peak scale and the end of inflation, $\Delta N_\text{peak}$, determine the value of the scalar spectral tilt on large scales, see eq.\eqref{polynomial modified}. We approximate $\Delta N_\text{peak}$ with the duration of the second phase of evolution, i.e. the number of e-folds that elapsed between the local maximum of $\epsilon_H$ and the end of inflation. We find that all the models we study are compatible at least at $95\%$ C.L. with the large-scale measurement \eqref{Planck ns 95} for appropriate choices of $\Delta \tilde N_\text{rh}$. For example, $\Delta \tilde N_\text{rh}\lesssim22$ yields $n_s\gtrsim0.9565$ for the model with $b_1=7.8$. We have also shown that for these models to be consistent with \eqref{Planck ns 95} at least at $95\%$ C.L., the inequality \eqref{k peak polynomial} follows, implying that viable models that satisfy $n_s$ constraints must lead to peaks in the scalar power spectrum beyond LISA scales, e.g.  on scales relevant to LIGO or the Einstein telescope. 

In \cite{Braglia:2020eai} the authors fix $\Delta N_\text{CMB}=50$, which would follow from $\Delta \tilde N_\text{rh}\simeq 29.2$. This value of $\Delta \tilde N_\text{rh}$, while being allowed by the upper bound $\Delta \tilde N_\text{rh,max}$, yields values of the scalar spectral tilt not compatible with \eqref{Planck ns 95}. Nevertheless, in the following we will adopt $\Delta N_\text{CMB}=50$ for simplicity and consistency with \cite{Braglia:2020eai}. This still serves as a useful benchmark to illustrate our conclusions regarding non-Gaussianity for these models, bearing in mind that while this choice leads to tensions with the $n_s$ measurements, there are appropriate choices of reheating duration (and therefore $\Delta N_\text{CMB}$) that make these models compatible with the \textit{Planck} measurement \eqref{Planck ns 95}, as explained above. 

\begin{figure}
\centering
\captionsetup[subfigure]{justification=centering}
\begin{subfigure}[b]{0.48\textwidth}
\includegraphics[width=\textwidth]{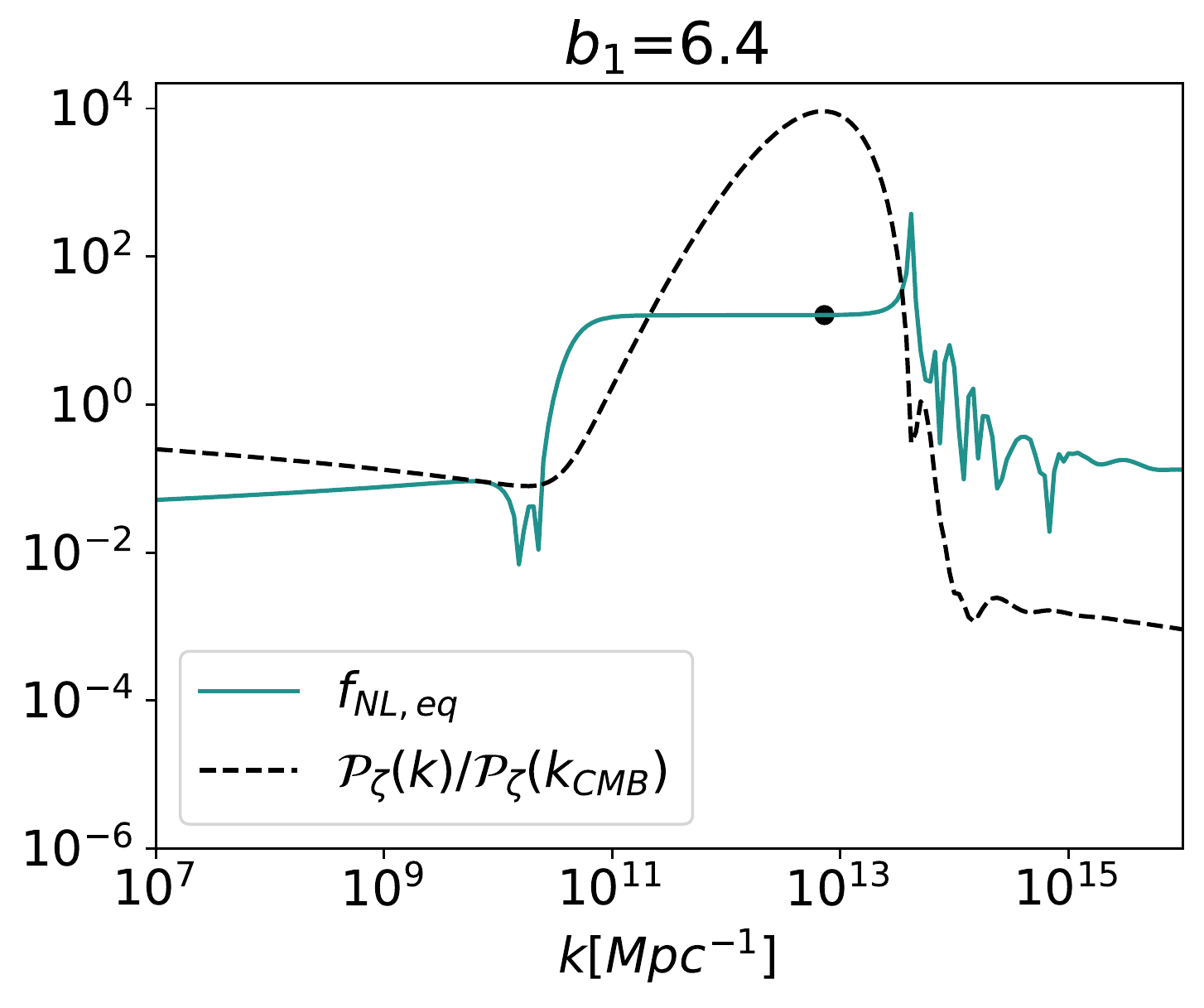}
\end{subfigure}
\begin{subfigure}[b]{0.48\textwidth}
\includegraphics[width=\textwidth]{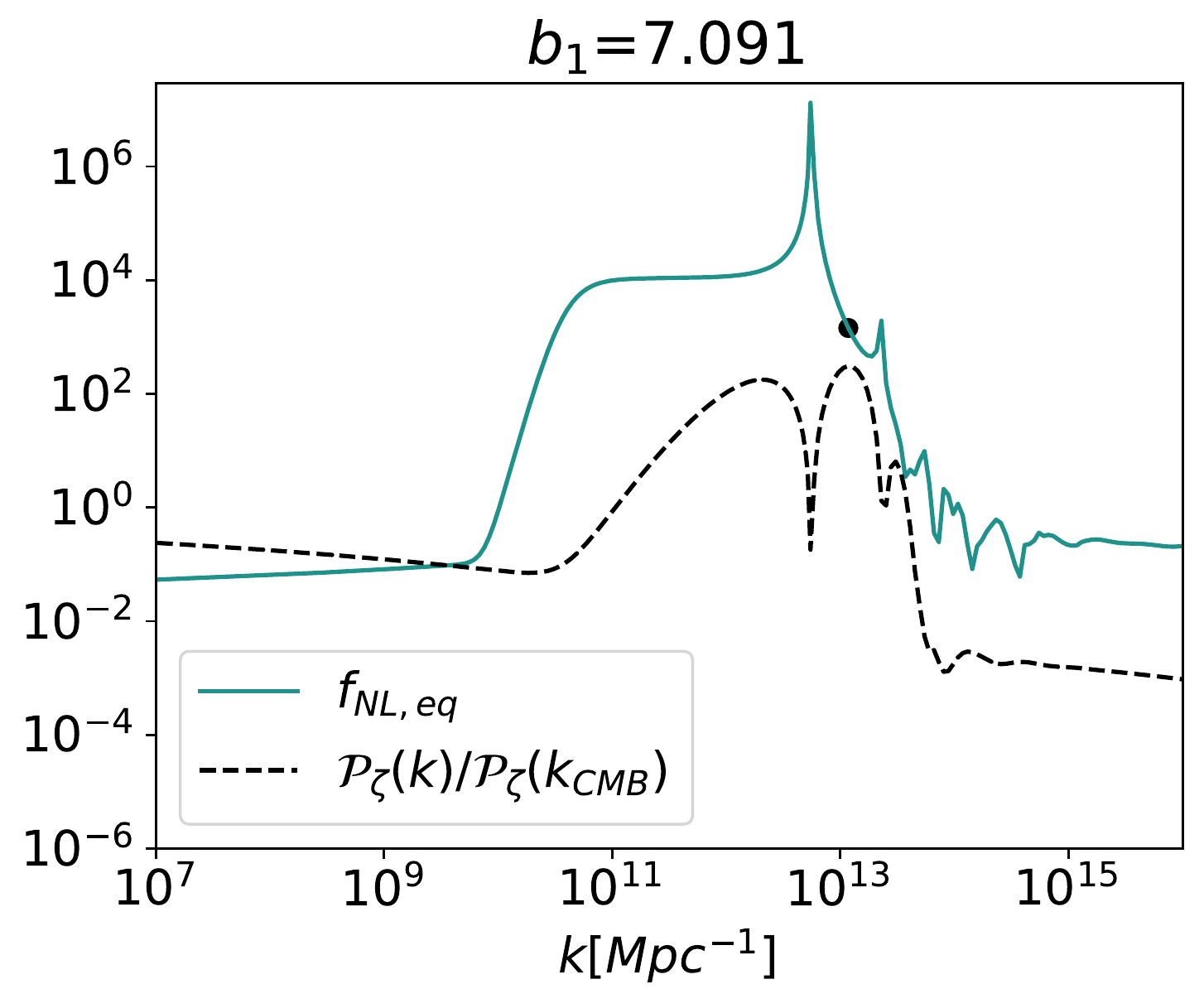}
\end{subfigure}
\begin{subfigure}[b]{0.48\textwidth}
\includegraphics[width=\textwidth]{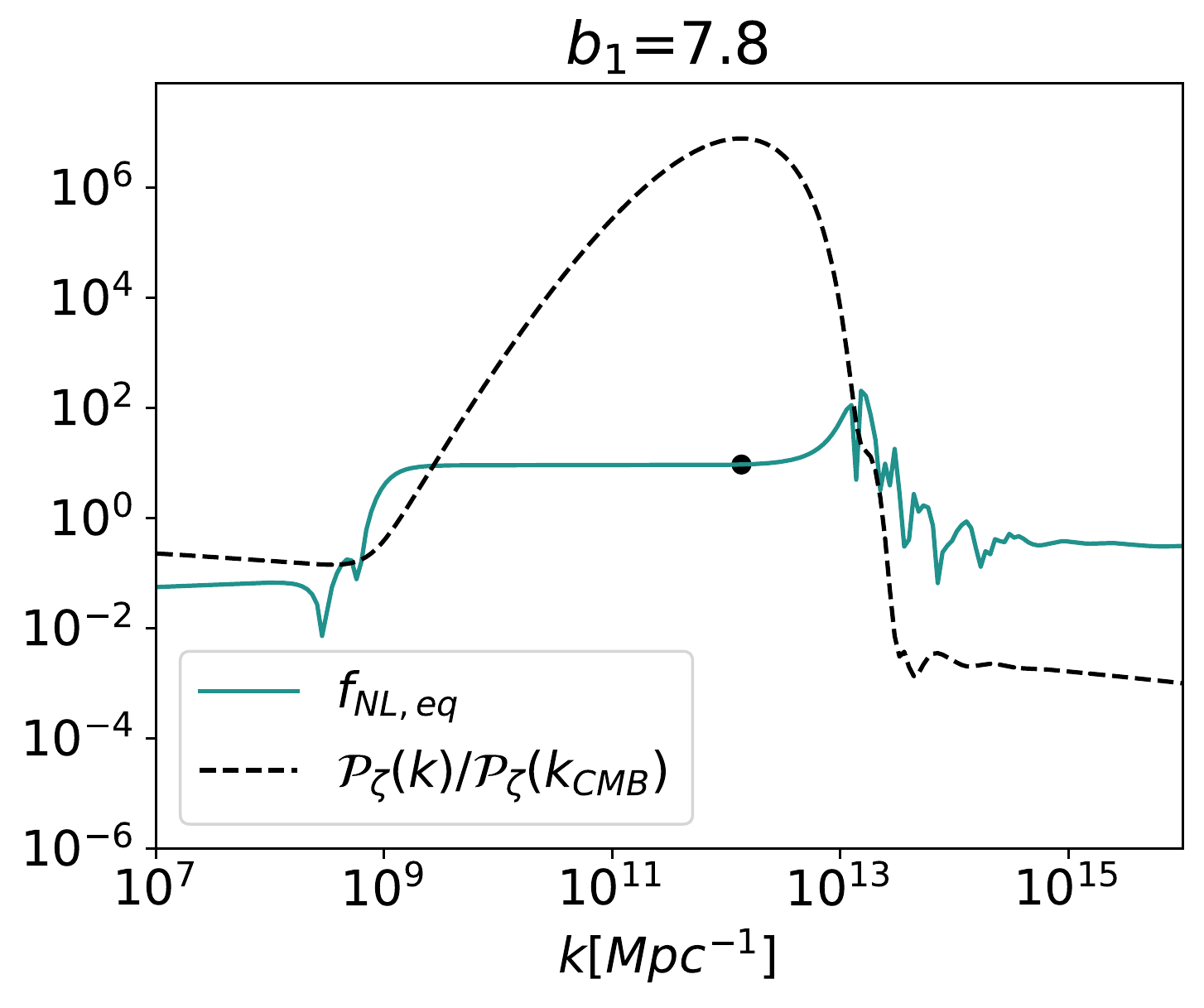}
\end{subfigure}
\caption{Scalar power spectrum, $\mathcal{P}_\zeta(k)/\mathcal{P}_\zeta(k_\text{CMB})$, and amplitude of the equilateral non-Gaussianity, $f_\text{NL,eq}$, for three models with initial conditions $\{\phi_\text{in}=7,\, \chi_\text{in}=7.31\}$ and different values of the geometrical parameter $b_1$. The horizontal axis has been cut to represent the peak region. 
The value of $f_\text{NL,eq}$ for the peak scale is highlighted with a black point.} 
\label{fig:fNL equilateral for diff curvature}
\end{figure}
We display in figure \ref{fig:fNL equilateral for diff curvature} results for $\mathcal{P}_\zeta(k)/\mathcal{P}_\zeta(k_\text{CMB})$ and the reduced bispectrum in the equilateral configuration, $k_1=k_2=k_3=k$, at scales around the peak region for the three models introduced in section \ref{subsec: model and background}. 

The enhancement of $\mathcal{P}_\zeta(k)$ relative to the CMB scale is $\mathcal{P}_\zeta(k_\text{peak})/ \mathcal{P}_\zeta(k_\text{CMB})\simeq \{9.2\times 10^3,\, 3\times10^2,\, 7.9\times10^6\}$ for each $b_1$ in increasing order. Interestingly the peak amplitude does not always increase for larger $b_1$, i.e. more strongly-curved field spaces. We confirm this by systematically exploring more $b_1$ values in the range $b_1\in[4,\,8]$, see appendix \ref{app: Pz fNL many b1}. In particular, figure \ref{fig:Pzeta kpeak vs b1} shows that for $b_1\gtrsim6.8$ the peak amplitude starts decreasing up until the critical value $b_1\sim7.09$, after which $\mathcal{P}_\zeta(k_\text{peak})$ increases again for increasing $b_1$. The presence of a critical value for the curvature divides the $b_1$ range into two regions, which motivates our choice of the three cases to work with.

In figure \ref{fig:fNL equilateral for diff curvature}, we see that for $b_1=6.4$ and $7.8$, $f_\text{NL,eq}$ is approximately flat, i.e. scale-independent, over the peak scales. However, for $b_1=7.091$, which is close to the critical value, the scalar power spectrum exhibits a two-peak structure with the second peak being the largest, and the non-Gaussianity at $k_\text{peak}$ is in a region of rapidly changing $f_\text{NL,eq}$, outside of the plateau. As demonstrated in appendix \ref{app: Pz fNL many b1} this is the exception, and all models with non-critical field-space curvature display a plateau in $f_\text{NL,eq}$. We also find that the amplitude of the equilateral non-Gaussianity at the peak scale is non-monotonic for increasing $b_1$, with $f_\text{NL,eq}(k_\text{peak})\simeq \{16.2,\, 1444.6,\, 9.4\}$ for each $b_1$ in increasing order. 

Excluding the critical value $b_1=7.091$, the approximate scale-independence of the equilateral non-Gaussianity amplitude over peak scales points to non-Gaussianity of the local shape. We confirm this for the case $b_1=7.8$ by explicitly looking at the shape dependence of $f_\text{NL}$ at fixed overall scale $k_1+k_2+k_3=3k_\text{peak}$. 
\begin{figure}
    \centering
    \includegraphics[width = 0.7\textwidth]{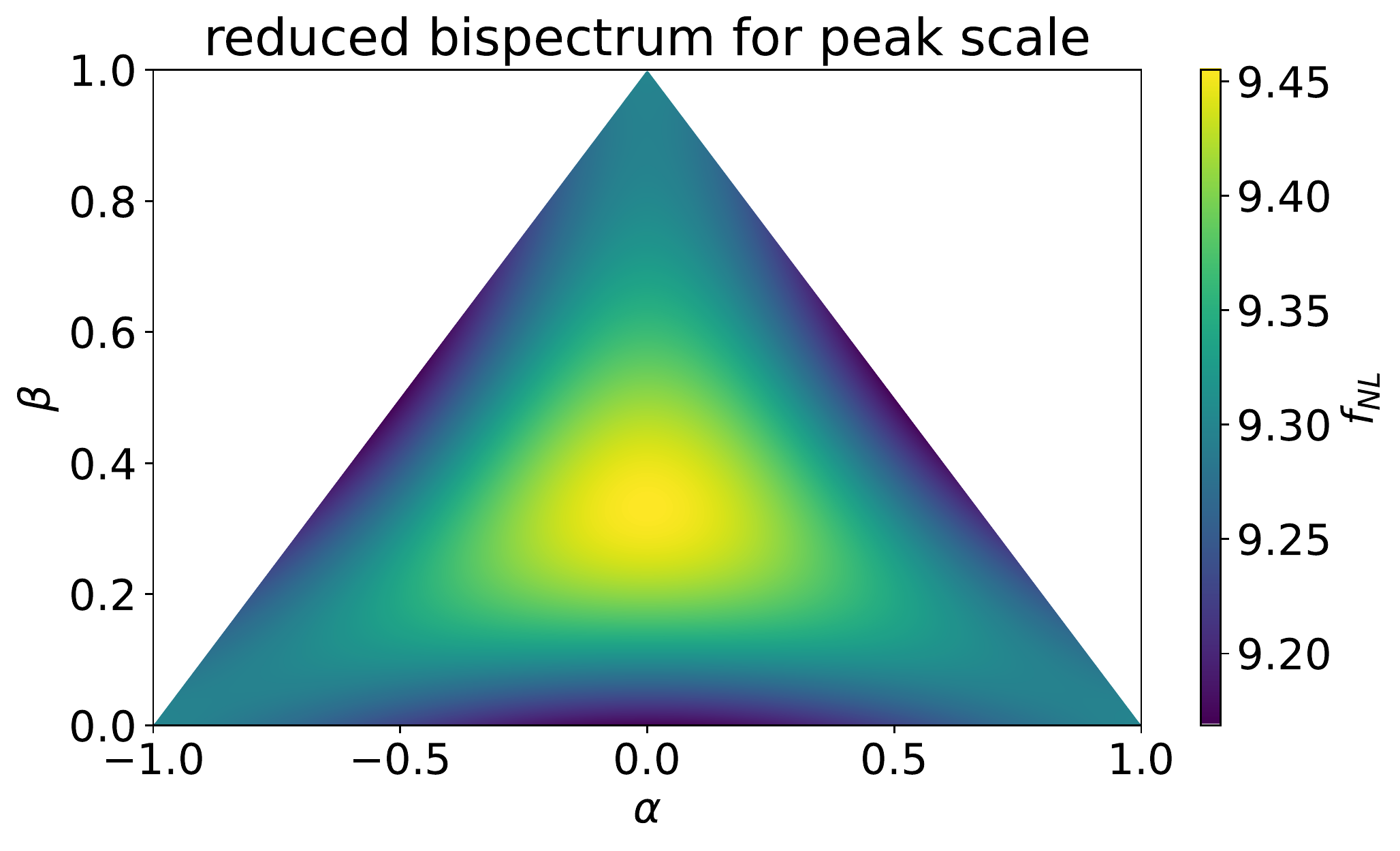}
    \caption{Reduced bispectrum $f_\text{NL}(k_1,\,k_2,\,k_3)$ as a function of $(\alpha,\,\beta)$, see the text for their definitions. Note that the squeezed configuration corresponds to the points $(\alpha, \,\beta)=\{(-1,0),\,(1,0),\,(0,1)\}$, while the equilateral configuration corresponds to $(\alpha, \,\beta)=(0,1/3)$.}
    \label{fig:triangle plot}
\end{figure}
In figure \ref{fig:triangle plot} we display the reduced bispectrum $f_\text{NL}(k_1,\,k_2,\,k_3)$ as a function of $(\alpha,\,\beta)$, defined as $k_1=3k_\text{peak}(1-\beta)/2$, $k_2=3k_\text{peak}(1+\alpha+\beta)/4$ and $k_3=3k_\text{peak}(1-\alpha+\beta)/4$. The range of values of the bispectrum shows that $f_\text{NL}$ is almost constant for all triangle configurations, meaning that the bispectrum is scale-independent and therefore of the local shape. We have checked that as we increase the resolution of the triangle plot, i.e. the closer we get to the squeezed configuration represented by the triangle points $(\alpha, \,\beta)=\{(-1,0),\,(1,0),\,(0,1)\}$, we see a drop in the value of $f_\text{NL}$, which is expected since $f_\text{NL,eq}$ does eventually display a scale dependence at scales outside of the plateau.


\section[\texorpdfstring{Calculating non-Gaussianities: the numerical $\bm{\delta N}$ approach}{Calculating non-Gaussianities: the numerical delta N approach}]{Calculating non-Gaussianities: the numerical $\bm{\delta N}$ approach}
\label{sec: delta N}
When non-Gaussianity is local, it is highly suggestive that it has been produced on super-horizon scales. In order to test this hypothesis, and to cross check the results obtained in section \ref{sec: non-gaussianities}, we develop here a second numerical approach, based on the $\delta N$ formalism \cite{Starobinsky:1985ibc, Sasaki:1995aw, Wands:2000dp, Lyth:2004gb, Lyth:2005fi}. 

The $\delta N$ formalism is a powerful tool to compute the non-linear evolution of cosmological perturbations on large scales, $k\ll aH$. In particular, the curvature perturbation is identified with the perturbed number of e-folds of evolution between a spatially-flat hypersurface at time $t=0$ and a uniform-density hypersurface at time $t$ 
\begin{equation}
\label{delta N 1}
    \delta N (\mathbf{x},\, t) \equiv \zeta (\mathbf{x}, \, t)\;,
\end{equation}
where $N \equiv \int \mathrm{d}t\, H(t)$. By Taylor expanding eq.\eqref{delta N 1} for two fields and by retaining only linear contributions\footnote{We have checked that truncating the expansion at the linear order is sufficient to justify the results obtained in section \ref{sec: non-gaussianities}, see e.g. figure \ref{fig:delta N power spectrum}.}, we get
\begin{equation}
    \label{zeta delta N}
    \zeta(\mathbf{x},\, t) = \frac{\partial N}{\partial \phi} \delta \phi(\mathbf{x},\, t) + \frac{\partial N}{\partial \chi} \delta \chi(\mathbf{x},\, t) +  \frac{\partial N}{\partial \phi'} \delta \phi'(\mathbf{x},\, t) + \frac{\partial N}{\partial \chi'} \delta \chi'(\mathbf{x},\, t)  \;.
\end{equation}
Eq.\eqref{zeta delta N} can be used to calculate the scalar power spectrum
\begin{equation}
    \begin{split}
    \label{Pzeta delta N}
    \mathcal{P}_\zeta(k) = & \left(N_\phi\right)^2 \mathcal{P}_{\phi \phi}(k) + N_\phi N_\chi \mathcal{P}_{\phi\chi}(k) + N_\phi N_{\phi'} \mathcal{P}_{\phi\phi'}(k) + N_\phi N_{\chi'} \mathcal{P}_{\phi\chi'}(k) \\
    &+ N_\chi N_\phi \mathcal{P}_{\chi\phi}(k) + \left(N_\chi\right)^2 \mathcal{P}_{\chi\chi}(k) + N_\chi N_{\phi'} \mathcal{P}_{\chi\phi'}(k) + N_\chi N_{\chi'} \mathcal{P}_{\chi\chi'}(k) \\
    &+ N_{\phi'} N_\phi \mathcal{P}_{\phi'\phi}(k) + N_{\phi'} N_\chi \mathcal{P}_{\phi'\chi}(k)+ \left(N_{\phi'}\right)^2 \mathcal{P}_{\phi'\phi'}(k)+ N_{\phi'} N_{\chi'} \mathcal{P}_{\phi'\chi'}(k) \\
    &+ N_{\chi'} N_\phi \mathcal{P}_{\chi'\phi}(k) + N_{\chi'} N_\chi \mathcal{P}_{\chi'\chi}(k) + N_{\chi'} N_{\phi'} \mathcal{P}_{\chi'\phi'}(k) + \left(N_{\chi'}\right)^2 \mathcal{P}_{\chi'\chi'}(k) \;,    
    \end{split}
\end{equation}
where $\langle \delta X_\mathbf{k_1} \delta Y_\mathbf{k_2}\rangle \equiv (2\pi)^3 \delta(\mathbf{k_1+k_2}) {2\pi^2}/{k^3}\mathcal{P}_{XY}(k_1)$ and $N_X\equiv \partial N /\partial X$. For each scale the 2-point correlators appearing in eq.\eqref{Pzeta delta N} are evaluated at horizon crossing, and the $\delta N$ formalism takes into account all the subsequent (and possibly complicated) evolution to produce the curvature power spectrum. 

We apply eq.\eqref{Pzeta delta N} to calculate super-horizon contributions to the scalar power spectrum and primordial scalar non-Gaussianities in the models analysed in section \ref{sec: non-gaussianities}. For this purpose, we calculate in appendix \ref{app: analytic 2pt corr} analytic expressions for the correlators appearing in \eqref{Pzeta delta N}, which need to be evaluated at horizon crossing. In particular, we employ the mode functions for non-interacting, massless fields in a quasi-de Sitter background, see eq.\eqref{correlator}.
\begin{figure}
\centering
\includegraphics[width=1\textwidth]{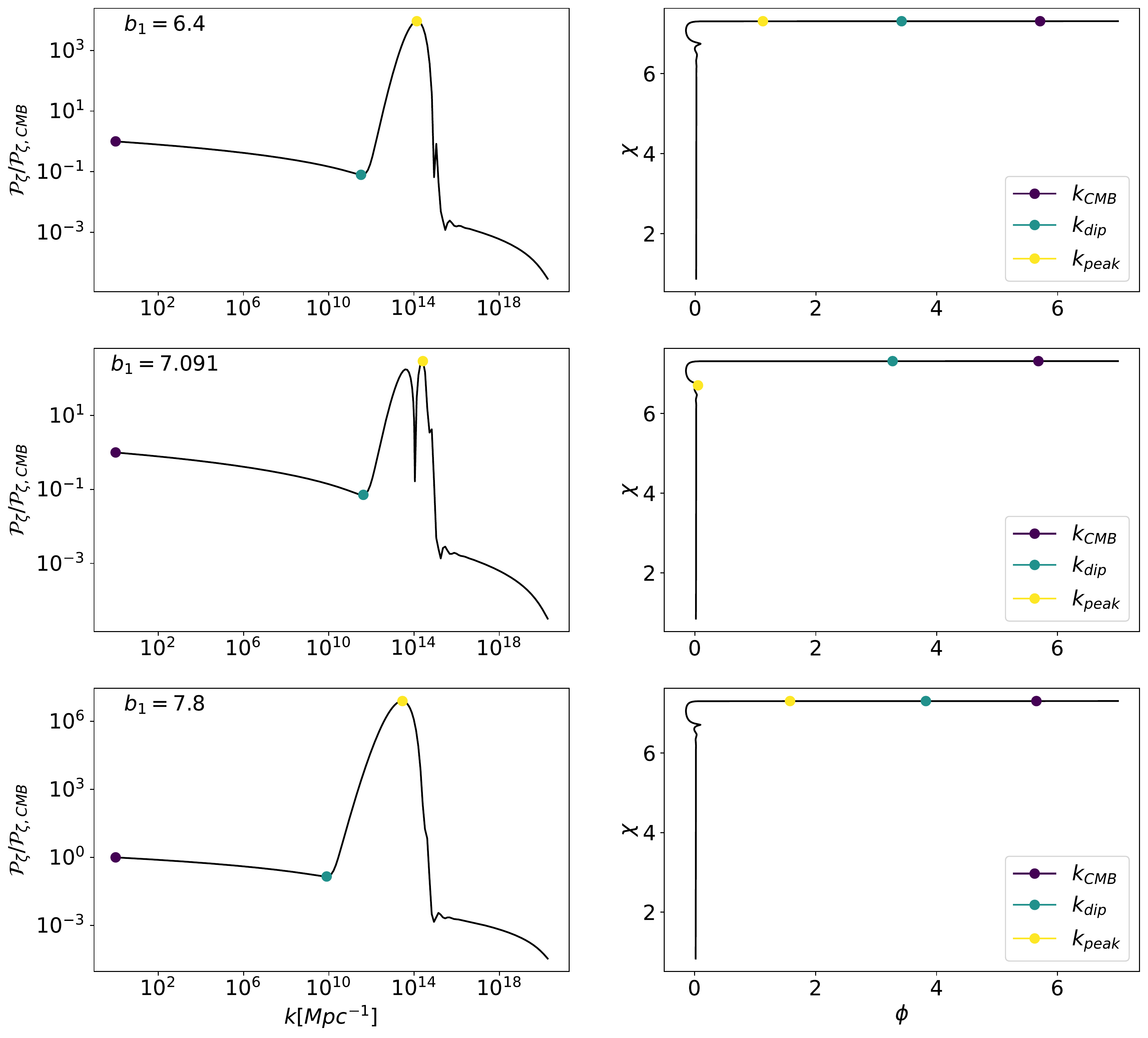}
\caption{The scalar power spectrum, $\mathcal{P}_\zeta(k)/\mathcal{P}_\zeta (k_\text{CMB})$, calculated with \texttt{PyTransport} (left) and field-space trajectory (right) for three models with $\{\phi_\text{in}=7,\, \chi_\text{in}=7.31\}$ and different values of the field-space curvature. The plots in each row correspond to the same $b_1$ value, and in each row we highlight the CMB, dip and peak scales (left) and the time during the fields' evolution when these scales cross the horizon (right).}
\label{fig:horizon_crossing}
\end{figure}
For the models with $b_1=6.4$ and $7.8$, the peak scale leaves the horizon before the turn in field space (e.g. before the isocurvature perturbation becomes unstable), see figure \ref{fig:horizon_crossing}. Here we represent on the left $\mathcal{P}_\zeta(k)/\mathcal{P}_\zeta(k_\text{CMB})$, with the CMB, dip and peak scales highlighted, and mark on the right the times at which these scales cross the horizon on top of the fields' evolution. In both cases, the peak scale leaves the horizon before the field-space trajectory turns and for this reason we expect our $\delta N$ approach, that relies on the correlator \eqref{correlator}, to well describe the perturbations in the peak region. When $b_1=7.091$, the scalar power spectrum exhibits a double peak, with the second peak larger than the first one and the peak scale exiting the horizon during the transition. For critical curvature values, $b_1\sim 7.09$, we therefore expect sub-horizon effects to be relevant. This implies that correlators at horizon crossing receive corrections from sub-horizon interactions, and using the correlators for non-interacting, massless fields on de Sitter, eq.\eqref{correlator}, is only appropriate as a first approximation. In this (exceptional) case we therefore expect our $\delta N$ approach not to fully account for the perturbations evolution. 

By using the results of appendix \ref{app: analytic 2pt corr}, eq.\eqref{Pzeta delta N} reduces to 
\begin{equation}
    \begin{split}
    \label{Pzeta delta N reduced}
    \mathcal{P}_\zeta(k)  = &\left(N_\phi \right)^2 \mathcal{P}_{\phi\phi}(k) + \left(N_\chi\right)^2 \mathcal{P}_{\chi\chi}(k) + \left(N_{\phi'}\right)^2 \mathcal{P}_{\phi'\phi'}(k) + \left(N_{\chi'}\right)^2 \mathcal{P}_{\chi'\chi'}(k) \\
    & + 2N_{\phi'} N_{\chi'} \mathcal{P}_{\phi'\chi'}(k) + 2 N_\chi N_{\chi'} \mathcal{P}_{\chi\chi'}(k)  + 2 N_\phi N_{\chi'} \mathcal{P}_{\phi\chi'}(k) + 2 N_\chi N_{\phi'}\mathcal{P}_{\chi\phi'}(k) \;,    
    \end{split}
\end{equation}
where the dimensionless correlators are listed in eqs.\eqref{first final corr}-\eqref{last final corr} and need to be evaluated for each scale at horizon crossing. 

\subsection{The scalar power spectrum}
\label{sec: delta N power}
With $\Delta N_\text{CMB}=50$, we adjust $V_0$ in eq.\eqref{potential} such that the scalar power spectrum amplitude on large scales satisfies the \textit{Planck} normalisation \cite{Planck:2018jri}. We numerically solve the background evolution with initial conditions $\{\phi_\text{in}=7,\, \chi_\text{in}=7.31, \,\phi_\text{in}'=0,\, \chi_\text{in}'=0 \}$\footnote{We have chosen the initial condition $\phi_\text{in}$ such that inflation lasts longer than $\Delta N_\text{CMB}$ and the background trajectory is on the slow-roll attractor when the CMB scale crossed the horizon.} and stop the evolution 
when $\epsilon_H=1$: the resulting set of numerical solutions $\mathcal{B}=\{\phi(N),\, \chi(N),\, \phi'(N),\, \chi'(N)\}$ constitutes our reference background evolution. The energy density at the end of inflation for the solutions $\mathcal{B}$, $\bar \rho_\text{end}= 1/2\,\dot\sigma^2+V(\phi,\,\chi)|_{N=N_\text{end}}$, defines a constant-energy-density hypersurface, which will be our reference when calculating $\delta N$ \eqref{delta N 1}. 

We slice the time-evolution during the observable window of inflation by considering $n$ equally spaced e-folding numbers, $N_\star$, each of which is associated with a comoving scale, $k_\star$, that crossed the horizon at that time
\begin{equation}
\label{k of N}
    k_\star \simeq e^{N_\star-N_\text{CMB}} \times 0.05\;\text{Mpc}^{-1} \;,
\end{equation}
where we have assumed $H=\text{const}$ and $N_\text{CMB}$ is defined in eq.\eqref{Nstar}. Using $\mathcal{B}$, we evaluate for each $N_\star$ the corresponding set $\mathcal{B}_\star = 
\{\phi(N_\star),\, \chi(N_\star),\, \phi'( N_\star),\, \chi'(N_\star)\}$, which can be regarded as the set of initial conditions at the time of horizon crossing of the scale $k_\star$. For each $k_\star$, we evaluate the analytic expressions for the 2-point correlators in eq.\eqref{Pzeta delta N reduced}, see appendix \ref{app: analytic 2pt corr}, at horizon crossing by using the set $\mathcal{B}_\star$. The correlators values will be used in evaluating eq.\eqref{Pzeta delta N reduced} for each scale $k_\star$. 

We calculate the derivatives in \eqref{Pzeta delta N reduced} in the same way, so we describe in detail just one of the cases, $N_\phi$. For each time $N_\star$, we consider the reference set of initial conditions $\mathcal{B}_\star$ and vary only the initial condition for $\phi$, 
\begin{equation}
    \label{new IC}
    \mathcal{B}_\star \rightarrow \mathcal{B}_\star+\delta \mathcal{B}_\star = \{\phi(N_\star) (1+\delta),\, \chi(N_\star),\, \phi'(N_\star),\, \chi'(N_\star)\} \;.
\end{equation}
In particular, we consider 61 new initial conditions, with $\delta \in [-30\%, +30\%]$. For each one of these, which we label with the perturbed value $\phi_{\delta_i}\equiv \phi(N_\star) (1+\delta_i)$, where $i=1,\cdots,61$, we solve the background evolution up until $\rho=\bar \rho_\text{end}$ and evaluate the number of e-folds this takes, $N_{\delta_i}$. We then fit the obtained numerical data $(\phi_{\delta_i},\, N_{\delta_i})_{i=1,\cdots,61}$ with the linear function 
\begin{equation}
    \label{linear fit}
    N_{\delta} = c_0 + c_1 \phi_{\delta} \;,
\end{equation}
and identify $N_\phi=c_1$ for the scale $k_\star$. The product of the square $(N_\phi)^2$ with the value of the 2-point correlator $\mathcal{P}_{\phi\phi}$ at horizon crossing constitutes the first contribution to \eqref{Pzeta delta N reduced} for the scale $k_\star$. We repeat the same procedure for the remaining initial conditions in $\mathcal{B}_\star$ to derive $N_\chi, \, N_{\phi'}$ and $N_{\chi'}$. By using these results, the correlators at horizon crossing in eqs.\eqref{first final corr}-\eqref{last final corr}, and eq.\eqref{Pzeta delta N reduced}, we obtain $\mathcal{P}_\zeta(k_\star)$. Repeating this for all the $n$ scales yields to the numerical $\delta N$ result for the scalar power spectrum. 

\begin{figure}
    \centering
    \includegraphics[width = \linewidth]{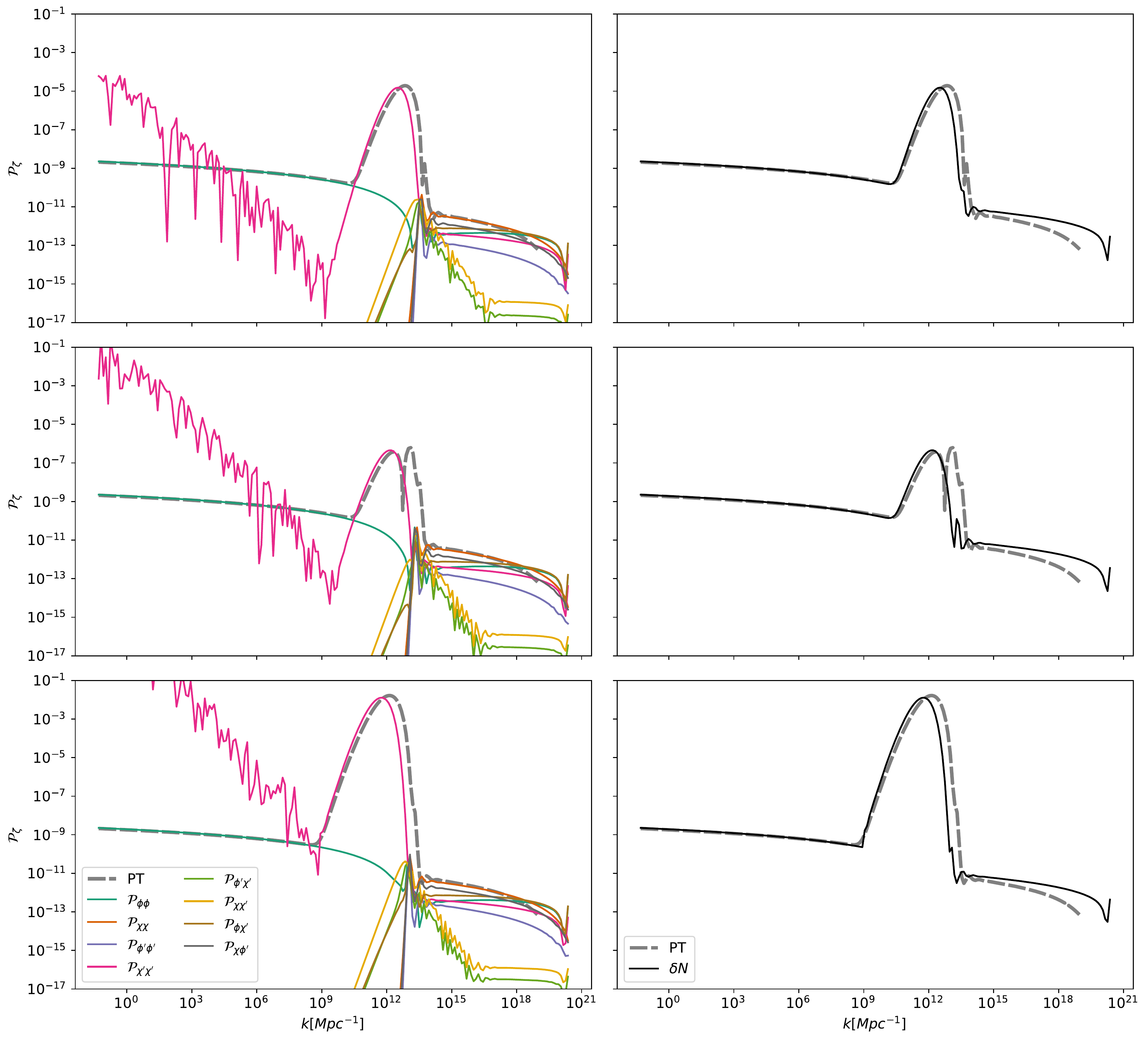}
    \caption{\textit{Left panel}: Comparison between numerical $\delta N$ results (colored lines) and the \texttt{PyTransport} scalar power spectrum (gray, dashed line) for models with $\{ \phi_\text{in}=7, \, \chi_\text{in}=7.31\}$, and different values of $b_1$. \textit{Right panel:} Total scalar power spectrum from the $\delta N$ calculation (black line), see eq.\eqref{Pzeta delta N reduced}, compared with the \texttt{PyTransport} result (gray, dashed line) for the same models. 
    }
    \label{fig:delta N power spectrum}
\end{figure}
We represent in the left panels of figure \ref{fig:delta N power spectrum} the results obtained for models with different values of $b_1$. In particular, the colored lines represent the eight contributions to the scalar power spectrum within the numerical $\delta N$ calculation, see eq.\eqref{Pzeta delta N reduced}, and the gray-dashed line displays the \texttt{PyTransport} result for comparison. 

\begin{figure}
    \centering
    \includegraphics[width = 0.7\linewidth]{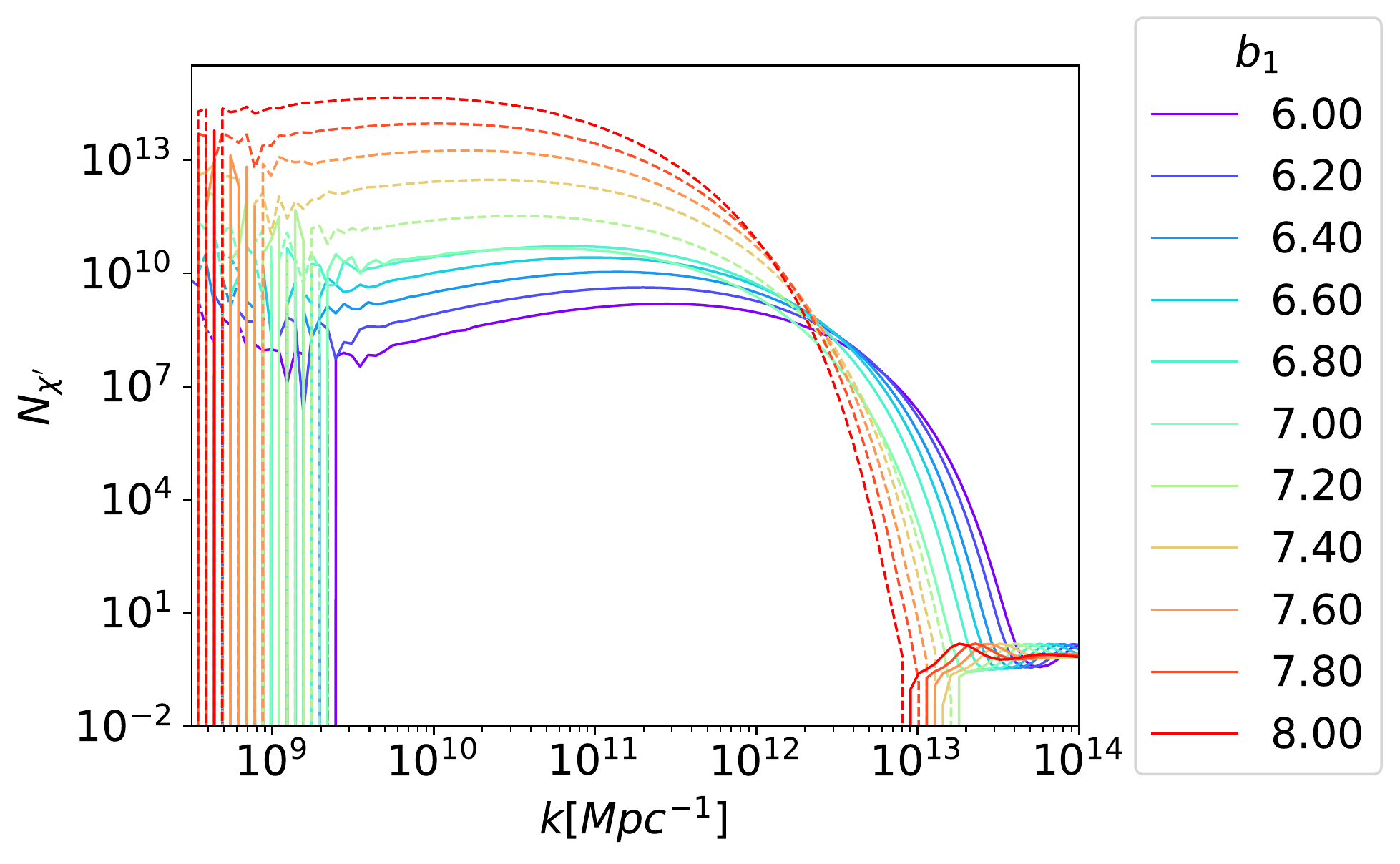}
    \caption{Values of the derivative $N_{\chi'}$ in the peak region for models with $\{ \phi_\text{in}=7, \, \chi_\text{in}=7.31\}$, and different values of $b_1$. We plot with a continuous (dashed) line positive (negative) values of $N_{\chi'}$. On large scales $N_{\chi'}$ is dominated by numerical noise.}
    \label{fig:N_chipr for vary b1}
\end{figure}

In each case the power spectrum is first dominated by perturbations in the $\phi$ field, which at these times is slowly rolling, while $\chi$ is frozen. The peak is produced by perturbations in the second field velocity, $\chi'$. On large scales, the $\chi'$ contribution is dominated by numerical noise\footnote{The unstable quantity here is $N_{\chi'}$, which rapidly oscillates between large positive and negative values, see figure \ref{fig:N_chipr for vary b1}. The information on the sign of $N_{\chi'}$ is lost in figure \ref{fig:delta N power spectrum}, due to the square taken, see eq.\eqref{Pzeta delta N reduced}.}. This is easily explained considering the fact that during the first stage of evolution $\chi$ is frozen, with $\chi'$ exponentially suppressed, and dealing with such extreme values leads to some instabilities in the numerics. The numerical $\delta N$ peak is slightly different with respect to the \texttt{PyTransport} result, especially when considering its position. This is probably due to the analytic correlators that we use to implement the $\delta N$ calculation, which rely on a series of approximations, see eq.\eqref{correlator}. 

After the peak region, the scalar power spectrum displays a second slow-roll plateau. The largest contribution at these scales is given by fluctuations in $\chi$, which is slowly rolling and dominates the second phase of evolution after the transition. Comparable to this contribution are those of mixed cross-correlators, $\mathcal{P}_{\chi\phi'}$ and $\mathcal{P}_{\phi\chi'}$. The sum of all contributions yields a plateau slightly larger than the \texttt{PyTransport} one, see the right panels in figure \ref{fig:delta N power spectrum}, where we compare the sum\footnote{On large scales we neglect the noise-dominated $\left( N_{\chi'}\right)^2\mathcal{P}_{\chi'\chi'}(k)$ contribution.} of the eight $\delta N$ contributions, see eq.\eqref{Pzeta delta N reduced}, with the \texttt{PyTransport} results. The discrepancy between the two approaches increases towards smaller scales. These differences might be explained by the fact that small scales cross the horizon towards the end of inflation, where the slow-roll approximation fails, and our initial conditions for the  $\delta N$ formalism are no longer appropriate.

While the agreement between the numerical $\delta N$ results and the output of \texttt{PyTransport} is quite remarkable for $b_1=6.4$ and $7.8$, in the case of $b_1=7.091$ the fluctuations in $\chi'$ explain only the first (secondary) peak in the scalar power spectrum. This is not surprising considering the fact that the peak scale crossed the horizon during the turn in field space, see figure \ref{fig:horizon_crossing}, signalling that sub-horizon effects (not accounted for by our analytic correlators at horizon crossing) need to be taken into account. 

As discussed in section \ref{sec: non Gaussianities PT}, the \texttt{PyTransport} results show a critical behavior in the scalar power spectrum for $b_1\simeq 7.09$, see also figure \ref{fig: Pz fNL many b1}. 
We find a sign of criticality also in quantities calculated with the $\delta N$ approach.
We display in figure \ref{fig:N_chipr for vary b1} values of $N_{\chi'}$ for scales in the peak region for different values of $b_1$. We distinguish positive (continuous) and negative (dashed) values of $N_{\chi'}$. The transition between positive and negative $N_{\chi'}$ in the peak region is located between $b_1=7$ and $b_1=7.2$, exactly where expected from the \texttt{PyTransport} results. This flags with a second independent approach the presence of critical values for the geometrical parameter around $b_1=7.09$. 

\subsection{Scalar non-Gaussianities}
\label{sec: delta N non Gauss}
The amplitude of scalar non-Gaussianities (more precisely the shape-independent part of $f_\text{NL}$) can be calculated within the $\delta N$ formalism as (see e.g. \cite{Vernizzi:2006ve, Byrnes:2008wi}) 
\begin{equation}
    \label{FNL delta N def}
    f_\text{NL} = \frac{5}{6}\frac{\sum_{IJ} N_{,IJ} N_{,I} N_{,J}}{\left( \sum_I N_{,I}^2\right)^2} \;,
\end{equation}
where $I=\{\phi, \, \chi, \, \phi',\, \chi'\}$. In section \ref{sec: delta N power} we have demonstrated that the peak in the scalar power spectrum is due to fluctuations in the second field velocity, $\chi'$. For this reason we expect the amplitude of non-Gaussianity around the scale $k_\text{peak}$ to be approximately given as
\begin{equation}
    \label{fNL delta N chi' only}
    f_\text{NL} \simeq \frac{5}{6} \frac{N_{\chi'\chi'}}{{N_{\chi'}}^2} \;.
\end{equation}
Since the numerical $\delta N$ approach does not reproduce the position of the peak in the scalar power spectrum precisely, see figure \ref{fig:delta N power spectrum}, we use the numerical results from \texttt{PyTransport} to derive the peak position, $k_\text{peak}$. We select 10 scales that crossed the horizon around the same time as $k_\text{peak}$, in the range
\begin{equation}
    \label{k min k max}
    10^{\log_{10}{k_\text{peak}-0.1}}\leq k_\star \leq 10^{\log_{10}{k_\text{peak}+0.1}} \;.
\end{equation}
For each of these scales, we use the reference background evolution to calculate the initial conditions at horizon crossing, $\mathcal{B}_\star$, and consider a small variation of the initial condition for the velocity $\chi'$, 
\begin{equation}
    \label{new IC chi}
    \mathcal{B}_\star\rightarrow \mathcal{B}_\star +\delta \mathcal{B}_\star=\{\phi(N_\star),\, \chi(N_\star),\, \phi'(N_\star) ,\, \chi'(N_\star)(1+\delta)\} \;.
\end{equation}
Following a procedure similar to what described in section \ref{sec: delta N power}, we derive the numerical data $(\chi'_{\delta_i},\, N_{\delta_i})_{i=1,\cdots,61}$ and obtain the derivatives $N_{\chi'}$ and $N_{\chi'\chi'}$ by using a linear and quadratic fit respectively to the data. We then combine these derivatives for each scale $k_\star$ as prescribed by eq.\eqref{fNL delta N chi' only}.

\begin{figure}
\centering
\includegraphics[width=0.7\textwidth]{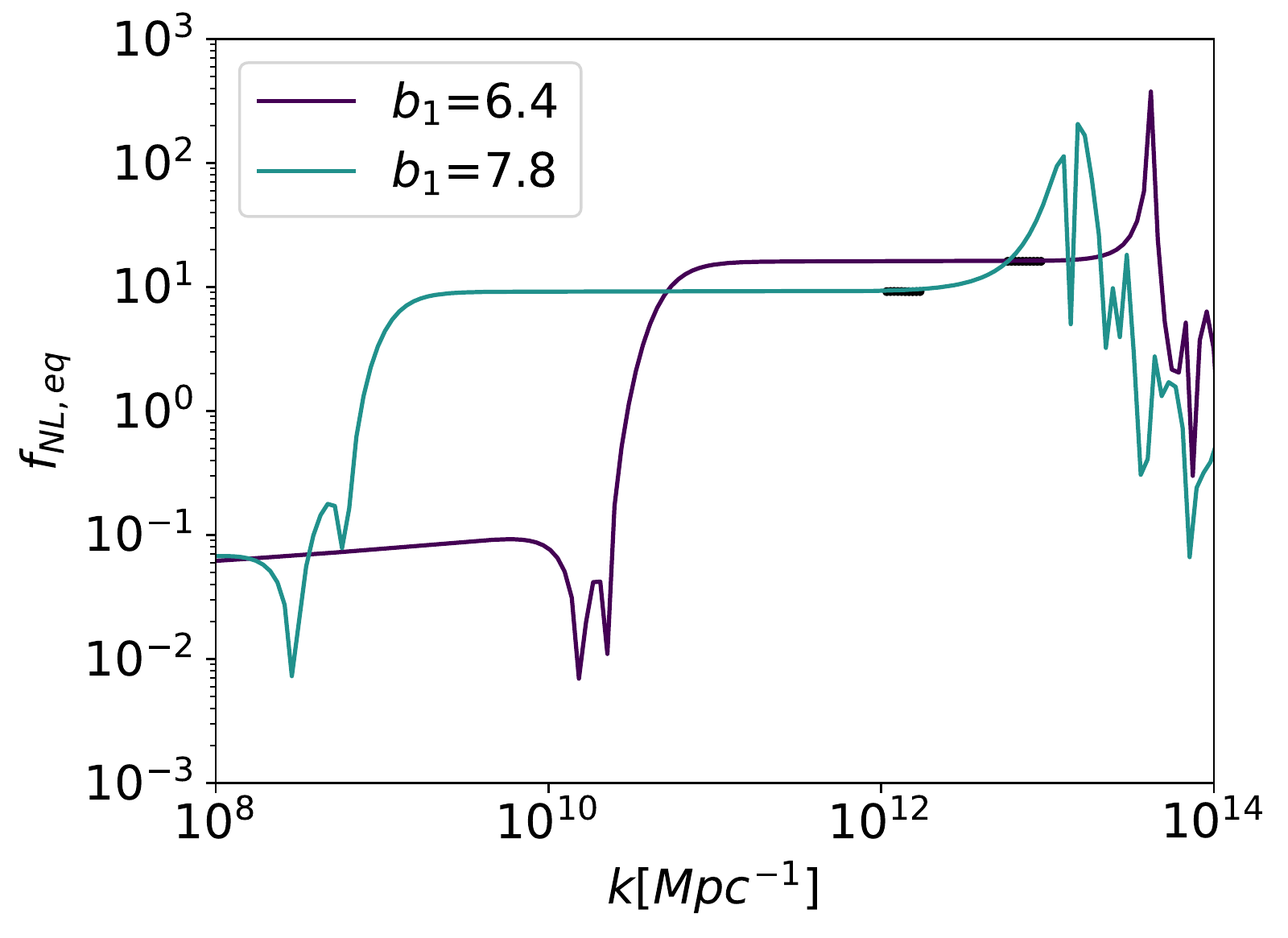}
\caption{Numerical $\delta N$ results for $f_\text{NL}$, eq.~\eqref{fNL delta N chi' only}, (black points) calculated for 10 scales around $k_\text{peak}$, which are defined in \eqref{k min k max}. We consider models with $\{\phi_\text{in}=7,\, \chi_\text{in}=7.31\}$ and two different values of $b_1$. For comparison, also the \texttt{PyTransport} results for the plateau region, see figure \ref{fig:fNL equilateral for diff curvature}, are plotted (colored lines).}
\label{fig:fNL delta N}
\end{figure}
We represent the results for our two\footnote{We do not expect the numerical $\delta N$ approach to yield the correct $f_\text{NL}$ for the model with $b_1=7.091$ as we have already seen that it fails at reproducing the principal peak in $\mathcal{P}_\zeta$, see the central panel in figure \ref{fig:delta N power spectrum}.} working examples with $b_1=6.4$ and $7.8$ in figure \ref{fig:fNL delta N}. For comparison, we show the numerical $\delta N$ values together with the \texttt{PyTransport} results. The non-Gaussianity amplitudes for scales around $k_\text{peak}$ calculated with the two different numerical approaches agree to a very good level. 

\section{The field-space geometry, non-Gaussianity and perturbativity}
\label{sec: perturbativity}
In section \ref{sec: non-gaussianities}, we calculated with \texttt{PyTransport} the amplitude of non-Gaussianity associated with large peaks in the scalar power spectrum for models of \cite{Braglia:2020eai,Kallosh:2022vha} with different values of the field-space curvature. In section \ref{sec: delta N} we have validated the results of section \ref{sec: non-gaussianities} by means of a second numerical approach, based on the $\delta N$ formalism. Here, we explore in more  detail the dependence of the amplitude of non-Gaussianity at peak scales on the geometrical parameter $b_1$, and use the results to assess the perturbativity of these models.

\begin{figure}
\centering
\includegraphics[width=0.7\textwidth]{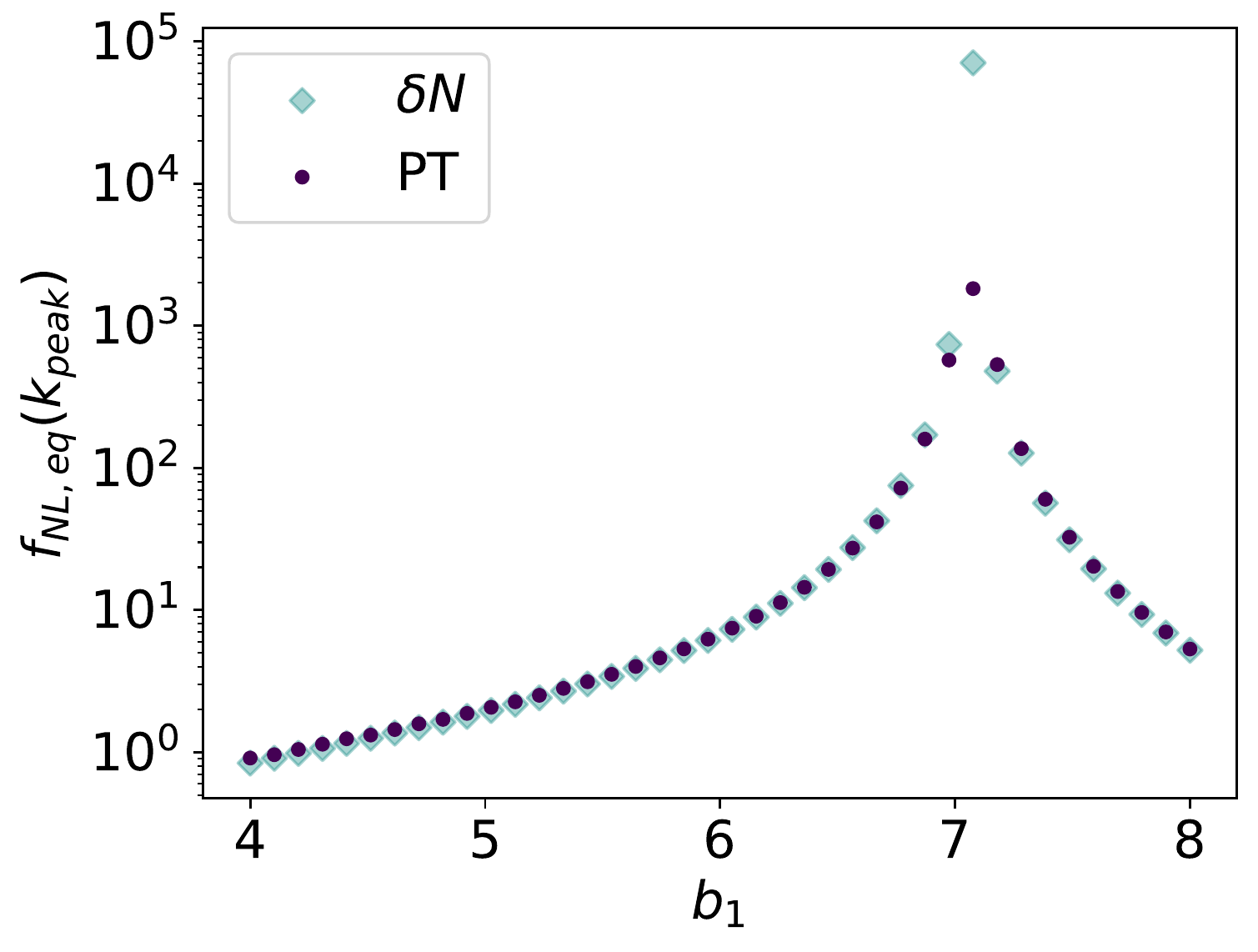}
\caption{Non-Gaussianity amplitude at the peak scale, $f_\text{NL,eq}(k_\text{peak})$, for models with $\{\phi_\text{in}=7,\, \chi_\text{in}=7.31\}$ and different values of $b_1$. We distinguish the results obtained with \texttt{PyTransport} (purple circles) and with the $\delta N$ approach (cyan diamonds).}
\label{fig:fNL vs b1}
\end{figure}
We display in figure \ref{fig:fNL vs b1} $f_\text{NL,eq}$ at $k_\text{peak}$ for models with different values of $b_1$ (for more details see appendix \ref{app: Pz fNL many b1}). By comparing the \texttt{PyTransport} and numerical $\delta N$ results, we see that overall the two series of points agree quite well, with deviations for cases with $b_1\simeq7.09$. These are expected considering the fact that $k_\text{peak}$ crossed the horizon during the turn in field space for these models (see figure \ref{fig:horizon_crossing}) and sub-horzion effects, not captured by the analytic correlators at horizon crossing that we use, are therefore expected to be relevant. The amplitude $f_\text{NL,eq}$ increases in magnitude for increasing $b_1$, with a faster and faster rate, peaks for $b_1\simeq 7.09$ and then decreases again for the remaining range of $b_1$. As is the case for the power spectrum amplitude at $k_\text{peak}$, see figure \ref{fig:Pzeta kpeak vs b1}, the amplitude of non-Gaussianity also displays a critical behavior at $b_1\simeq 7.09$. 

As shown in section \ref{sec: non-gaussianities}, the non-Gaussianity at peak scales is approximately of the local type for models with $b_1$ different from the critical value. For local non-Gaussianity, when one field or field velocity perturbation at horizon crossing is the dominant contribution to $\zeta$ (as is the case here), the curvature perturbation in real space can be expanded as (see for example \cite{Wands:2010af})\footnote{See \cite{Meng:2022ixx} for a calculation including the cubic non-Gaussianity parameter.}
\begin{equation}
    \zeta(\mathbf{x}) = \zeta_G(\mathbf{x})+\frac{3}{5} f_\text{NL} \left(\zeta_G(\mathbf{x})^2 - \langle \zeta_G(\mathbf{x})^2\rangle \right)\,,
    \label{eq:localfnl}
\end{equation}
where $\zeta_G(\mathbf{x})$ is the Gaussian contribution to the curvature perturbation $\zeta(\mathbf{x})$ and $\langle \zeta_G(\mathbf{x})^2\rangle$ is its variance. Beginning from eq.\eqref{eq:localfnl} and moving to Fourier space, we find that
\begin{equation}
    \mathcal{P}_\zeta(k) = \mathcal{P}_{\zeta_\text{G}}(k) + 
     \frac{9}{25}\, f_\text{NL}^2 \,\frac{k^3}{2 \pi} \int_{L^{-1}} {\rm d}^3{p}\;\frac{\mathcal{P}_{\zeta_\text{G}}(p)\,\mathcal{P}_{\zeta_\text{G}}(|\mathbf{p}-\mathbf{k}|)}{p^3 |\mathbf{p}-\mathbf{k}|^3}\,,
\end{equation}
where $L$ is an infrared cutoff and $\mathcal{P}_{\zeta_\text{G}}$ the Gaussian power spectrum that comes from including only the leading term in eq.\eqref{eq:localfnl}. When the Gaussian power spectrum is scale invariant, the integral above can be performed exactly (see for example \cite{Lyth:2006gd,Kumar:2009ge}), and the ratio between the one-loop power spectrum and the Gaussian one is 
\begin{equation}
\label{criterion 1}
    \frac{\mathcal{P}^\text{loop}_{\zeta_\text{G}}}{\mathcal{P}_{\zeta_\text{G}}} = 
   \frac{36}{9} f_\text{NL}^2  \mathcal{P}_{\zeta_\text{G}}  \ln{k L} \,.
\end{equation}
One can take $L$ to be the scale over which the 
power spectrum is being measured, and hence can be taken to be close to $k^{-1}$ for our purposes, such that $\ln{k L}\sim {\cal O}(1)$. Given our power-spectrum is not scale invariant over the peak scales, this expression can only be taken as an indication of when perturbativity breaks down. By requiring ${\mathcal{P}^{\rm loop}_\zeta}/{\mathcal{P}_{\zeta_{\rm G}}}\ll1$, eq.\eqref{criterion 1} yields 
\begin{equation}
    \label{criterion}
    f_\text{NL}^2  \mathcal{P}_{\zeta_\text{G}} \ll 1 \;,
\end{equation}
which we will use in the following as a criterion to assess the perturbativity of these models, given the amplitude of the scalar power spectrum and non-Gaussianity at the peak scales. Note that \eqref{criterion} is an upper limit, and perturbativity, or least the accuracy of results, may be in question well before the bound is saturated.

\begin{figure}
    \centering
    \includegraphics[width = 0.7\textwidth]{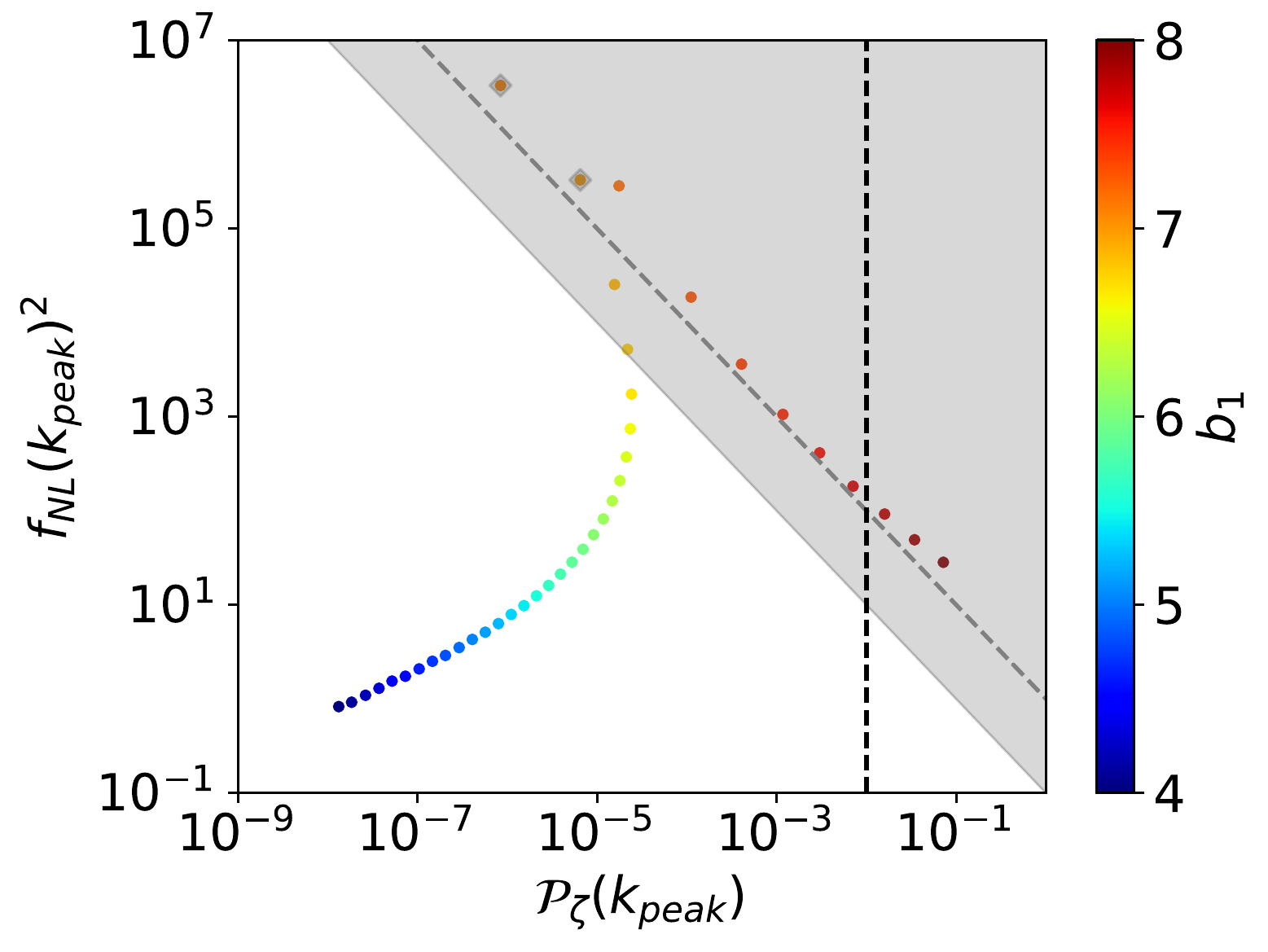}
    \caption{Values of ${f_\text{NL}(k_\text{peak})}^2$ and $\mathcal{P}_\zeta(k_\text{peak})$ for models with $\{\phi_\text{in}=7,\, \chi_\text{in}=7.31\}$ and different values of $b_1$ (see the bar legend). The gray area highlights ${f_\text{NL}(k_\text{peak})}^2$ and $\mathcal{P}_\zeta(k_\text{peak})$ values for which $\mathcal{P}_\zeta(k_\text{peak}){f_\text{NL}(k_\text{peak})}^2 \gtrsim 0.1$, with the dashed-gray line corresponding to ${f_\text{NL}(k_\text{peak})}^2 \mathcal{P}_\zeta(k_\text{peak}) =1$. We mark with a dashed, vertical line $\mathcal{P}_\zeta(k_\text{peak})=10^{-2}$, the approximate benchmark value for PBH production when the curvature perturbation is Gaussian.}
    \label{fig:perturbativity checks}
\end{figure}
In figure \ref{fig:perturbativity checks} we represent ${f_\text{NL}(k_\text{peak})}^2$ against $\mathcal{P}_\zeta(k_\text{peak})$ for the same models considered in figure \ref{fig:fNL vs b1}. The gray area highlights regions for which $\mathcal{P}_\zeta(k_\text{peak}){f_\text{NL}(k_\text{peak})}^2 \gtrsim 0.1$, which according to \eqref{criterion} signals a breakdown of perturbativity. The points highlighted with a gray diamond correspond to models with critical values of the field-space curvature, therefore we should not apply the criterion \eqref{criterion}, valid for local non-Gaussianity, to these points. With these points excluded, all models with $b_1\gtrsim 6.9$ violate \eqref{criterion}. In particular, models that could lead to PBH production (the usual criteria is $\mathcal{P}_\zeta(k_\text{peak})\simeq10^{-2}$ for Gaussian perturbations which drops to $\mathcal{P}_\zeta(k_\text{peak})\simeq10^{-3}$ when the smoothed density contrast, rather than $\zeta$, is used to calculate the abundance of PBHs \cite{DeLuca:2022rfz}) and/or large second-order GWs are all flawed by perturbativity issues.

Figure \ref{fig:perturbativity checks} illustrates again the fact that the peak amplitude $\mathcal{P}_\zeta(k_\text{peak})$ does not always increase by increasing the value of the geometrical parameter, as well as how ${f_\text{NL,eq}(k_\text{peak})}^2$ behaves. This has important consequences for power spectra with smaller peaks, $10^{-6}\lesssim \mathcal{P}_\zeta(k_\text{peak})\lesssim 3\times 10^{-5}$: they can produced within models with different values of $b_1$, but each model  corresponds to a different non-Gaussianity amplitude, some of which violate the criterion \eqref{criterion}.

\section{Discussion}
\label{sec: discussion}
In this work we have investigated multi-fields models of inflation with hyperbolic field space. We focus on the polynomial $\alpha$-attractors introduced in \cite{Braglia:2020eai}, which can deliver a large peak in the scalar power spectrum on small scales due to geometrical destabilisation and turning trajectories. In some cases the scalar power spectrum could lead to large second-order GWs and possibly result into PBH production. We show that peaks at scales $k_\text{peak}\gtrsim 4.7\times 10^{13}\,\text{Mpc}^{-1}$ are consistent at least at $95\%$ C.L. with large-scale measurements of the scalar spectral tilt, and therefore these models could provide a relevant source to the stochastic background of GWs at interferometer scales, and deserve further investigation. 

Up to now the models of \cite{Braglia:2020eai}, and other similar ones where multi-field effects are responsible for the peak in $\mathcal{P}_\zeta(k)$ (see e.g. \cite{Palma:2020ejf,Fumagalli:2020adf, Iacconi:2021ltm}) have been investigated only at the linear level, i.e. by employing the linear equations of motion for the perturbations. In this work we make a step towards investigating non-linear effects, and calculate the scalar bispectrum at peak scales. We find that the amplitude of non-Gaussianity, as well as the scalar power spectrum peak, depends non-monotonically on the field-space curvature and on the initial condition for the second field, which encourages us to identify critical values for these parameters. For models with non-critical values of the parameters, we show that the peak in the scalar power spectrum results from super-horizon effects and that the scalar non-Gaussianity is of the local type at peak scales, with a plateau region highly reminiscent of that found in single-field models (see e.g. \cite{Davies:2021loj}). This is an important result since the effect of non-Gaussianity on gravitational wave production has been investigated in detail only for local non-Gaussianity. Our result indicates that the framework of, e.g., \cite{Adshead:2021hnm}, which is based on local non-Gaussianity, can immediately be applied to models such as the one studied here, at least as a first approximation.

We derive and cross check our results by employing two different numerical tools, the code \texttt{PyTransport} and a newly developed approach based on the $\delta N$ formalism, whose applicability is due to the super-horizon evolution of peak-scales modes. The numerical $\delta N$ results allow us to also establish that the scalar power spectrum peak originates from fluctuations in the second field velocity.

We employ our results to assess the perturbativity of these models. Since the non-Gaussianity is approximately local, we use as a diagnostic tool the quantity $f_\text{NL}(k_\text{peak})^2 \,\mathcal{P}_\zeta(k_\text{peak})$ and find that many realisations of these models are flawed by perturbativity issues, including phenomenologically-interesting cases with large peaks in the scalar power spectrum, $10^{-3}\lesssim \mathcal{P}_\zeta(k_\text{peak})\lesssim 10^{-2}$, see figures \ref{fig:perturbativity checks} and \ref{fig:perturbativity IC}. To our knowledge this work provides the first attempt to assess the perturbativity of a multi-field model with a peak in the scalar power spectrum. 

Non-Gaussianity arising in two-field models of inflation with hyperbolic field space and non-geodesic motion has been the subject of several investigations \cite{Garcia-Saenz:2018vqf, Fumagalli:2019noh, Garcia-Saenz:2019njm, Bjorkmo:2019qno, Ferreira:2020qkf}\footnote{For recent efforts in computing non-Gaussianity in multi-field models within the cosmological bootstrap program see e.g. \cite{Wang:2022eop}, and \cite{Werth:2023pfl} for the cosmological flow framework.}. Previous studies address models where the entropic perturbation instability is triggered on sub-horizon scales\footnote{More recently, super-horizon effects are taken into account by means of an analytical $\delta N$ calculation, showing that rapidly turning trajectories can produce potentially large bispectra, with contributions from many shapes \cite{Iarygina:2023msy}.} and the curvature perturbation exponentially grows around horizon crossing, getting amplified at all scales; for these models the non-Gaussianity is enhanced in the flattend configuration. As discussed above, in this work we consider a different class of models, focusing on hyperbolic models delivering amplified scalar fluctuations on small scales. 

There are many possible directions for future work. Similarly to what has been done for single-field models leading to enhanced fluctuations \cite{Inomata:2022yte,Kristiano:2022maq,Riotto:2023hoz, Kristiano:2023scm, Riotto:2023gpm, Choudhury:2023jlt, Firouzjahi:2023aum, Motohashi:2023syh}, one could check the viability of multi-field models by calculating the one-loop correction to the tree-level scalar power spectrum. Note that this approach does not rely on the assumption of the non-Gaussianity being of the local type. We also plan to expand on these results by comparing them with the non-Gaussianity produced within multi-field models where the peak is produced from sub-horizon effects. In addition, it would be interesting to understand why, for the models considered in this work, sub-horizon effects become relevant for critical values of the field-space curvature and initial conditions.

We conclude by noting that there are two important lessons to draw from our work, which likely extend to other models. First, non-Gaussianity can be large over peak scales, potentially leading to important effects in PBH formation rates, and in the spectrum of scalar induced gravitational waves. Secondly, non-Gaussianity can potentially be too large for strongly curved field space metrics, which at a minimum renders perturbative calculations invalid in this regime.

\section*{Acknowledgments}
The authors would like to thank David Wands for many interesting discussions on related topics and for very useful comments on the manuscript. DJM is supported by a Royal Society University Research Fellowship and LI by a Royal Society funded  postdoctoral position.

\appendix

\section[\texorpdfstring{Models with varying $\bm{\chi_\text{in}}$}{Models with varying chi in}]{Varying $\bm{\chi_\text{in}}$}
\label{app: vary chi in}
We illustrate here the impact that changes in the initial condition of the second field, $\chi_\text{in}$, have on the scalar power spectrum and non-Gaussianity. We consider $20$ models with $\phi_\text{in}=7$ and $b_1=7.6$ (chosen such that the scalar power spectrum reaches $\mathcal{P}_\zeta(k_\text{peak})\simeq 10^{-2}$ at LIGO scales), and vary $\chi_\text{in}\in [5.5,\,9.3]$. 
\begin{figure}
    \centering
    \includegraphics[width = 0.7\linewidth]{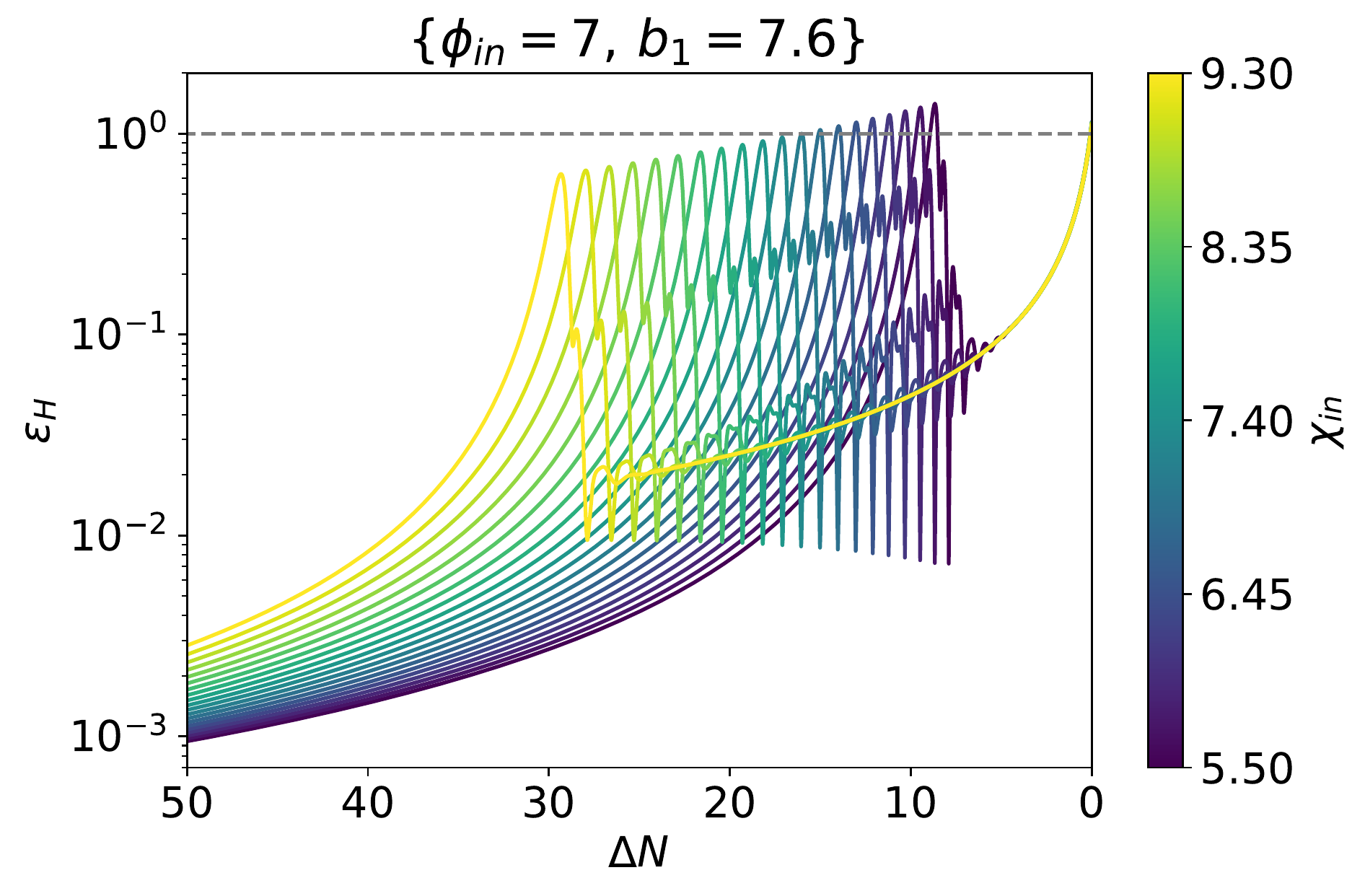}
    \caption{Numerical evolution of $\epsilon_H$, see eq.\eqref{epsilon H}, displayed against $\Delta N \equiv N_\text{end}-N$ for models with $\{\phi_\text{in}=7,\, b_1=7.6\}$ and different values of the initial condition $\chi_\text{in}$.}
    \label{fig:background chi_in}
\end{figure}
In figure \ref{fig:background chi_in} the time-dependence of $\epsilon_H$ shows two phases of inflation, separated by a transition with $\epsilon_H\sim1$, with the duration of the second phase of evolution, which is driven by $\chi$, set by the value of $\chi_\text{in}$. We note that the exact value of $\epsilon_H$ at the transition also depends on $\chi_\text{in}$, and for some models inflation is briefly interrupted. 

The duration of the second phase of evolution is important in determining the value of the scalar spectral tilt on large scales. This is shown in eq.\eqref{polynomial modified}, where we can approximate $\Delta N_\text{peak}$ with the duration of the second phase. Assuming instant reheating, we iteratively solve eq.\eqref{Nstar} with $V_0$ values compatible with \textit{Planck} measurements of the amplitude of scalar perturbations and find $\Delta N_\text{CMB,inst rh}\simeq 61.4$ for these models. We also find that the maximum allowed duration of reheating, see eq.\eqref{max rh}, is $\Delta \tilde N_\text{rh,max}\simeq 44.7$. These models are compatible at least at $95\%$ C.L. with \eqref{Planck ns 95} for initial conditions $\chi_\text{in}<9.1$ and appropriate choices of $\Delta \tilde N_\text{rh}$, with larger $\chi_\text{in}$ corresponding to shorter reheating stages. We also find that initial conditions $6\lesssim \chi_\text{in}\lesssim 8$ could yield a peak in the scalar power spectrum at ET or LIGO scales, whilst complying with \eqref{Planck ns 95} at least at $95\%$ C.L.. By using eq.\eqref{Nstar}, it is easy to show that $\Delta N_\text{CMB}=50$ (used in \cite{Braglia:2020eai}) is not compatible with the values obtained for $\Delta N_\text{CMB,inst rh}$ and $\Delta \tilde N_\text{rh,max}$. Nevertheless, constraining the parameter space of these models in view of the compatibility with large-scale measurements is not the focus of this work, so we will employ $\Delta N_\text{CMB}=50$ here, as was used in \cite{Braglia:2020eai} and in the main body of the paper for the set of models where we varied $b_1$. 

\begin{figure}
    \centering
    \includegraphics[width = 1\textwidth]{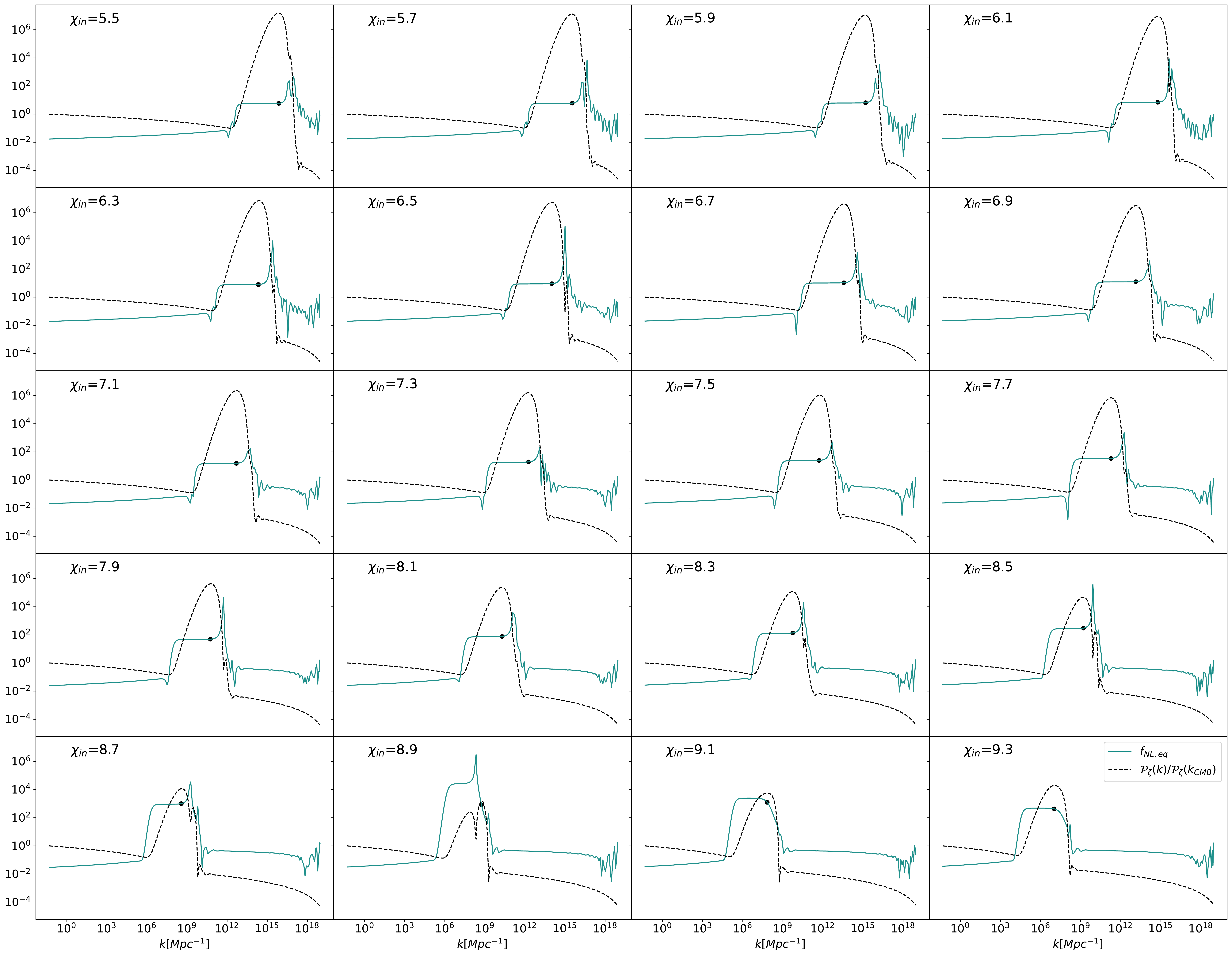}
    \caption{Scalar power spectrum, $\mathcal{P}_\zeta(k)/\mathcal{P}_\zeta(k_\text{CMB})$, and equilateral non-Gaussianity amplitude, $f_\text{NL,eq}(k)$, for 20 models with $\{\phi_\text{in}=7, \, b_1=7.6\}$ and increasing values of $\chi_\text{in}$ in the range $\chi_\text{in}\in[5.5,\,9.3]$. These results have been obtained with \texttt{PyTransport}. In each case, we highlight the value of $f_\text{NL,eq}$ corresponding to the peak scale with a black dot.
    }
    \label{fig:Pzeta fNL ICs}
\end{figure}
We display in figure \ref{fig:Pzeta fNL ICs} results for $\mathcal{P}_\zeta(k)/\mathcal{P}_\zeta(k_\text{CMB})$ and the reduced bispectrum in the equilateral configuration, $k_1=k_2=k_3=k$, at scales around the peak region for the same models of figure \ref{fig:background chi_in}. For larger $\chi_\text{in}$ the second phase of background evolution lasts longer, see figure \ref{fig:background chi_in}, which explains why the peak is located on larger scales. The peak region is characterised by an almost flat profile for $f_\text{NL,eq}$, as for the models analysed in section \ref{sec: non-gaussianities}, hinting at non-Gaussianity of the local type. This is true for all cases except for $\chi_\text{in}=8.9$, where $f_\text{NL,eq}(k_\text{peak})$ is outside of the plateau region, in a zone of rapidly varying $f_\text{NL,eq}$. Also note that in this case the scalar power spectrum exhibits a two-peak structure, with the second peak being the largest. 
\begin{figure}
    \centering
    \includegraphics[width=1\textwidth]{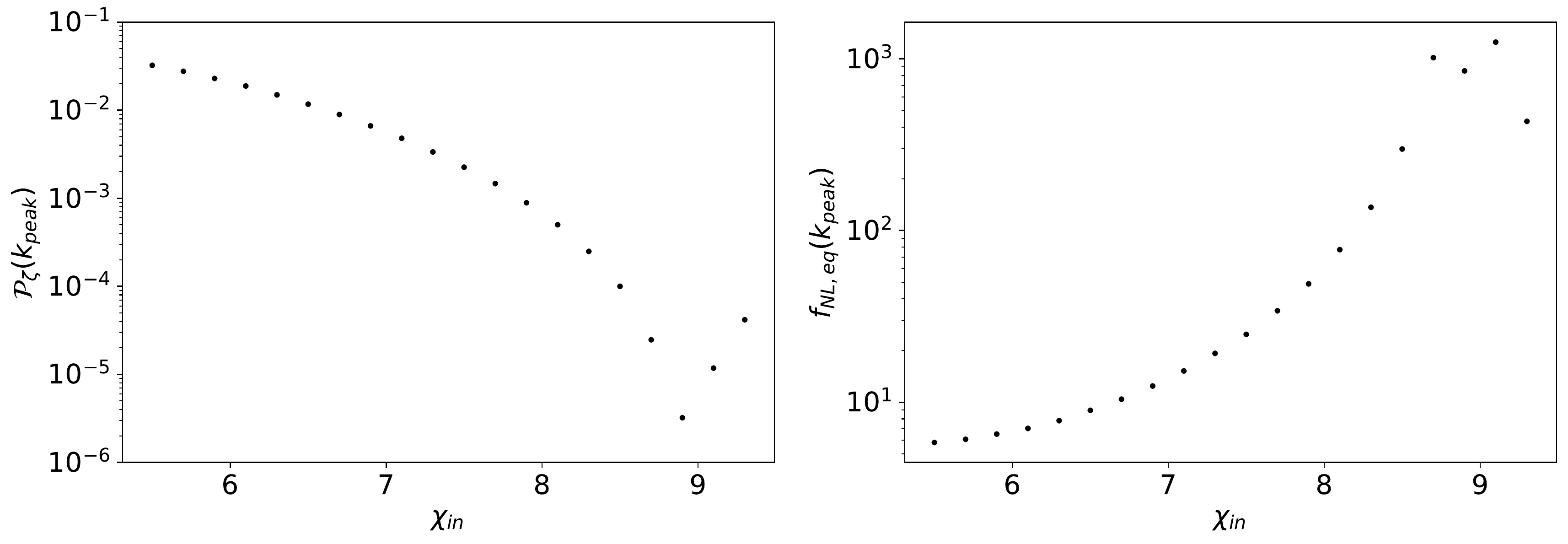}
    \caption{Amplitude of the peak in the scalar power spectrum (left) and equilateral non-Gaussianity at the peak scale (right) displayed against $\chi_\text{in}$ for the 20 models of figure \ref{fig:Pzeta fNL ICs}.}
    \label{fig:kpeak IC critical}
\end{figure}
Considering this and the behavior of $\mathcal{P}_\zeta(k_\text{peak})$ and $f_\text{NL,eq}(k_\text{peak})$ against $\chi_\text{in}$, see figure \ref{fig:kpeak IC critical}, we conclude that models in which  $\chi_\text{in}$ is varied also display signs of criticality. 

\begin{figure}
    \centering
    \includegraphics[width=0.8\textwidth]{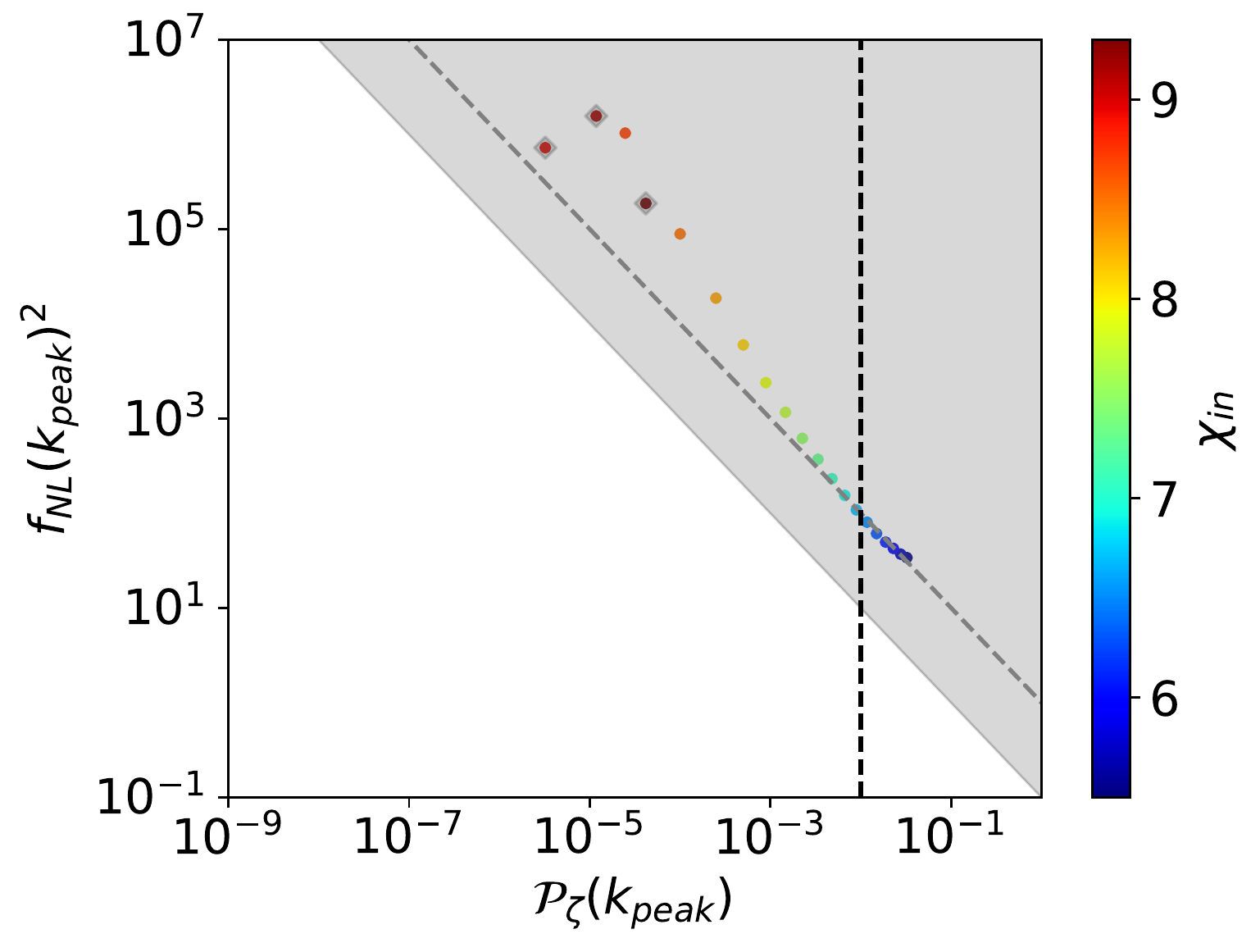}
    \caption{Values of ${f_\text{NL}(k_\text{peak})}^2$ and $\mathcal{P}_\zeta(k_\text{peak})$ for models with $\{\phi_\text{in}=7,\, b_1=7.6\}$ and different values of $\chi_\text{in}$. The gray area highlights ${f_\text{NL}(k_\text{peak})}^2$ and $\mathcal{P}_\zeta(k_\text{peak})$ values for which $\mathcal{P}_\zeta(k_\text{peak}){f_\text{NL}(k_\text{peak})}^2 \gtrsim 0.1$, with the dashed-gray line corresponding to ${f_\text{NL}(k_\text{peak})}^2 \mathcal{P}_\zeta(k_\text{peak})=1$. We mark with a dashed, vertical line $\mathcal{P}_\zeta(k_\text{peak})=10^{-2}$, the approximate benchmark value for PBH production when the curvature perturbation is Gaussian.}
    \label{fig:perturbativity IC}
\end{figure}
Similarly to what has been done for the set of models where we varied $b_1$, see figure \ref{fig:perturbativity checks}, we display in figure \ref{fig:perturbativity IC} values of ${f_\text{NL}(k_\text{peak})}^2$ against $\mathcal{P}_\zeta(k_\text{peak})$. The points highlighted with gray diamonds correspond to critical values of $\chi_\text{in}$, in which case the non-Gaussianity is not of the local type. For all the other models, the criterion \eqref{criterion} applies and figure \ref{fig:perturbativity IC} shows that they are all flawed by perturbativity issues, including those delivering a large peak at LIGO and ET scales. 

\section[\texorpdfstring{Models with varying $\bm{b_1}$}{Models with varying b1}]{Varying $\bm{b_1}$}
\label{app: Pz fNL many b1}
To motivate our choice of values for the field-space geometrical parameter $b_1$ in section \ref{sec: non-gaussianities}, we include here results for the scalar power spectrum and amplitude of equilateral non-Gaussianity derived for additional $b_1$ values in the range $b_1\in[4,\,8]$.  
\begin{figure}
    \centering
    \includegraphics[width=0.9\linewidth]{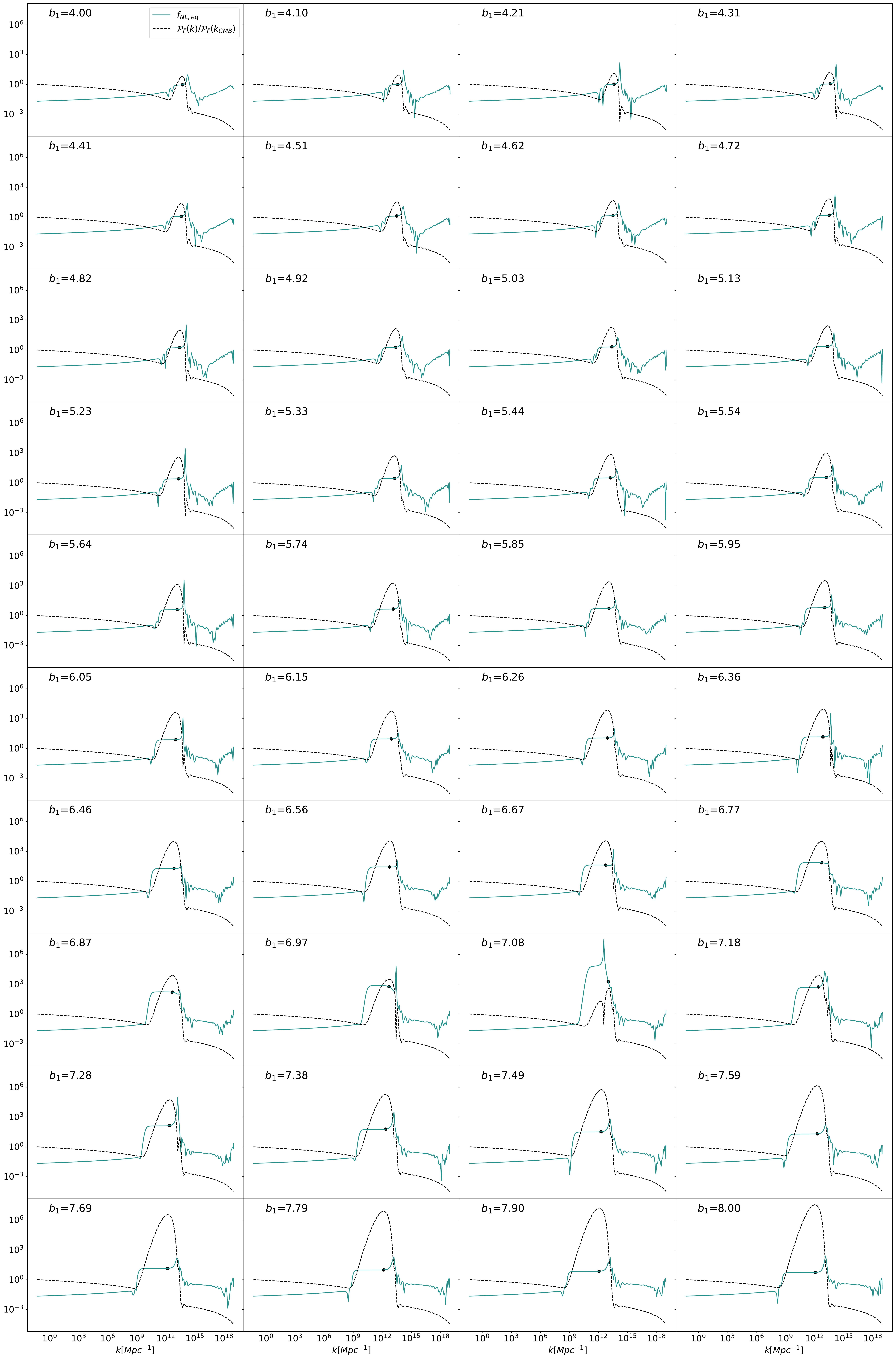}
    \caption{Scalar power spectrum, $\mathcal{P}_\zeta(k)/\mathcal{P}_\zeta(k_\text{CMB})$, and equilateral non-Gaussianity amplitude, $f_\text{NL,eq}(k)$, for 40 models with $\{\phi_\text{in}=7, \, \chi_\text{in}=7.31\}$ and increasing values of $b_1\in[4,\,8]$. These results have been obtained with \texttt{PyTransport}. In each case, we highlight the value of $f_\text{NL,eq}$ corresponding to the peak scale with a black dot.}
    \label{fig: Pz fNL many b1}
\end{figure}
We display in figure \ref{fig: Pz fNL many b1} numerical results for $\mathcal{P}_\zeta(k)/\mathcal{P}_\zeta(k_\text{CMB})$ and $f_\text{NL,eq}$, calculated using \texttt{PyTransport}, for 40 values of $b_1$. When $b_1= 7.08$, the scalar power spectrum is characterised by a two-peak structure, with the second peak being also the principal one. Also, while in all the other cases $f_\text{NL,eq}(k_\text{peak})$ (black dots in figure \ref{fig: Pz fNL many b1}) lies in the plateau region, for $b_1=7.08$ this is not true.
\begin{figure}
    \centering
    \includegraphics[width = 0.7\textwidth]{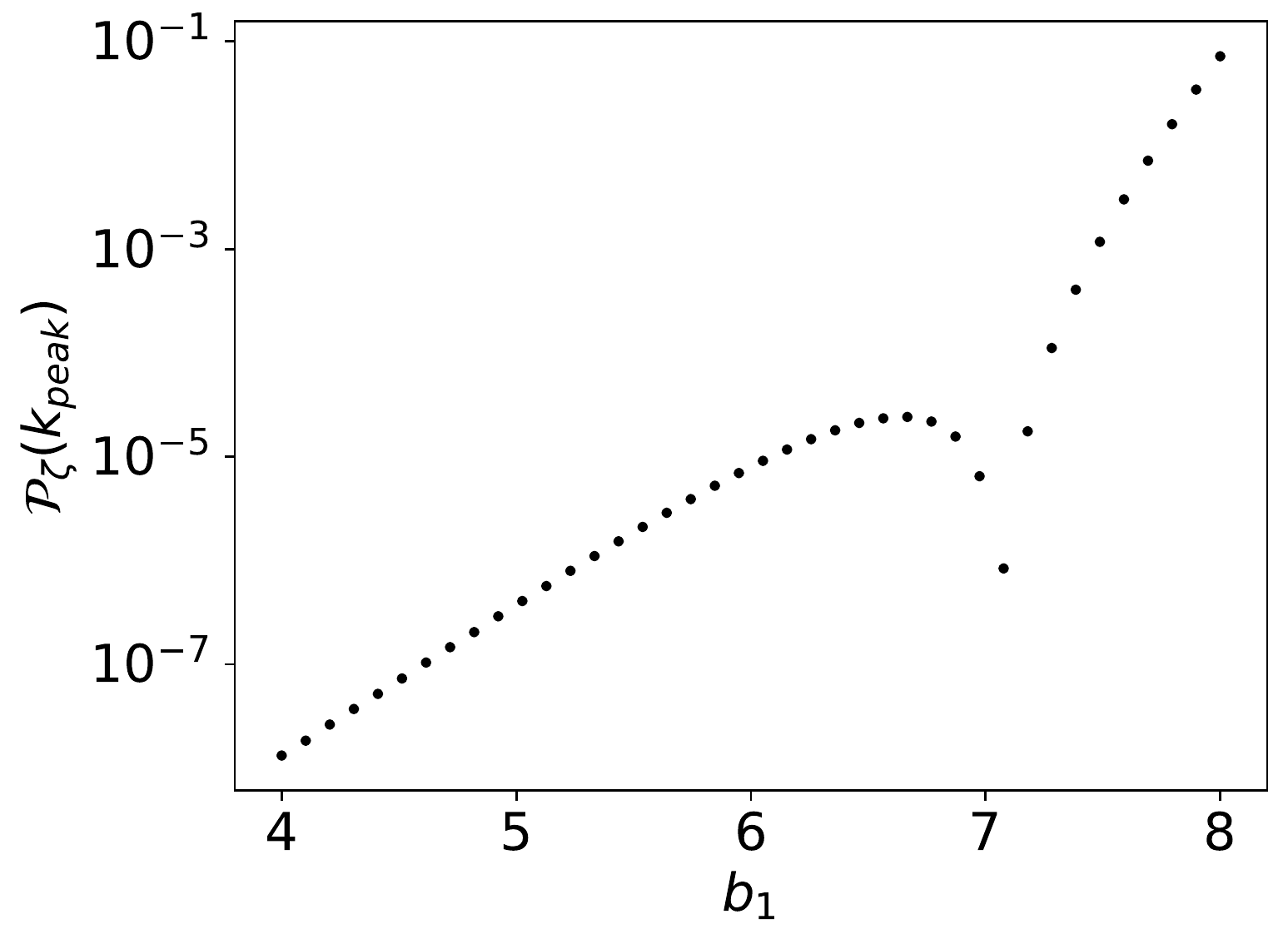}
    \caption{Amplitude of the peak in the scalar power spectrum displayed against $b_1$ for the 40 models of figure \ref{fig: Pz fNL many b1}.}
    \label{fig:Pzeta kpeak vs b1}
\end{figure}
We show in figure \ref{fig:Pzeta kpeak vs b1} values of $\mathcal{P}_\zeta(k_\text{peak})$ against $b_1$, which clearly display a non-monotonic behavior. 

The results represented in figures \ref{fig: Pz fNL many b1} and \ref{fig:Pzeta kpeak vs b1} show a change of behavior for the power spectrum and equilateral non-Gaussianity amplitudes at the peak scale when $b_1\sim7.09$. For this reason, in section \ref{sec: non-gaussianities} we choose to discuss the three values $b_1=\{6.4, \, 7.091,\, 7.8\}$, respectively smaller than, similar to and larger than the critical value. 

\section{Analytic 2-point correlators at horizon crossing}
\label{app: analytic 2pt corr}
Here we derive expressions for the 2-point correlation function of the fields and their velocities used in the numerical $\delta N$ computation described in section \ref{sec: delta N}. In particular, we follow the approach of \cite{Dias:2016rjq}. 

Assuming that the fields are massless and non-interacting at the time in which to evaluate the 2-point correlators, i.e. at horizon crossing, the 2-point correlator at unequal times in nearly de Sitter spacetime reads\footnote{Besides requiring the fields to be non-interacting and effectively massless, eq.\eqref{correlator} also relies on the slow-roll approximation, $\epsilon_H<1$. This holds if the scales of interest exit the horizon before the turn in field space, which is true for peak scales in all the field-space curvature cases considered, except for the critical values $b_1\simeq7.09$, see section \ref{sec: delta N}.}
\begin{multline}
\label{correlator}
    \langle Q^I(\mathbf{k_1}, \eta_1) Q^J(\mathbf{k_2}, \eta_2)\rangle =  (2\pi)^3  \delta(\mathbf{k_1}+\mathbf{k_2})\;\left(\frac{2\pi^2}{k_1^3} \right) \,\Pi^{IJ}(\eta_1, \eta_2) \; \times \\ 
    \times \frac{H(\eta_1)H(\eta_2)}{(2\pi)^2} (1-ik_1 \eta_1)(1+ik_1\eta_2) \exp{\left\{ik_1 (\eta_1-\eta_2)\right\}} \quad\text{with}\quad \eta_1<\eta_2 \;,
\end{multline}
where we use conformal time, $\mathrm{d}\eta \equiv \mathrm{d}t/a(t)$. At linear-order $Q^I(\mathbf{k_1}, \eta_1)$ stands for the field perturbation, with the indices $I=\{\phi, \, \chi\}$ such that, e.g., $Q^\phi=\delta \phi$. In eq.\eqref{correlator}, $\Pi^{IJ}(\eta_1, \eta_2)$ is defined as \cite{Elliston:2012ab}
\begin{multline}
\label{Pi definition}
    \Pi  ^{IJ}(\eta_1, \eta_2) \equiv \underbrace{\mathcal{T} \exp{ \left\{-\int_{\eta_1}^{\eta_2} \mathrm{d}\eta \; \Gamma^I_{KL}\; \frac{\mathrm{d}\phi^K}{\mathrm{d}\eta}\right\}}}_{\equiv \mathcal{E}^I_L(\eta_1,\,\eta_2)} \;\mathcal{G}^{LJ}(\eta_1) \\
    = \left(\delta^I_L - \int_{\eta_1}^{\eta_2} \mathrm{d}\eta' \; \Gamma^I_{KL}(\eta')\; \frac{\mathrm{d}\phi^K}{\mathrm{d}\eta'} + \int_{\eta_1}^{\eta_2} \mathrm{d}\eta' \, \int_{\eta_1}^{\eta'} \mathrm{d}\eta'' \; \Gamma^I_{KM}(\eta')\; \frac{\mathrm{d}\phi^K}{\mathrm{d}\eta'}\, \Gamma^M_{NL}(\eta'')\; \frac{\mathrm{d}\phi^N}{\mathrm{d}\eta''} + \cdots \right)\;\mathcal{G}^{LJ}(\eta_1) \;, 
\end{multline}
where we have defined the exponential operator $\mathcal{E}^I_L(\eta_1,\,\eta_2)$ and given the first three terms of its Taylor expansion. Eq.\eqref{Pi definition} shows that in the limit $\eta_2\to\eta_1\to\eta$, $\mathcal{E}^I_L(\eta_1,\,\eta_2)\to \delta^I_L$ and therefore $\Pi^{IJ}$ coincides with $\mathcal{G}^{IJ}$. 

Using \eqref{correlator} and \eqref{Pi definition}, it is straightforward to show that the fields 2-point correlator at equal times is given by 
\begin{equation}
\label{field corr at equal times}
   \langle Q^I(\mathbf{k_1}, \eta_1) Q^J(\mathbf{k_2}, \eta_2)\rangle|_{\eta_2\to\eta_1\to\eta} = (2\pi)^3  \delta(\mathbf{k_1}+\mathbf{k_2})\,  \left(\frac{2\pi^2}{k_1^3} \right) \left(\frac{H}{2\pi} \right)^2\,\mathcal{G}^{IJ}\, \left(1+{k_1}^2\eta^2 \right) \;. 
\end{equation}
In order to derive correlators involving field velocities, it is useful to first calculate derivatives of $\Pi^{IJ}(\eta_1,\,\eta_2)$ with respect to $\eta_1$ and $\eta_2$. By using eq.\eqref{Pi definition} and applying the Leibniz integral rule, we obtain
\begin{align}
    \label{Pi der eta1}
    \frac{\mathrm{d}}{\mathrm{d}\eta_1}\Pi^{IJ}(\eta_1,\,\eta_2) & = \mathcal{E}^I_N (\eta_1, \, \eta_2) \;\Gamma^N_{LK}(\eta_1)\; \frac{\mathrm{d}\phi^K}{\mathrm{d}\eta_1}\;  \mathcal{G}^{LJ}(\eta_1) + \mathcal{E}^I_L(\eta_1,\, \eta_2) \;\frac{\mathrm{d}\mathcal{G}^{LJ}(\eta_1)}{\mathrm{d}\eta_1} \;, \\
    \label{Pi der eta2}
    \frac{\mathrm{d}}{\mathrm{d}\eta_2}\Pi^{IJ}(\eta_1,\,\eta_2) & = -\Gamma^I_{KM}(\eta_2)\;\frac{\mathrm{d}\phi^K}{\mathrm{d}\eta_2} \;\mathcal{E}^M_L (\eta_1, \, \eta_2) \;\mathcal{G}^{LJ}(\eta_1)\;,
\end{align}
and 
\begin{multline}
\label{Pi der eta1 eta2}
    \frac{\mathrm{d}^2}{\mathrm{d}\eta_1 \mathrm{d}\eta_2}\Pi^{IJ}(\eta_1,\,\eta_2)  = -\Gamma^I_{KM}(\eta_2)\;\frac{\mathrm{d}\phi^K}{\mathrm{d}\eta_2} \\
    \times \left( \mathcal{E}^M_N (\eta_1, \, \eta_2) \;\Gamma^N_{LP}(\eta_1)\; \frac{\mathrm{d}\phi^P}{\mathrm{d}\eta_1}\;  \mathcal{G}^{LJ}(\eta_1) + \mathcal{E}^M_L(\eta_1,\, \eta_2) \;\frac{\mathrm{d}\mathcal{G}^{LJ}(\eta_1)}{\mathrm{d}\eta_1} \right) \;. 
\end{multline}
By taking the limit $\eta_2\to\eta_1\to\eta$, eqs.\eqref{Pi der eta1}-\eqref{Pi der eta1 eta2} reduce to 
\begin{align}
    \label{Pi der eta1 same time}
    \frac{\mathrm{d}}{\mathrm{d}\eta_1}\Pi^{IJ}(\eta_1,\,\eta_2)|_{\eta_2\to\eta_1\to\eta} & = \Gamma^I_{KL}\; \frac{\mathrm{d}\phi^K}{\mathrm{d}\eta}\;  \mathcal{G}^{LJ} + \frac{\mathrm{d}\mathcal{G}^{IJ}}{\mathrm{d}\eta} \;, \\
    \label{Pi der eta2 same time}
    \frac{\mathrm{d}}{\mathrm{d}\eta_2}\Pi^{IJ}(\eta_1,\,\eta_2)|_{\eta_2\to\eta_1\to\eta} & = -\Gamma^I_{KL}\;\frac{\mathrm{d}\phi^K}{\mathrm{d}\eta} \;\mathcal{G}^{LJ}\;, \\
    \label{Pi der eta1 eta2 same time}
    \frac{\mathrm{d}^2}{\mathrm{d}\eta_1 \mathrm{d}\eta_2}\Pi^{IJ}(\eta_1,\,\eta_2)|_{\eta_2\to\eta_1\to\eta} &  = -\Gamma^I_{KM}\;\frac{\mathrm{d}\phi^K}{\mathrm{d}\eta}\left( \Gamma^M_{LP}\; \frac{\mathrm{d}\phi^P}{\mathrm{d}\eta}\;  \mathcal{G}^{LJ} + \frac{\mathrm{d}\mathcal{G}^{MJ}}{\mathrm{d}\eta} \right) \;. 
\end{align}
By using eq.\eqref{correlator} and eqs.\eqref{Pi der eta1 same time}-\eqref{Pi der eta1 eta2 same time}, one can then derive the equal-times field-velocity cross-correlators 
\begin{multline}
    \label{field vel corr at equal times} 
    \langle Q^I(\mathbf{k_1}, \eta_1) \frac{\mathrm{d}}{\mathrm{d}\eta_2}Q^J(\mathbf{k_2}, \eta_2)\rangle|_{\eta_2\to\eta_1\to\eta} 
    = (2\pi)^3  \delta(\mathbf{k_1}+\mathbf{k_2})\, \left(\frac{2\pi^2}{k_1^3} \right)\,  \left(\frac{H}{2\pi} \right)^2 \\ \times \left[ -\Gamma^I_{KL} \frac{\mathrm{d}\phi^K}{\mathrm{d}\eta} \mathcal{G}^{LJ} \left(1+{k_1}^2\eta^2 \right) + \mathcal{G}^{IJ} {k_1}^2\eta \,(1-ik_1\eta) \right]  \;,
\end{multline}
\begin{multline}
    \label{vel field corr at equal times}
    \langle \frac{\mathrm{d}}{\mathrm{d}\eta_1} Q^I(\mathbf{k_1}, \eta_1) Q^J(\mathbf{k_2}, \eta_2)\rangle|_{\eta_2\to\eta_1\to\eta} 
    = (2\pi)^3  \delta(\mathbf{k_1}+\mathbf{k_2})\, \left(\frac{2\pi^2}{k_1^3} \right)\,  \left(\frac{H}{2\pi} \right)^2 \\ \times \left\{ \left[\Gamma^I_{KL} \frac{\mathrm{d}\phi^K}{\mathrm{d}\eta} \mathcal{G}^{LJ} + \frac{\mathrm{d} \mathcal{G}^{IJ}}{\mathrm{d}\eta}\right]\left(1+{k_1}^2\eta^2 \right) + \mathcal{G}^{IJ} {k_1}^2\eta\, (1+ik_1\eta) \right\}
\end{multline}
and the equal-times velocity-velocity correlators 
\begin{multline}
    \label{vel vel corr at equal times}
    \langle \frac{\mathrm{d}}{\mathrm{d}\eta_1} Q^I(\mathbf{k_1}, \eta_1)\frac{\mathrm{d}}{\mathrm{d}\eta_2}  Q^J(\mathbf{k_2}, \eta_2)\rangle|_{\eta_2\to\eta_1\to\eta} 
    = (2\pi)^3  \delta(\mathbf{k_1}+\mathbf{k_2})\, \left(\frac{2\pi^2}{k_1^3} \right)\,   \left(\frac{H}{2\pi} \right)^2 \\ \times \left\{ - \Gamma^I_{KM} \frac{\mathrm{d}\phi^K}{\mathrm{d}\eta} \left[\Gamma^M_{LP} \frac{\mathrm{d}\phi^P}{\mathrm{d}\eta} \mathcal{G}^{LJ} + \frac{\mathrm{d} \mathcal{G}^{MJ}}{\mathrm{d}\eta}\right] \left(1+{k_1}^2\eta^2 \right) + \mathcal{G}^{IJ} {k_1}^4\eta^2 \right\}\;. 
\end{multline}
Note that in deriving eqs.\eqref{field vel corr at equal times}-\eqref{vel vel corr at equal times}, we do not take time-derivatives of the Hubble rate since eq.\eqref{correlator} relies on the slow-roll approximation, $\epsilon_H<1$.

For the purpose of the numerical $\delta N$ calculation of section \ref{sec: delta N}, we need the time-derivatives of the fields to be calculated with respect to $N$, where $\mathrm{d}N =a H \mathrm{d}\eta$. Note that the prime symbol stands for a derivative with respect to $N$. Also, on de Sitter $\eta \simeq -1/\left(aH \right)$. By keeping this into account and using the metric \eqref{action} and the Christoffel symbols \eqref{christoffel} in eqs.\eqref{field corr at equal times}, \eqref{field vel corr at equal times}-\eqref{vel vel corr at equal times}, we get 
\begin{align}
\label{first eq}
    & \mathcal{P}_{\phi\phi} (k_1) = \left( \frac{H}{2\pi}\right)^2 \left[1+\left(\frac{k_1}{aH}\right)^2 \right] \;, \\
    & \mathcal{P}_{\chi\chi} (k_1) = \left( \frac{H}{2\pi}\right)^2 e^{-2b_1 \phi} \left[1+\left(\frac{k_1}{aH}\right)^2 \right] \;,\\
    \label{phi chi corr}
    & \mathcal{P}_{\phi\chi} (k_1) = \mathcal{P}_{\chi\phi} (k_1) = 0 \;, \\
    & \mathcal{P}_{\phi'\phi'} (k_1) = \left( \frac{H}{2\pi}\right)^2  \left\{ e^{2b_1 \phi} \left(b_1 \chi'\right)^2 \left[1+\left(\frac{k_1}{aH}\right)^2 \right] + \left(\frac{k_1}{aH}\right)^4 \right\} \;, \\
    & \mathcal{P}_{\chi'\chi'} (k_1) =\left( \frac{H}{2\pi}\right)^2  e^{-2b_1\phi}  \left\{ \left[\left(b_1 \phi'\right)^2 + \left(b_1 \chi' e^{b_1\phi}\right)^2 \right] \left[1+\left( \frac{k_1}{aH}\right)^2\right]+ \left(\frac{k_1}{aH}\right)^4  \right\} \;, \\
    & \mathcal{P}_{\chi'\phi'} (k_1) = \mathcal{P}_{\phi'\chi'} (k_1) =\left( \frac{H}{2\pi}\right)^2 \left\{ -{b_1}^2 \chi' \phi' \left[1+\left(\frac{k_1}{aH}\right)^2 \right]\right\}  \;, \\
    & \mathcal{P}_{\phi'\phi} (k_1) =  \mathcal{P}_{\phi\phi'} (k_1)^* =  -\left( \frac{H}{2\pi}\right)^2 \left(\frac{k_1}{aH} \right)^2 \left(1-i\frac{k_1}{aH} \right) \;, \\
    & \mathcal{P}_{\chi'\chi} (k_1) =  \mathcal{P}_{\chi\chi'} (k_1)^* = -\left( \frac{H}{2\pi}\right)^2 \, e^{-2b_1\phi} \left\{ b_1 \phi'\left[1+\left(\frac{k_1}{aH}\right)^2 \right] + \left(\frac{k_1}{aH} \right)^2 \left(1-i\frac{k_1}{aH} \right) \right\} \;, \\
    & \mathcal{P}_{\chi'\phi} (k_1) =  \mathcal{P}_{\phi\chi'} (k_1) = \left( \frac{H}{2\pi}\right)^2 b_1 \chi' \left[1+\left(\frac{k_1}{aH}\right)^2 \right] \;, \\
    \label{last eq}
    & \mathcal{P}_{\phi'\chi} (k_1) =  \mathcal{P}_{\chi\phi'} (k_1) = -\left( \frac{H}{2\pi}\right)^2 b_1 \chi'\left[1+\left(\frac{k_1}{aH}\right)^2 \right] \;, 
\end{align}
where we simplify the notation by defining the dimensionless power spectrum $\mathcal{P}_{\delta X \delta Y}\equiv \mathcal{P}_{XY} $. 

\begin{figure}
\begin{subfigure}{.5\textwidth}
  \centering
  \includegraphics[width=\textwidth]{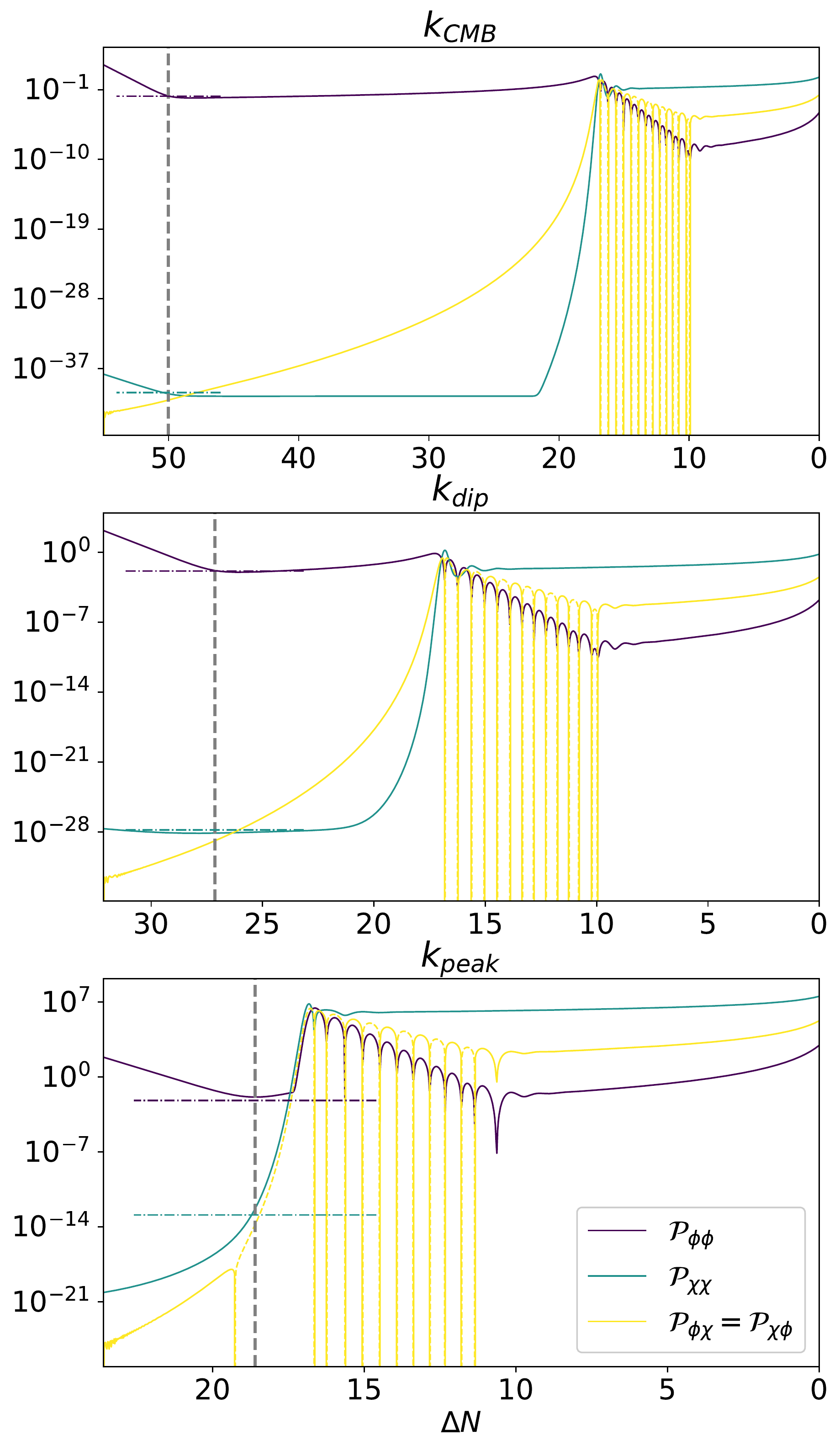}
\end{subfigure}
\begin{subfigure}{.5\textwidth}
  \centering
  \includegraphics[width=\textwidth]{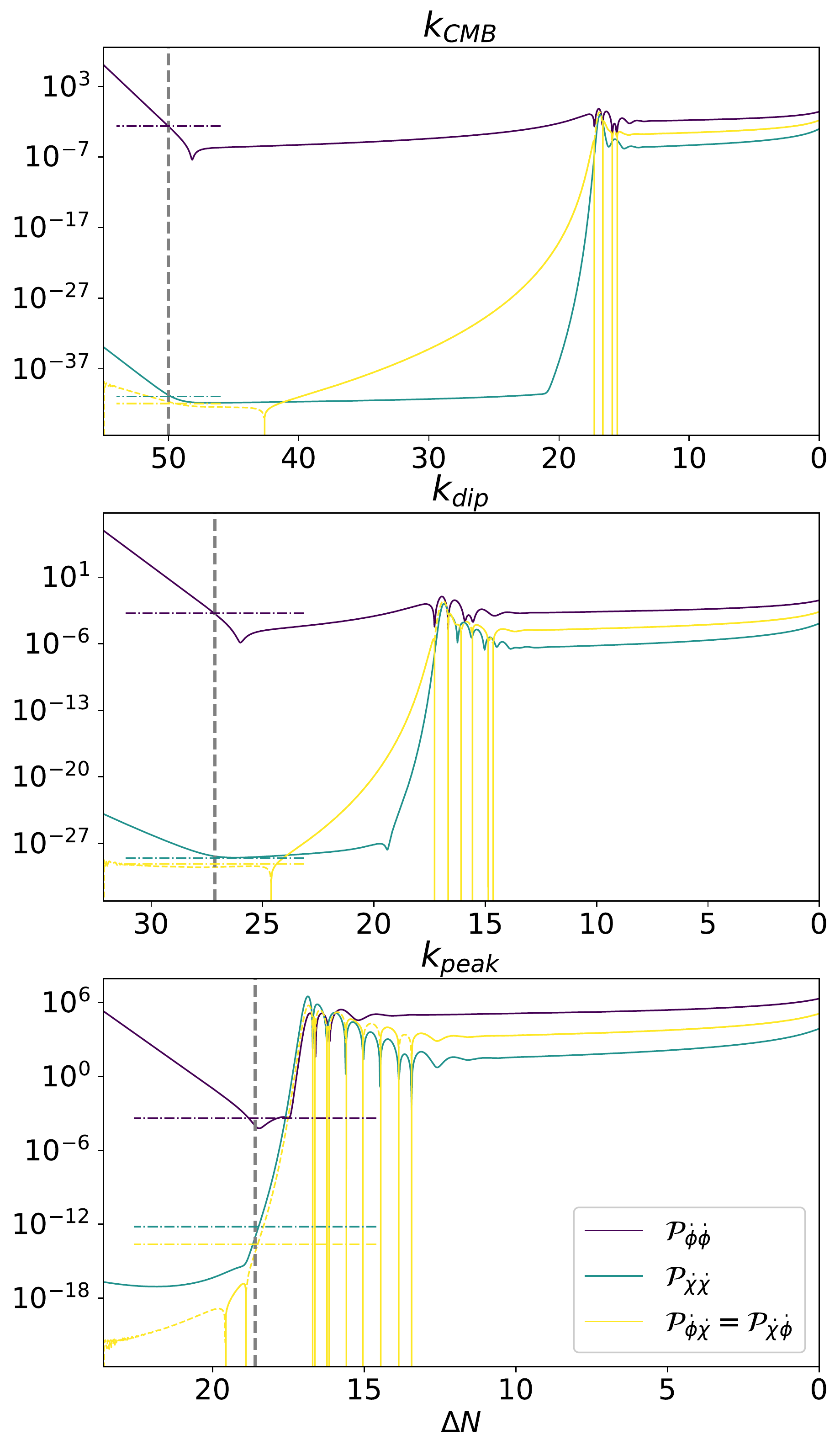}
\end{subfigure}
\caption{Evolution of the field (left) and  velocities (right) correlators represented against $\Delta N \equiv N_\text{end}-N$ as produced numerically with \texttt{PyTransport} for a model with $\{\phi_\text{in}=7, \, \chi_\text{in}=7.31, \, b_1=7.8\}$. Each panel corresponds to a specific scale, namely the CMB, dip and peak scales from top to bottom. The dashed, vertical line in each panel represents the e-folding time when the corresponding scale crossed the horizon during inflation. We represent with dot-dashed, horizontal lines the (absolute value) of the results obtained by evaluating at horizon crossing the analytic expressions derived above. We only plot these values for times 4 e-folds before and after horizon crossing, as they are intended for comparison around horizon crossing only.}
    \label{fig:2pt_correlators_1}
\end{figure}
\begin{figure}
\begin{subfigure}{.5\textwidth}
  \centering
  \includegraphics[width=\textwidth]{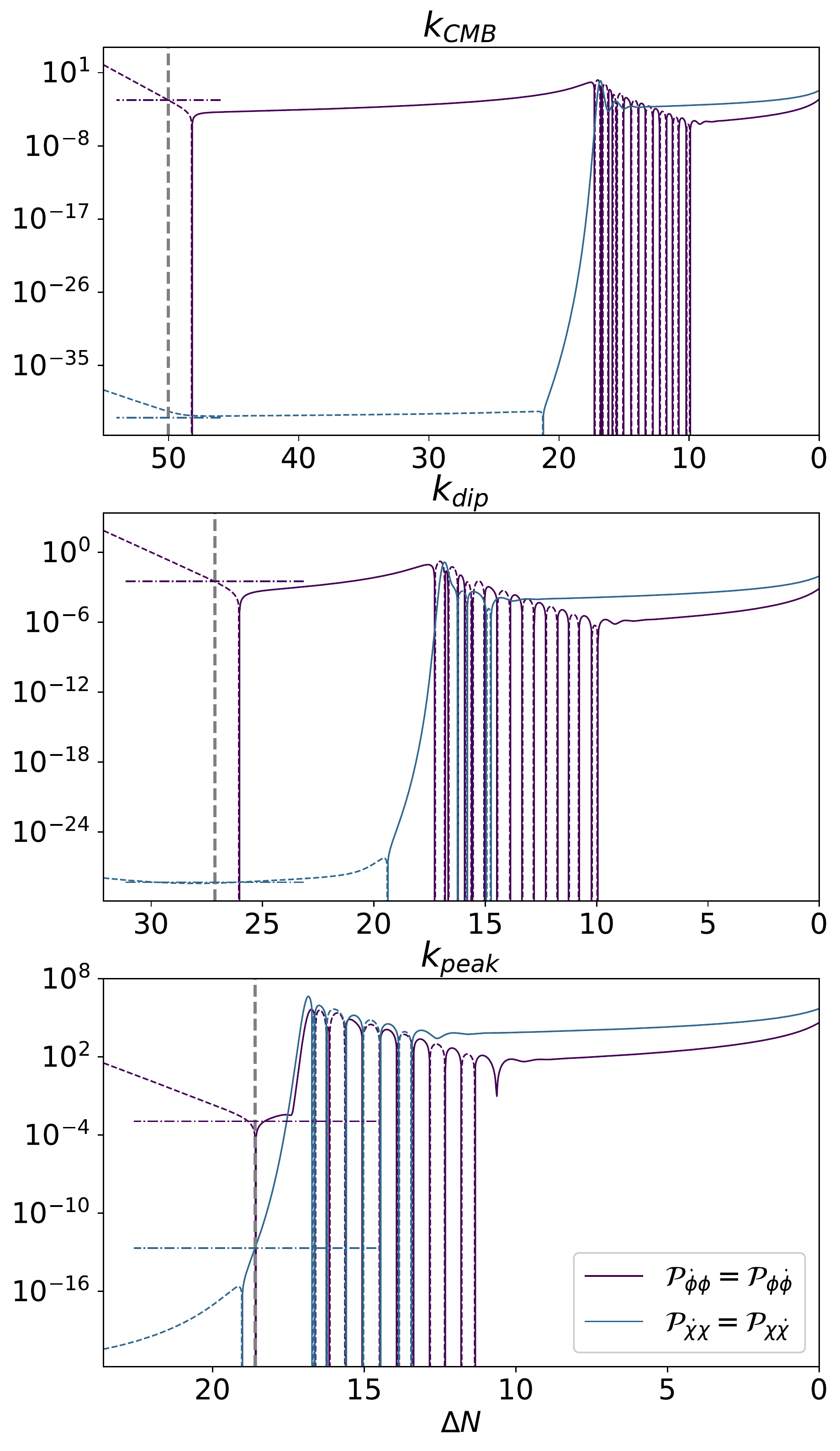}
\end{subfigure}
\begin{subfigure}{.5\textwidth}
  \centering
  \includegraphics[width=\textwidth]{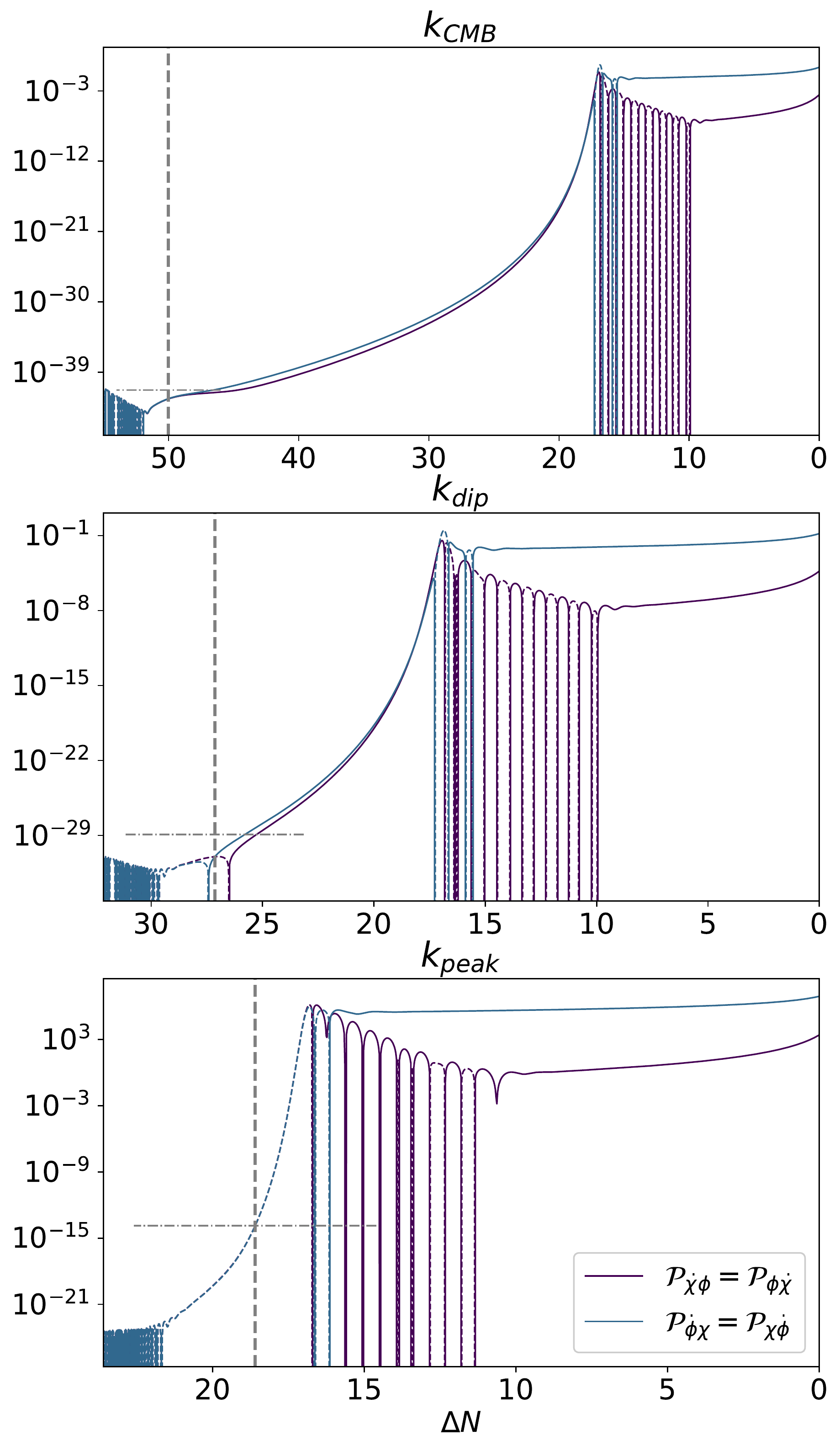}
\end{subfigure}
\caption{Evolution of correlators involving one time derivative represented against $\Delta N \equiv N_\text{end}-N$ as produced numerically with \texttt{PyTransport} for a model with $\{\phi_\text{in}=7, \, \chi_\text{in}=7.31, \, b_1=7.8\}$. Each panel corresponds to a specific scale, namely the CMB, dip and peak scales from top to bottom. The dashed, vertical line in each panel represents the e-folding time when the corresponding scale crossed the horizon during inflation. We represent with dot-dashed, horizontal lines the (absolute value) of the results obtained by evaluating at horizon crossing the analytic expressions derived above. We only plot these values for times 4 e-folds before and after horizon crossing, as they are intended for comparison around horizon crossing only.}
    \label{fig:2pt_correlators_2}
\end{figure}
In figures \ref{fig:2pt_correlators_1} and \ref{fig:2pt_correlators_2} we provide an explicit check of the expressions derived above. For this purpose we transform derivatives with respect to $N$ into derivatives with respect to cosmic time $t$. We compare the numerical evolution of each correlator, calculated with \texttt{PyTransport}, with the corresponding dot-dashed line, representing the analytic expressions derived above evaluated at horizon crossing ($k_1=aH$). If two correlators are equal numerically, e.g. $\mathcal{P}_{\chi'\phi'}$ and $\mathcal{P}_{\phi'\chi'}$, we only represent one of the two. In the left panel of figure \ref{fig:2pt_correlators_1} there is no line corresponding to the analytic correlator $\mathcal{P}_{\chi\phi} = \mathcal{P}_{\phi\chi}$ because it is zero according to the corresponding analytic expression \eqref{phi chi corr}, due to the fact that the metric is diagonal. Figures \ref{fig:2pt_correlators_1} and \ref{fig:2pt_correlators_2} show that the analytical correlators match remarkably well the numerical values at horizon crossing in most cases, with slight deviations for the mixed field-velocity cross-correlators, see the right panel of figure \ref{fig:2pt_correlators_2}.

In eqs.\eqref{first eq}-\eqref{last eq}, the leading terms in the $k_1/(aH)\ll1$ limit represent the homogeneous growing mode, to be used in the $\delta N$ calculation, while terms proportional to $k_1/(aH)$ constitute gradient corrections. By retaining only the homogeneous growing mode, the 2-point correlators at horizon crossing are
\begin{align}
    \label{first final corr}
    & \mathcal{P}_{\phi \phi}(k_1)= \left( \frac{H}{2\pi}\right)^2 \;,\\
    \label{chi chi corr}
    &\mathcal{P}_{\chi \chi}(k_1)= \left( \frac{H}{2\pi}\right)^2 e^{-2b_1\phi}  \;,\\
    & \mathcal{P}_{\phi \chi}(k_1)= \mathcal{P}_{\chi \phi}(k_1)= 0 \;, \\
    & \mathcal{P}_{\phi' \phi'}(k_1)=  \left( \frac{H}{2\pi}\right)^2  e^{2b_1 \phi} \left(b_1 \chi'\right)^2 \;, 
\end{align}
\begin{align}
    &\mathcal{P}_{\chi' \chi'}(k_1)= \left( \frac{H}{2\pi}\right)^2  e^{-2b_1\phi}  \left[\left(b_1 \phi'\right)^2 + \left(b_1 \chi' e^{b_1\phi}\right)^2 \right] \;, \\
    & \mathcal{P}_{\phi' \chi'}(k_1)= \mathcal{P}_{\chi' \phi'}(k_1)= -\left( \frac{H}{2\pi}\right)^2 {b_1}^2 \chi' \phi'  \;, \\
    & \mathcal{P}_{\phi' \phi}(k_1)=\mathcal{P}_{\phi \phi'}(k_1)= 0 \;, \\
    & \mathcal{P}_{\chi' \chi}(k_1)=\mathcal{P}_{\chi \chi'}(k_1)= -\left( \frac{H}{2\pi}\right)^2 \, e^{-2b_1\phi}\,  b_1 \phi'  \;, \\
    & \mathcal{P}_{\chi' \phi}(k_1)=\mathcal{P}_{\phi \chi'}(k_1)= \left( \frac{H}{2\pi}\right)^2 \,b_1 \chi' \;, 
\end{align}
\begin{equation}
    \label{last final corr}
    \mathcal{P}_{\phi' \chi}(k_1)=\mathcal{P}_{\chi \phi'}(k_1)= -\left( \frac{H}{2\pi}\right)^2 \,b_1 \chi' \;, 
\end{equation}
where the time-dependent functions appearing in these expressions are to be evaluated at horizon crossing, $k_1/(aH)=1$. These correlators represent the initial conditions for the $\delta N$ calculation of section \ref{sec: delta N}. 

\bibliography{refs} 

\providecommand{\href}[2]{#2}\begingroup\raggedright\begin{thebibliography}{100}

\bibitem{Planck:2018jri}
{\scshape Planck} collaboration, \emph{{Planck 2018 results. X. Constraints on
  inflation}}, \href{https://doi.org/10.1051/0004-6361/201833887}{\emph{Astron.
  Astrophys.} {\bfseries 641} (2020) A10}
  [\href{https://arxiv.org/abs/1807.06211}{{\ttfamily 1807.06211}}].

\bibitem{10.1093/mnras/168.2.399}
B.~J. Carr and S.~W. Hawking, \emph{{Black Holes in the Early Universe}},
  \href{https://doi.org/10.1093/mnras/168.2.399}{\emph{Monthly Notices of the
  Royal Astronomical Society} {\bfseries 168} (1974) 399}.

\bibitem{Sasaki:2018dmp}
M.~Sasaki, T.~Suyama, T.~Tanaka and S.~Yokoyama, \emph{{Primordial black
  holes\textemdash{}perspectives in gravitational wave astronomy}},
  \href{https://doi.org/10.1088/1361-6382/aaa7b4}{\emph{Class. Quant. Grav.}
  {\bfseries 35} (2018) 063001}
  [\href{https://arxiv.org/abs/1801.05235}{{\ttfamily 1801.05235}}].

\bibitem{Carr:2020gox}
B.~Carr, K.~Kohri, Y.~Sendouda and J.~Yokoyama, \emph{{Constraints on
  primordial black holes}},
  \href{https://doi.org/10.1088/1361-6633/ac1e31}{\emph{Rept. Prog. Phys.}
  {\bfseries 84} (2021) 116902}
  [\href{https://arxiv.org/abs/2002.12778}{{\ttfamily 2002.12778}}].

\bibitem{Bird:2016dcv}
S.~Bird, I.~Cholis, J.~B. Mu\~noz, Y.~Ali-Ha\"\i{}moud, M.~Kamionkowski, E.~D.
  Kovetz et~al., \emph{{Did LIGO detect dark matter?}},
  \href{https://doi.org/10.1103/PhysRevLett.116.201301}{\emph{Phys. Rev. Lett.}
  {\bfseries 116} (2016) 201301}
  [\href{https://arxiv.org/abs/1603.00464}{{\ttfamily 1603.00464}}].

\bibitem{Bertone:2018krk}
G.~Bertone and T.~Tait, M.~P., \emph{{A new era in the search for dark
  matter}}, \href{https://doi.org/10.1038/s41586-018-0542-z}{\emph{Nature}
  {\bfseries 562} (2018) 51}
  [\href{https://arxiv.org/abs/1810.01668}{{\ttfamily 1810.01668}}].

\bibitem{Bartolo:2018evs}
N.~Bartolo, V.~De~Luca, G.~Franciolini, A.~Lewis, M.~Peloso and A.~Riotto,
  \emph{{Primordial Black Hole Dark Matter: LISA Serendipity}},
  \href{https://doi.org/10.1103/PhysRevLett.122.211301}{\emph{Phys. Rev. Lett.}
  {\bfseries 122} (2019) 211301}
  [\href{https://arxiv.org/abs/1810.12218}{{\ttfamily 1810.12218}}].

\bibitem{Ananda:2006af}
K.~N. Ananda, C.~Clarkson and D.~Wands, \emph{{The Cosmological gravitational
  wave background from primordial density perturbations}},
  \href{https://doi.org/10.1103/PhysRevD.75.123518}{\emph{Phys. Rev. D}
  {\bfseries 75} (2007) 123518}
  [\href{https://arxiv.org/abs/gr-qc/0612013}{{\ttfamily gr-qc/0612013}}].

\bibitem{Baumann:2007zm}
D.~Baumann, P.~J. Steinhardt, K.~Takahashi and K.~Ichiki, \emph{{Gravitational
  Wave Spectrum Induced by Primordial Scalar Perturbations}},
  \href{https://doi.org/10.1103/PhysRevD.76.084019}{\emph{Phys. Rev. D}
  {\bfseries 76} (2007) 084019}
  [\href{https://arxiv.org/abs/hep-th/0703290}{{\ttfamily hep-th/0703290}}].

\bibitem{Saito:2008em}
R.~Saito, J.~Yokoyama and R.~Nagata, \emph{{Single-field inflation, anomalous
  enhancement of superhorizon fluctuations, and non-Gaussianity in primordial
  black hole formation}},
  \href{https://doi.org/10.1088/1475-7516/2008/06/024}{\emph{JCAP} {\bfseries
  06} (2008) 024} [\href{https://arxiv.org/abs/0804.3470}{{\ttfamily
  0804.3470}}].

\bibitem{Saito:2009jt}
R.~Saito and J.~Yokoyama, \emph{{Gravitational-Wave Constraints on the
  Abundance of Primordial Black Holes}},
  \href{https://doi.org/10.1143/PTP.126.351}{\emph{Prog. Theor. Phys.}
  {\bfseries 123} (2010) 867}
  [\href{https://arxiv.org/abs/0912.5317}{{\ttfamily 0912.5317}}].

\bibitem{Fumagalli:2021cel}
J.~Fumagalli, S.~Renaux-Petel and L.~T. Witkowski, \emph{{Resonant features in
  the stochastic gravitational wave background}},
  \href{https://arxiv.org/abs/2105.06481}{{\ttfamily 2105.06481}}.

\bibitem{Witkowski:2021raz}
L.~T. Witkowski, G.~Dom\`enech, J.~Fumagalli and S.~Renaux-Petel,
  \emph{{Expansion history-dependent oscillations in the scalar-induced
  gravitational wave background}},
  \href{https://arxiv.org/abs/2110.09480}{{\ttfamily 2110.09480}}.

\bibitem{Braglia:2020taf}
M.~Braglia, X.~Chen and D.~K. Hazra, \emph{{Probing Primordial Features with
  the Stochastic Gravitational Wave Background}},
  \href{https://doi.org/10.1088/1475-7516/2021/03/005}{\emph{JCAP} {\bfseries
  03} (2021) 005} [\href{https://arxiv.org/abs/2012.05821}{{\ttfamily
  2012.05821}}].

\bibitem{Motohashi:2017kbs}
H.~Motohashi and W.~Hu, \emph{{Primordial Black Holes and Slow-Roll
  Violation}}, \href{https://doi.org/10.1103/PhysRevD.96.063503}{\emph{Phys.
  Rev. D} {\bfseries 96} (2017) 063503}
  [\href{https://arxiv.org/abs/1706.06784}{{\ttfamily 1706.06784}}].

\bibitem{Garcia-Bellido:2017mdw}
J.~Garcia-Bellido and E.~Ruiz~Morales, \emph{{Primordial black holes from
  single field models of inflation}},
  \href{https://doi.org/10.1016/j.dark.2017.09.007}{\emph{Phys. Dark Univ.}
  {\bfseries 18} (2017) 47} [\href{https://arxiv.org/abs/1702.03901}{{\ttfamily
  1702.03901}}].

\bibitem{Germani:2017bcs}
C.~Germani and T.~Prokopec, \emph{{On primordial black holes from an inflection
  point}}, \href{https://doi.org/10.1016/j.dark.2017.09.001}{\emph{Phys. Dark
  Univ.} {\bfseries 18} (2017) 6}
  [\href{https://arxiv.org/abs/1706.04226}{{\ttfamily 1706.04226}}].

\bibitem{Ballesteros:2017fsr}
G.~Ballesteros and M.~Taoso, \emph{{Primordial black hole dark matter from
  single field inflation}},
  \href{https://doi.org/10.1103/PhysRevD.97.023501}{\emph{Phys. Rev. D}
  {\bfseries 97} (2018) 023501}
  [\href{https://arxiv.org/abs/1709.05565}{{\ttfamily 1709.05565}}].

\bibitem{Cicoli:2018asa}
M.~Cicoli, V.~A. Diaz and F.~G. Pedro, \emph{{Primordial Black Holes from
  String Inflation}},
  \href{https://doi.org/10.1088/1475-7516/2018/06/034}{\emph{JCAP} {\bfseries
  06} (2018) 034} [\href{https://arxiv.org/abs/1803.02837}{{\ttfamily
  1803.02837}}].

\bibitem{Dalianis:2018frf}
I.~Dalianis, A.~Kehagias and G.~Tringas, \emph{{Primordial black holes from
  \ensuremath{\alpha}-attractors}},
  \href{https://doi.org/10.1088/1475-7516/2019/01/037}{\emph{JCAP} {\bfseries
  01} (2019) 037} [\href{https://arxiv.org/abs/1805.09483}{{\ttfamily
  1805.09483}}].

\bibitem{Passaglia:2018ixg}
S.~Passaglia, W.~Hu and H.~Motohashi, \emph{{Primordial black holes and local
  non-Gaussianity in canonical inflation}},
  \href{https://doi.org/10.1103/PhysRevD.99.043536}{\emph{Phys. Rev. D}
  {\bfseries 99} (2019) 043536}
  [\href{https://arxiv.org/abs/1812.08243}{{\ttfamily 1812.08243}}].

\bibitem{Bhaumik:2019tvl}
N.~Bhaumik and R.~K. Jain, \emph{{Primordial black holes dark matter from
  inflection point models of inflation and the effects of reheating}},
  \href{https://doi.org/10.1088/1475-7516/2020/01/037}{\emph{JCAP} {\bfseries
  01} (2020) 037} [\href{https://arxiv.org/abs/1907.04125}{{\ttfamily
  1907.04125}}].

\bibitem{Balaji:2022rsy}
S.~Balaji, J.~Silk and Y.-P. Wu, \emph{{Induced gravitational waves from the
  cosmic coincidence}},
  \href{https://doi.org/10.1088/1475-7516/2022/06/008}{\emph{JCAP} {\bfseries
  06} (2022) 008} [\href{https://arxiv.org/abs/2202.00700}{{\ttfamily
  2202.00700}}].

\bibitem{Ragavendra:2023ret}
H.~V. Ragavendra and L.~Sriramkumar, \emph{{Observational Imprints of Enhanced
  Scalar Power on Small Scales in Ultra Slow Roll Inflation and Associated
  Non-Gaussianities}},
  \href{https://doi.org/10.3390/galaxies11010034}{\emph{Galaxies} {\bfseries
  11} (2023) 34} [\href{https://arxiv.org/abs/2301.08887}{{\ttfamily
  2301.08887}}].

\bibitem{Geller:2022nkr}
S.~R. Geller, W.~Qin, E.~McDonough and D.~I. Kaiser, \emph{{Primordial black
  holes from multifield inflation with nonminimal couplings}},
  \href{https://doi.org/10.1103/PhysRevD.106.063535}{\emph{Phys. Rev. D}
  {\bfseries 106} (2022) 063535}
  [\href{https://arxiv.org/abs/2205.04471}{{\ttfamily 2205.04471}}].

\bibitem{Qin:2023lgo}
W.~Qin, S.~R. Geller, S.~Balaji, E.~McDonough and D.~I. Kaiser, \emph{{Planck
  Constraints and Gravitational Wave Forecasts for Primordial Black Hole Dark
  Matter Seeded by Multifield Inflation}},
  \href{https://arxiv.org/abs/2303.02168}{{\ttfamily 2303.02168}}.

\bibitem{Baumann:2014nda}
D.~Baumann and L.~McAllister, \emph{{Inflation and String Theory}}, Cambridge
  Monographs on Mathematical Physics. Cambridge University Press, 2015,
  \href{https://doi.org/10.1017/CBO9781316105733}{10.1017/CBO9781316105733},
  [\href{https://arxiv.org/abs/1404.2601}{{\ttfamily 1404.2601}}].

\bibitem{Fumagalli:2020adf}
J.~Fumagalli, S.~Renaux-Petel, J.~W. Ronayne and L.~T. Witkowski,
  \emph{{Turning in the landscape: a new mechanism for generating Primordial
  Black Holes}},  \href{https://arxiv.org/abs/2004.08369}{{\ttfamily
  2004.08369}}.

\bibitem{Palma:2020ejf}
G.~A. Palma, S.~Sypsas and C.~Zenteno, \emph{{Seeding primordial black holes in
  multifield inflation}},
  \href{https://doi.org/10.1103/PhysRevLett.125.121301}{\emph{Phys. Rev. Lett.}
  {\bfseries 125} (2020) 121301}
  [\href{https://arxiv.org/abs/2004.06106}{{\ttfamily 2004.06106}}].

\bibitem{Braglia:2020eai}
M.~Braglia, D.~K. Hazra, F.~Finelli, G.~F. Smoot, L.~Sriramkumar and A.~A.
  Starobinsky, \emph{{Generating PBHs and small-scale GWs in two-field models
  of inflation}},
  \href{https://doi.org/10.1088/1475-7516/2020/08/001}{\emph{JCAP} {\bfseries
  08} (2020) 001} [\href{https://arxiv.org/abs/2005.02895}{{\ttfamily
  2005.02895}}].

\bibitem{Iacconi:2021ltm}
L.~Iacconi, H.~Assadullahi, M.~Fasiello and D.~Wands, \emph{{Revisiting
  small-scale fluctuations in \ensuremath{\alpha}-attractor models of
  inflation}}, \href{https://doi.org/10.1088/1475-7516/2022/06/007}{\emph{JCAP}
  {\bfseries 06} (2022) 007}
  [\href{https://arxiv.org/abs/2112.05092}{{\ttfamily 2112.05092}}].

\bibitem{Kallosh:2022vha}
R.~Kallosh and A.~Linde, \emph{{Dilaton-axion inflation with PBHs and GWs}},
  \href{https://doi.org/10.1088/1475-7516/2022/08/037}{\emph{JCAP} {\bfseries
  08} (2022) 037} [\href{https://arxiv.org/abs/2203.10437}{{\ttfamily
  2203.10437}}].

\bibitem{Braglia:2022phb}
M.~Braglia, A.~Linde, R.~Kallosh and F.~Finelli, \emph{{Hybrid
  $\alpha$-attractors, primordial black holes and gravitational wave
  backgrounds}},  \href{https://arxiv.org/abs/2211.14262}{{\ttfamily
  2211.14262}}.

\bibitem{Kallosh:2013hoa}
R.~Kallosh and A.~Linde, \emph{{Universality Class in Conformal Inflation}},
  \href{https://doi.org/10.1088/1475-7516/2013/07/002}{\emph{JCAP} {\bfseries
  07} (2013) 002} [\href{https://arxiv.org/abs/1306.5220}{{\ttfamily
  1306.5220}}].

\bibitem{Kallosh:2013daa}
R.~Kallosh and A.~Linde, \emph{{Multi-field Conformal Cosmological
  Attractors}},
  \href{https://doi.org/10.1088/1475-7516/2013/12/006}{\emph{JCAP} {\bfseries
  12} (2013) 006} [\href{https://arxiv.org/abs/1309.2015}{{\ttfamily
  1309.2015}}].

\bibitem{Ferrara:2013rsa}
S.~Ferrara, R.~Kallosh, A.~Linde and M.~Porrati, \emph{{Minimal Supergravity
  Models of Inflation}},
  \href{https://doi.org/10.1103/PhysRevD.88.085038}{\emph{Phys. Rev. D}
  {\bfseries 88} (2013) 085038}
  [\href{https://arxiv.org/abs/1307.7696}{{\ttfamily 1307.7696}}].

\bibitem{Kallosh:2013pby}
R.~Kallosh and A.~Linde, \emph{{Superconformal generalization of the chaotic
  inflation model $\frac{\lambda}{4} \phi^{4} - \frac{\xi}{2} \phi^{2}R$}},
  \href{https://doi.org/10.1088/1475-7516/2013/06/027}{\emph{JCAP} {\bfseries
  06} (2013) 027} [\href{https://arxiv.org/abs/1306.3211}{{\ttfamily
  1306.3211}}].

\bibitem{Kallosh:2013lkr}
R.~Kallosh and A.~Linde, \emph{{Superconformal generalizations of the
  Starobinsky model}},
  \href{https://doi.org/10.1088/1475-7516/2013/06/028}{\emph{JCAP} {\bfseries
  06} (2013) 028} [\href{https://arxiv.org/abs/1306.3214}{{\ttfamily
  1306.3214}}].

\bibitem{Kallosh:2013maa}
R.~Kallosh and A.~Linde, \emph{{Non-minimal Inflationary Attractors}},
  \href{https://doi.org/10.1088/1475-7516/2013/10/033}{\emph{JCAP} {\bfseries
  10} (2013) 033} [\href{https://arxiv.org/abs/1307.7938}{{\ttfamily
  1307.7938}}].

\bibitem{Kallosh:2013tua}
R.~Kallosh, A.~Linde and D.~Roest, \emph{{Universal Attractor for Inflation at
  Strong Coupling}},
  \href{https://doi.org/10.1103/PhysRevLett.112.011303}{\emph{Phys. Rev. Lett.}
  {\bfseries 112} (2014) 011303}
  [\href{https://arxiv.org/abs/1310.3950}{{\ttfamily 1310.3950}}].

\bibitem{Kallosh:2013yoa}
R.~Kallosh, A.~Linde and D.~Roest, \emph{{Superconformal Inflationary
  $\alpha$-Attractors}},
  \href{https://doi.org/10.1007/JHEP11(2013)198}{\emph{JHEP} {\bfseries 11}
  (2013) 198} [\href{https://arxiv.org/abs/1311.0472}{{\ttfamily 1311.0472}}].

\bibitem{Kallosh:2015zsa}
R.~Kallosh and A.~Linde, \emph{{Escher in the Sky}},
  \href{https://doi.org/10.1016/j.crhy.2015.07.004}{\emph{Comptes Rendus
  Physique} {\bfseries 16} (2015) 914}
  [\href{https://arxiv.org/abs/1503.06785}{{\ttfamily 1503.06785}}].

\bibitem{Carrasco:2015uma}
J.~J.~M. Carrasco, R.~Kallosh, A.~Linde and D.~Roest, \emph{{Hyperbolic
  geometry of cosmological attractors}},
  \href{https://doi.org/10.1103/PhysRevD.92.041301}{\emph{Phys. Rev. D}
  {\bfseries 92} (2015) 041301}
  [\href{https://arxiv.org/abs/1504.05557}{{\ttfamily 1504.05557}}].

\bibitem{Galante:2014ifa}
M.~Galante, R.~Kallosh, A.~Linde and D.~Roest, \emph{{Unity of Cosmological
  Inflation Attractors}},
  \href{https://doi.org/10.1103/PhysRevLett.114.141302}{\emph{Phys. Rev. Lett.}
  {\bfseries 114} (2015) 141302}
  [\href{https://arxiv.org/abs/1412.3797}{{\ttfamily 1412.3797}}].

\bibitem{Fumagalli:2016sof}
J.~Fumagalli, \emph{{Renormalization Group independence of Cosmological
  Attractors}},
  \href{https://doi.org/10.1016/j.physletb.2017.04.017}{\emph{Phys. Lett. B}
  {\bfseries 769} (2017) 451}
  [\href{https://arxiv.org/abs/1611.04997}{{\ttfamily 1611.04997}}].

\bibitem{Kallosh:2022feu}
R.~Kallosh and A.~Linde, \emph{{Polynomial \ensuremath{\alpha}-attractors}},
  \href{https://doi.org/10.1088/1475-7516/2022/04/017}{\emph{JCAP} {\bfseries
  04} (2022) 017} [\href{https://arxiv.org/abs/2202.06492}{{\ttfamily
  2202.06492}}].

\bibitem{Brown:2017osf}
A.~R. Brown, \emph{{Hyperbolic Inflation}},
  \href{https://doi.org/10.1103/PhysRevLett.121.251601}{\emph{Phys. Rev. Lett.}
  {\bfseries 121} (2018) 251601}
  [\href{https://arxiv.org/abs/1705.03023}{{\ttfamily 1705.03023}}].

\bibitem{Mizuno:2017idt}
S.~Mizuno and S.~Mukohyama, \emph{{Primordial perturbations from inflation with
  a hyperbolic field-space}},
  \href{https://doi.org/10.1103/PhysRevD.96.103533}{\emph{Phys. Rev. D}
  {\bfseries 96} (2017) 103533}
  [\href{https://arxiv.org/abs/1707.05125}{{\ttfamily 1707.05125}}].

\bibitem{Achucarro:2017ing}
A.~Ach\'ucarro, R.~Kallosh, A.~Linde, D.-G. Wang and Y.~Welling,
  \emph{{Universality of multi-field $\alpha$-attractors}},
  \href{https://doi.org/10.1088/1475-7516/2018/04/028}{\emph{JCAP} {\bfseries
  04} (2018) 028} [\href{https://arxiv.org/abs/1711.09478}{{\ttfamily
  1711.09478}}].

\bibitem{Linde:2018hmx}
A.~Linde, D.-G. Wang, Y.~Welling, Y.~Yamada and A.~Ach\'ucarro,
  \emph{{Hypernatural inflation}},
  \href{https://doi.org/10.1088/1475-7516/2018/07/035}{\emph{JCAP} {\bfseries
  07} (2018) 035} [\href{https://arxiv.org/abs/1803.09911}{{\ttfamily
  1803.09911}}].

\bibitem{Christodoulidis:2018qdw}
P.~Christodoulidis, D.~Roest and E.~I. Sfakianakis, \emph{{Angular inflation in
  multi-field $\alpha$-attractors}},
  \href{https://doi.org/10.1088/1475-7516/2019/11/002}{\emph{JCAP} {\bfseries
  11} (2019) 002} [\href{https://arxiv.org/abs/1803.09841}{{\ttfamily
  1803.09841}}].

\bibitem{Bullock:1996at}
J.~S. Bullock and J.~R. Primack, \emph{{NonGaussian fluctuations and primordial
  black holes from inflation}},
  \href{https://doi.org/10.1103/PhysRevD.55.7423}{\emph{Phys. Rev. D}
  {\bfseries 55} (1997) 7423}
  [\href{https://arxiv.org/abs/astro-ph/9611106}{{\ttfamily
  astro-ph/9611106}}].

\bibitem{Yokoyama:1998pt}
J.~Yokoyama, \emph{{Chaotic new inflation and formation of primordial black
  holes}}, \href{https://doi.org/10.1103/PhysRevD.58.083510}{\emph{Phys. Rev.
  D} {\bfseries 58} (1998) 083510}
  [\href{https://arxiv.org/abs/astro-ph/9802357}{{\ttfamily
  astro-ph/9802357}}].

\bibitem{Byrnes:2012yx}
C.~T. Byrnes, E.~J. Copeland and A.~M. Green, \emph{{Primordial black holes as
  a tool for constraining non-Gaussianity}},
  \href{https://doi.org/10.1103/PhysRevD.86.043512}{\emph{Phys. Rev. D}
  {\bfseries 86} (2012) 043512}
  [\href{https://arxiv.org/abs/1206.4188}{{\ttfamily 1206.4188}}].

\bibitem{Young:2013oia}
S.~Young and C.~T. Byrnes, \emph{{Primordial black holes in non-Gaussian
  regimes}}, \href{https://doi.org/10.1088/1475-7516/2013/08/052}{\emph{JCAP}
  {\bfseries 08} (2013) 052} [\href{https://arxiv.org/abs/1307.4995}{{\ttfamily
  1307.4995}}].

\bibitem{Bugaev:2013vba}
E.~V. Bugaev and P.~A. Klimai, \emph{{Primordial black hole constraints for
  curvaton models with predicted large non-Gaussianity}},
  \href{https://doi.org/10.1142/S021827181350034X}{\emph{Int. J. Mod. Phys. D}
  {\bfseries 22} (2013) 1350034}
  [\href{https://arxiv.org/abs/1303.3146}{{\ttfamily 1303.3146}}].

\bibitem{Young:2014oea}
S.~Young and C.~T. Byrnes, \emph{{Long-short wavelength mode coupling tightens
  primordial black hole constraints}},
  \href{https://doi.org/10.1103/PhysRevD.91.083521}{\emph{Phys. Rev. D}
  {\bfseries 91} (2015) 083521}
  [\href{https://arxiv.org/abs/1411.4620}{{\ttfamily 1411.4620}}].

\bibitem{Young:2015cyn}
S.~Young, D.~Regan and C.~T. Byrnes, \emph{{Influence of large local and
  non-local bispectra on primordial black hole abundance}},
  \href{https://doi.org/10.1088/1475-7516/2016/02/029}{\emph{JCAP} {\bfseries
  02} (2016) 029} [\href{https://arxiv.org/abs/1512.07224}{{\ttfamily
  1512.07224}}].

\bibitem{Franciolini:2018vbk}
G.~Franciolini, A.~Kehagias, S.~Matarrese and A.~Riotto, \emph{{Primordial
  Black Holes from Inflation and non-Gaussianity}},
  \href{https://doi.org/10.1088/1475-7516/2018/03/016}{\emph{JCAP} {\bfseries
  03} (2018) 016} [\href{https://arxiv.org/abs/1801.09415}{{\ttfamily
  1801.09415}}].

\bibitem{Atal:2018neu}
V.~Atal and C.~Germani, \emph{{The role of non-gaussianities in Primordial
  Black Hole formation}},
  \href{https://doi.org/10.1016/j.dark.2019.100275}{\emph{Phys. Dark Univ.}
  {\bfseries 24} (2019) 100275}
  [\href{https://arxiv.org/abs/1811.07857}{{\ttfamily 1811.07857}}].

\bibitem{DeLuca:2019qsy}
V.~De~Luca, G.~Franciolini, A.~Kehagias, M.~Peloso, A.~Riotto and C.~\"Unal,
  \emph{{The Ineludible non-Gaussianity of the Primordial Black Hole
  Abundance}}, \href{https://doi.org/10.1088/1475-7516/2019/07/048}{\emph{JCAP}
  {\bfseries 07} (2019) 048}
  [\href{https://arxiv.org/abs/1904.00970}{{\ttfamily 1904.00970}}].

\bibitem{Ozsoy:2021qrg}
O.~\"Ozsoy and G.~Tasinato, \emph{{CMB \ensuremath{\mu}T cross correlations as
  a probe of primordial black hole scenarios}},
  \href{https://doi.org/10.1103/PhysRevD.104.043526}{\emph{Phys. Rev. D}
  {\bfseries 104} (2021) 043526}
  [\href{https://arxiv.org/abs/2104.12792}{{\ttfamily 2104.12792}}].

\bibitem{Taoso:2021uvl}
M.~Taoso and A.~Urbano, \emph{{Non-gaussianities for primordial black hole
  formation}}, \href{https://doi.org/10.1088/1475-7516/2021/08/016}{\emph{JCAP}
  {\bfseries 08} (2021) 016}
  [\href{https://arxiv.org/abs/2102.03610}{{\ttfamily 2102.03610}}].

\bibitem{Davies:2021loj}
M.~W. Davies, P.~Carrilho and D.~J. Mulryne, \emph{{Non-Gaussianity in
  inflationary scenarios for primordial black holes}},
  \href{https://arxiv.org/abs/2110.08189}{{\ttfamily 2110.08189}}.

\bibitem{Ferrante:2022mui}
G.~Ferrante, G.~Franciolini, A.~Iovino, Junior. and A.~Urbano,
  \emph{{Primordial non-Gaussianity up to all orders: Theoretical aspects and
  implications for primordial black hole models}},
  \href{https://doi.org/10.1103/PhysRevD.107.043520}{\emph{Phys. Rev. D}
  {\bfseries 107} (2023) 043520}
  [\href{https://arxiv.org/abs/2211.01728}{{\ttfamily 2211.01728}}].

\bibitem{Gow:2022jfb}
A.~D. Gow, H.~Assadullahi, J.~H.~P. Jackson, K.~Koyama, V.~Vennin and D.~Wands,
  \emph{{Non-perturbative non-Gaussianity and primordial black holes}},
  \href{https://arxiv.org/abs/2211.08348}{{\ttfamily 2211.08348}}.

\bibitem{Domenech:2021ztg}
G.~Dom\`enech, \emph{{Scalar induced gravitational waves review}},
  \href{https://arxiv.org/abs/2109.01398}{{\ttfamily 2109.01398}}.

\bibitem{Cai:2018dig}
R.-g. Cai, S.~Pi and M.~Sasaki, \emph{{Gravitational Waves Induced by
  non-Gaussian Scalar Perturbations}},
  \href{https://doi.org/10.1103/PhysRevLett.122.201101}{\emph{Phys. Rev. Lett.}
  {\bfseries 122} (2019) 201101}
  [\href{https://arxiv.org/abs/1810.11000}{{\ttfamily 1810.11000}}].

\bibitem{Unal:2018yaa}
C.~Unal, \emph{{Imprints of Primordial Non-Gaussianity on Gravitational Wave
  Spectrum}}, \href{https://doi.org/10.1103/PhysRevD.99.041301}{\emph{Phys.
  Rev. D} {\bfseries 99} (2019) 041301}
  [\href{https://arxiv.org/abs/1811.09151}{{\ttfamily 1811.09151}}].

\bibitem{Yuan:2020iwf}
C.~Yuan and Q.-G. Huang, \emph{{Gravitational waves induced by the local-type
  non-Gaussian curvature perturbations}},
  \href{https://doi.org/10.1016/j.physletb.2021.136606}{\emph{Phys. Lett. B}
  {\bfseries 821} (2021) 136606}
  [\href{https://arxiv.org/abs/2007.10686}{{\ttfamily 2007.10686}}].

\bibitem{Atal:2021jyo}
V.~Atal and G.~Dom\`enech, \emph{{Probing non-Gaussianities with the high
  frequency tail of induced gravitational waves}},
  \href{https://doi.org/10.1088/1475-7516/2021/06/001}{\emph{JCAP} {\bfseries
  06} (2021) 001} [\href{https://arxiv.org/abs/2103.01056}{{\ttfamily
  2103.01056}}].

\bibitem{Adshead:2021hnm}
P.~Adshead, K.~D. Lozanov and Z.~J. Weiner, \emph{{Non-Gaussianity and the
  induced gravitational wave background}},
  \href{https://doi.org/10.1088/1475-7516/2021/10/080}{\emph{JCAP} {\bfseries
  10} (2021) 080} [\href{https://arxiv.org/abs/2105.01659}{{\ttfamily
  2105.01659}}].

\bibitem{Ragavendra:2020sop}
H.~V. Ragavendra, P.~Saha, L.~Sriramkumar and J.~Silk, \emph{{Primordial black
  holes and secondary gravitational waves from ultraslow roll and punctuated
  inflation}}, \href{https://doi.org/10.1103/PhysRevD.103.083510}{\emph{Phys.
  Rev. D} {\bfseries 103} (2021) 083510}
  [\href{https://arxiv.org/abs/2008.12202}{{\ttfamily 2008.12202}}].

\bibitem{Garcia-Saenz:2022tzu}
S.~Garcia-Saenz, L.~Pinol, S.~Renaux-Petel and D.~Werth, \emph{{No-go theorem
  for scalar-trispectrum-induced gravitational waves}},
  \href{https://doi.org/10.1088/1475-7516/2023/03/057}{\emph{JCAP} {\bfseries
  03} (2023) 057} [\href{https://arxiv.org/abs/2207.14267}{{\ttfamily
  2207.14267}}].

\bibitem{Inomata:2022yte}
K.~Inomata, M.~Braglia and X.~Chen, \emph{{Questions on calculation of
  primordial power spectrum with large spikes: the resonance model case}},
  \href{https://arxiv.org/abs/2211.02586}{{\ttfamily 2211.02586}}.

\bibitem{Kristiano:2022maq}
J.~Kristiano and J.~Yokoyama, \emph{{Ruling Out Primordial Black Hole Formation
  From Single-Field Inflation}},
  \href{https://arxiv.org/abs/2211.03395}{{\ttfamily 2211.03395}}.

\bibitem{Riotto:2023hoz}
A.~Riotto, \emph{{The Primordial Black Hole Formation from Single-Field
  Inflation is Not Ruled Out}},
  \href{https://arxiv.org/abs/2301.00599}{{\ttfamily 2301.00599}}.

\bibitem{Choudhury:2023vuj}
S.~Choudhury, M.~R. Gangopadhyay and M.~Sami, \emph{{No-go for the formation of
  heavy mass Primordial Black Holes in Single Field Inflation}},
  \href{https://arxiv.org/abs/2301.10000}{{\ttfamily 2301.10000}}.

\bibitem{Choudhury:2023jlt}
S.~Choudhury, S.~Panda and M.~Sami, \emph{{No-go for PBH formation in EFT of
  single field inflation}},  \href{https://arxiv.org/abs/2302.05655}{{\ttfamily
  2302.05655}}.

\bibitem{Kristiano:2023scm}
J.~Kristiano and J.~Yokoyama, \emph{{Response to criticism on ''Ruling Out
  Primordial Black Hole Formation From Single-Field Inflation'': A note on
  bispectrum and one-loop correction in single-field inflation with primordial
  black hole formation}},  \href{https://arxiv.org/abs/2303.00341}{{\ttfamily
  2303.00341}}.

\bibitem{Riotto:2023gpm}
A.~Riotto, \emph{{The Primordial Black Hole Formation from Single-Field
  Inflation is Still Not Ruled Out}},
  \href{https://arxiv.org/abs/2303.01727}{{\ttfamily 2303.01727}}.

\bibitem{Firouzjahi:2023aum}
H.~Firouzjahi, \emph{{One-loop Corrections in Power Spectrum in Single Field
  Inflation}},  \href{https://arxiv.org/abs/2303.12025}{{\ttfamily
  2303.12025}}.

\bibitem{Motohashi:2023syh}
H.~Motohashi and Y.~Tada, \emph{{Squeezed bispectrum and one-loop corrections
  in transient constant-roll inflation}},
  \href{https://arxiv.org/abs/2303.16035}{{\ttfamily 2303.16035}}.

\bibitem{Choudhury:2023rks}
S.~Choudhury, S.~Panda and M.~Sami, \emph{{Quantum loop effects on the power
  spectrum and constraints on primordial black holes}},
  \href{https://arxiv.org/abs/2303.06066}{{\ttfamily 2303.06066}}.

\bibitem{Choudhury:2023hvf}
S.~Choudhury, S.~Panda and M.~Sami, \emph{{Galileon inflation evades the no-go
  for PBH formation in the single-field framework}},
  \href{https://arxiv.org/abs/2304.04065}{{\ttfamily 2304.04065}}.

\bibitem{Firouzjahi:2023ahg}
H.~Firouzjahi and A.~Riotto, \emph{{Primordial Black Holes and Loops in
  Single-Field Inflation}},  \href{https://arxiv.org/abs/2304.07801}{{\ttfamily
  2304.07801}}.

\bibitem{Firouzjahi:2023btw}
H.~Firouzjahi, \emph{{Loop Corrections in Gravitational Wave Spectrum in Single
  Field Inflation}},  \href{https://arxiv.org/abs/2305.01527}{{\ttfamily
  2305.01527}}.

\bibitem{Cheng:2021lif}
S.-L. Cheng, D.-S. Lee and K.-W. Ng, \emph{{Power spectrum of primordial
  perturbations during ultra-slow-roll inflation with back reaction effects}},
  \href{https://doi.org/10.1016/j.physletb.2022.136956}{\emph{Phys. Lett. B}
  {\bfseries 827} (2022) 136956}
  [\href{https://arxiv.org/abs/2106.09275}{{\ttfamily 2106.09275}}].

\bibitem{Mulryne:2016mzv}
D.~J. Mulryne and J.~W. Ronayne, \emph{{PyTransport: A Python package for the
  calculation of inflationary correlation functions}},
  \href{https://doi.org/10.21105/joss.00494}{\emph{J. Open Source Softw.}
  {\bfseries 3} (2018) 494} [\href{https://arxiv.org/abs/1609.00381}{{\ttfamily
  1609.00381}}].

\bibitem{Ronayne:2017qzn}
J.~W. Ronayne and D.~J. Mulryne, \emph{{Numerically evaluating the bispectrum
  in curved field-space\textemdash{} with PyTransport 2.0}},
  \href{https://doi.org/10.1088/1475-7516/2018/01/023}{\emph{JCAP} {\bfseries
  01} (2018) 023} [\href{https://arxiv.org/abs/1708.07130}{{\ttfamily
  1708.07130}}].

\bibitem{Gong:2011uw}
J.-O. Gong and T.~Tanaka, \emph{{A covariant approach to general field space
  metric in multi-field inflation}},
  \href{https://doi.org/10.1088/1475-7516/2012/02/E01}{\emph{JCAP} {\bfseries
  03} (2011) 015} [\href{https://arxiv.org/abs/1101.4809}{{\ttfamily
  1101.4809}}].

\bibitem{Gordon:2000hv}
C.~Gordon, D.~Wands, B.~A. Bassett and R.~Maartens, \emph{{Adiabatic and
  entropy perturbations from inflation}},
  \href{https://doi.org/10.1103/PhysRevD.63.023506}{\emph{Phys. Rev. D}
  {\bfseries 63} (2000) 023506}
  [\href{https://arxiv.org/abs/astro-ph/0009131}{{\ttfamily
  astro-ph/0009131}}].

\bibitem{GrootNibbelink:2001qt}
S.~Groot~Nibbelink and B.~J.~W. van Tent, \emph{{Scalar perturbations during
  multiple field slow-roll inflation}},
  \href{https://doi.org/10.1088/0264-9381/19/4/302}{\emph{Class. Quant. Grav.}
  {\bfseries 19} (2002) 613}
  [\href{https://arxiv.org/abs/hep-ph/0107272}{{\ttfamily hep-ph/0107272}}].

\bibitem{Renaux-Petel:2015mga}
S.~Renaux-Petel and K.~Turzy\'nski, \emph{{Geometrical Destabilization of
  Inflation}},
  \href{https://doi.org/10.1103/PhysRevLett.117.141301}{\emph{Phys. Rev. Lett.}
  {\bfseries 117} (2016) 141301}
  [\href{https://arxiv.org/abs/1510.01281}{{\ttfamily 1510.01281}}].

\bibitem{Turzynski:2014tza}
S.~Renaux-Petel and K.~Turzynski, \emph{{On reaching the adiabatic limit in
  multi-field inflation}},
  \href{https://doi.org/10.1088/1475-7516/2015/06/010}{\emph{JCAP} {\bfseries
  06} (2015) 010} [\href{https://arxiv.org/abs/1405.6195}{{\ttfamily
  1405.6195}}].

\bibitem{Renaux-Petel:2017dia}
S.~Renaux-Petel, K.~Turzy\'nski and V.~Vennin, \emph{{Geometrical
  destabilization, premature end of inflation and Bayesian model selection}},
  \href{https://doi.org/10.1088/1475-7516/2017/11/006}{\emph{JCAP} {\bfseries
  11} (2017) 006} [\href{https://arxiv.org/abs/1706.01835}{{\ttfamily
  1706.01835}}].

\bibitem{Garcia-Saenz:2018ifx}
S.~Garcia-Saenz, S.~Renaux-Petel and J.~Ronayne, \emph{{Primordial fluctuations
  and non-Gaussianities in sidetracked inflation}},
  \href{https://doi.org/10.1088/1475-7516/2018/07/057}{\emph{JCAP} {\bfseries
  1807} (2018) 057} [\href{https://arxiv.org/abs/1804.11279}{{\ttfamily
  1804.11279}}].

\bibitem{Garcia-Saenz:2018vqf}
S.~Garcia-Saenz and S.~Renaux-Petel, \emph{{Flattened non-Gaussianities from
  the effective field theory of inflation with imaginary speed of sound}},
  \href{https://doi.org/10.1088/1475-7516/2018/11/005}{\emph{JCAP} {\bfseries
  11} (2018) 005} [\href{https://arxiv.org/abs/1805.12563}{{\ttfamily
  1805.12563}}].

\bibitem{Grocholski:2019mot}
O.~Grocholski, M.~Kalinowski, M.~Kolanowski, S.~Renaux-Petel, K.~Turzy\'nski
  and V.~Vennin, \emph{{On backreaction effects in geometrical destabilisation
  of inflation}},
  \href{https://doi.org/10.1088/1475-7516/2019/05/008}{\emph{JCAP} {\bfseries
  05} (2019) 008} [\href{https://arxiv.org/abs/1901.10468}{{\ttfamily
  1901.10468}}].

\bibitem{Dias:2016rjq}
M.~Dias, J.~Frazer, D.~J. Mulryne and D.~Seery, \emph{{Numerical evaluation of
  the bispectrum in multiple field inflation\textemdash{}the transport approach
  with code}}, \href{https://doi.org/10.1088/1475-7516/2016/12/033}{\emph{JCAP}
  {\bfseries 12} (2016) 033}
  [\href{https://arxiv.org/abs/1609.00379}{{\ttfamily 1609.00379}}].

\bibitem{Mulryne:2013uka}
D.~J. Mulryne, \emph{{Transporting non-Gaussianity from sub to super-horizon
  scales}}, \href{https://doi.org/10.1088/1475-7516/2013/09/010}{\emph{JCAP}
  {\bfseries 09} (2013) 010} [\href{https://arxiv.org/abs/1302.3842}{{\ttfamily
  1302.3842}}].

\bibitem{Seery:2012vj}
D.~Seery, D.~J. Mulryne, J.~Frazer and R.~H. Ribeiro, \emph{{Inflationary
  perturbation theory is geometrical optics in phase space}},
  \href{https://doi.org/10.1088/1475-7516/2012/09/010}{\emph{JCAP} {\bfseries
  09} (2012) 010} [\href{https://arxiv.org/abs/1203.2635}{{\ttfamily
  1203.2635}}].

\bibitem{Mulryne:2010rp}
D.~J. Mulryne, D.~Seery and D.~Wesley, \emph{{Moment transport equations for
  the primordial curvature perturbation}},
  \href{https://doi.org/10.1088/1475-7516/2011/04/030}{\emph{JCAP} {\bfseries
  04} (2011) 030} [\href{https://arxiv.org/abs/1008.3159}{{\ttfamily
  1008.3159}}].

\bibitem{Mulryne:2009kh}
D.~J. Mulryne, D.~Seery and D.~Wesley, \emph{{Moment transport equations for
  non-Gaussianity}},
  \href{https://doi.org/10.1088/1475-7516/2010/01/024}{\emph{JCAP} {\bfseries
  01} (2010) 024} [\href{https://arxiv.org/abs/0909.2256}{{\ttfamily
  0909.2256}}].

\bibitem{Dias:2015rca}
M.~Dias, J.~Frazer and D.~Seery, \emph{{Computing observables in curved
  multifield models of inflation\textemdash{}A guide (with code) to the
  transport method}},
  \href{https://doi.org/10.1088/1475-7516/2015/12/030}{\emph{JCAP} {\bfseries
  12} (2015) 030} [\href{https://arxiv.org/abs/1502.03125}{{\ttfamily
  1502.03125}}].

\bibitem{Seery:2016lko}
D.~Seery, \emph{{CppTransport: a platform to automate calculation of
  inflationary correlation functions}},
  \href{https://arxiv.org/abs/1609.00380}{{\ttfamily 1609.00380}}.

\bibitem{Butchers:2018hds}
S.~Butchers and D.~Seery, \emph{{Numerical evaluation of inflationary 3-point
  functions on curved field space\textemdash{}with the transport method
  \textbackslash{}\& CppTransport}},
  \href{https://doi.org/10.1088/1475-7516/2018/07/031}{\emph{JCAP} {\bfseries
  07} (2018) 031} [\href{https://arxiv.org/abs/1803.10563}{{\ttfamily
  1803.10563}}].

\bibitem{Dias:2014msa}
M.~Dias, J.~Elliston, J.~Frazer, D.~Mulryne and D.~Seery, \emph{{The curvature
  perturbation at second order}},
  \href{https://doi.org/10.1088/1475-7516/2015/02/040}{\emph{JCAP} {\bfseries
  02} (2015) 040} [\href{https://arxiv.org/abs/1410.3491}{{\ttfamily
  1410.3491}}].

\bibitem{Starobinsky:1985ibc}
A.~A. Starobinsky, \emph{{Multicomponent de Sitter (Inflationary) Stages and
  the Generation of Perturbations}}, {\emph{JETP Lett.} {\bfseries 42} (1985)
  152}.

\bibitem{Sasaki:1995aw}
M.~Sasaki and E.~D. Stewart, \emph{{A General analytic formula for the spectral
  index of the density perturbations produced during inflation}},
  \href{https://doi.org/10.1143/PTP.95.71}{\emph{Prog. Theor. Phys.} {\bfseries
  95} (1996) 71} [\href{https://arxiv.org/abs/astro-ph/9507001}{{\ttfamily
  astro-ph/9507001}}].

\bibitem{Wands:2000dp}
D.~Wands, K.~A. Malik, D.~H. Lyth and A.~R. Liddle, \emph{{A New approach to
  the evolution of cosmological perturbations on large scales}},
  \href{https://doi.org/10.1103/PhysRevD.62.043527}{\emph{Phys. Rev. D}
  {\bfseries 62} (2000) 043527}
  [\href{https://arxiv.org/abs/astro-ph/0003278}{{\ttfamily
  astro-ph/0003278}}].

\bibitem{Lyth:2004gb}
D.~H. Lyth, K.~A. Malik and M.~Sasaki, \emph{{A General proof of the
  conservation of the curvature perturbation}},
  \href{https://doi.org/10.1088/1475-7516/2005/05/004}{\emph{JCAP} {\bfseries
  05} (2005) 004} [\href{https://arxiv.org/abs/astro-ph/0411220}{{\ttfamily
  astro-ph/0411220}}].

\bibitem{Lyth:2005fi}
D.~H. Lyth and Y.~Rodriguez, \emph{{The Inflationary prediction for primordial
  non-Gaussianity}},
  \href{https://doi.org/10.1103/PhysRevLett.95.121302}{\emph{Phys. Rev. Lett.}
  {\bfseries 95} (2005) 121302}
  [\href{https://arxiv.org/abs/astro-ph/0504045}{{\ttfamily
  astro-ph/0504045}}].

\bibitem{Vernizzi:2006ve}
F.~Vernizzi and D.~Wands, \emph{{Non-gaussianities in two-field inflation}},
  \href{https://doi.org/10.1088/1475-7516/2006/05/019}{\emph{JCAP} {\bfseries
  05} (2006) 019} [\href{https://arxiv.org/abs/astro-ph/0603799}{{\ttfamily
  astro-ph/0603799}}].

\bibitem{Byrnes:2008wi}
C.~T. Byrnes, K.-Y. Choi and L.~M.~H. Hall, \emph{{Conditions for large
  non-Gaussianity in two-field slow-roll inflation}},
  \href{https://doi.org/10.1088/1475-7516/2008/10/008}{\emph{JCAP} {\bfseries
  10} (2008) 008} [\href{https://arxiv.org/abs/0807.1101}{{\ttfamily
  0807.1101}}].

\bibitem{Wands:2010af}
D.~Wands, \emph{{Local non-Gaussianity from inflation}},
  \href{https://doi.org/10.1088/0264-9381/27/12/124002}{\emph{Class. Quant.
  Grav.} {\bfseries 27} (2010) 124002}
  [\href{https://arxiv.org/abs/1004.0818}{{\ttfamily 1004.0818}}].

\bibitem{Meng:2022ixx}
D.-S. Meng, C.~Yuan and Q.-g. Huang, \emph{{One-loop correction to the enhanced
  curvature perturbation with local-type non-Gaussianity for the formation of
  primordial black holes}},
  \href{https://doi.org/10.1103/PhysRevD.106.063508}{\emph{Phys. Rev. D}
  {\bfseries 106} (2022) 063508}
  [\href{https://arxiv.org/abs/2207.07668}{{\ttfamily 2207.07668}}].

\bibitem{Lyth:2006gd}
D.~H. Lyth, \emph{{Non-gaussianity and cosmic uncertainty in curvaton-type
  models}}, \href{https://doi.org/10.1088/1475-7516/2006/06/015}{\emph{JCAP}
  {\bfseries 06} (2006) 015}
  [\href{https://arxiv.org/abs/astro-ph/0602285}{{\ttfamily
  astro-ph/0602285}}].

\bibitem{Kumar:2009ge}
J.~Kumar, L.~Leblond and A.~Rajaraman, \emph{{Scale Dependent Local
  Non-Gaussianity from Loops}},
  \href{https://doi.org/10.1088/1475-7516/2010/04/024}{\emph{JCAP} {\bfseries
  04} (2010) 024} [\href{https://arxiv.org/abs/0909.2040}{{\ttfamily
  0909.2040}}].

\bibitem{DeLuca:2022rfz}
V.~De~Luca and A.~Riotto, \emph{{A note on the abundance of primordial black
  holes: Use and misuse of the metric curvature perturbation}},
  \href{https://doi.org/10.1016/j.physletb.2022.137035}{\emph{Phys. Lett. B}
  {\bfseries 828} (2022) 137035}
  [\href{https://arxiv.org/abs/2201.09008}{{\ttfamily 2201.09008}}].

\bibitem{Fumagalli:2019noh}
J.~Fumagalli, S.~Garcia-Saenz, L.~Pinol, S.~Renaux-Petel and J.~Ronayne,
  \emph{{Hyper-Non-Gaussianities in Inflation with Strongly Nongeodesic
  Motion}}, \href{https://doi.org/10.1103/PhysRevLett.123.201302}{\emph{Phys.
  Rev. Lett.} {\bfseries 123} (2019) 201302}
  [\href{https://arxiv.org/abs/1902.03221}{{\ttfamily 1902.03221}}].

\bibitem{Garcia-Saenz:2019njm}
S.~Garcia-Saenz, L.~Pinol and S.~Renaux-Petel, \emph{{Revisiting
  non-Gaussianity in multifield inflation with curved field space}},
  \href{https://arxiv.org/abs/1907.10403}{{\ttfamily 1907.10403}}.

\bibitem{Bjorkmo:2019qno}
T.~Bjorkmo, R.~Z. Ferreira and M.~C.~D. Marsh, \emph{{Mild Non-Gaussianities
  under Perturbative Control from Rapid-Turn Inflation Models}},
  \href{https://doi.org/10.1088/1475-7516/2019/12/036}{\emph{JCAP} {\bfseries
  12} (2019) 036} [\href{https://arxiv.org/abs/1908.11316}{{\ttfamily
  1908.11316}}].

\bibitem{Ferreira:2020qkf}
R.~Z. Ferreira, \emph{{Non-Gaussianities in models of inflation with large and
  negative entropic masses}},
  \href{https://doi.org/10.1088/1475-7516/2020/08/034}{\emph{JCAP} {\bfseries
  08} (2020) 034} [\href{https://arxiv.org/abs/2003.13410}{{\ttfamily
  2003.13410}}].

\bibitem{Wang:2022eop}
D.-G. Wang, G.~L. Pimentel and A.~Ach\'ucarro, \emph{{Bootstrapping Multi-Field
  Inflation: non-Gaussianities from light scalars revisited}},
  \href{https://arxiv.org/abs/2212.14035}{{\ttfamily 2212.14035}}.

\bibitem{Werth:2023pfl}
D.~Werth, L.~Pinol and S.~Renaux-Petel, \emph{{Cosmological Flow of Primordial
  Correlators}},  \href{https://arxiv.org/abs/2302.00655}{{\ttfamily
  2302.00655}}.

\bibitem{Iarygina:2023msy}
O.~Iarygina, M.~C.~D. Marsh and G.~Salinas, \emph{{Non-Gaussianity in
  rapid-turn multi-field inflation}},
  \href{https://arxiv.org/abs/2303.14156}{{\ttfamily 2303.14156}}.

\bibitem{Elliston:2012ab}
J.~Elliston, D.~Seery and R.~Tavakol, \emph{{The inflationary bispectrum with
  curved field-space}},
  \href{https://doi.org/10.1088/1475-7516/2012/11/060}{\emph{JCAP} {\bfseries
  11} (2012) 060} [\href{https://arxiv.org/abs/1208.6011}{{\ttfamily
  1208.6011}}].

\end{thebibliography}\endgroup
\bibliographystyle{JHEP}

\end{document}